\documentclass[submission,copyright,creativecommons]{eptcs}
\providecommand{\event}{arxiv} 
\usepackage{quiver}
\usepackage{breakurl}             

\title{
	Cartesian institutions with evidence: 
	Data and system modelling with diagrammatic constraints and generalized sketches
}
\author{Zinovy Diskin 
	\institute{McMaster University, Ontario, Canada}
	\email{diskinz@mcmaster.ca}
}
\def\titlerunning{Generalized sketches}
\def\authorrunning{Z. Diskin}

\usepackage{bm}
\DeclareMathAlphabet{\mathantt}{OT1}{antt}{li}{it}
\DeclareMathAlphabet{\mathpzc}{OT1}{pzc}{m}{it}

\usepackage{amsmath,mleftright,mathtools}
\DeclareMathAlphabet\mathbfcal{OMS}{cmsy}{b}{n}
\usepackage[mathscr]{euscript}
 \let\mathscr\relax
\usepackage[scr]{rsfso}

\DeclareMathSymbol{\cmemptyset}{\mathord}{symbols}{59}   

\usepackage[utf8]{inputenc}
\usepackage[T1]{fontenc}
\usepackage[english]{babel}

\usepackage{color}
\usepackage{multirow}
\usepackage{tabularx}
\usepackage{booktabs}

\usepackage{longfbox}

\usepackage{amsmath}
\usepackage{amsthm}
\usepackage{amssymb}
\usepackage{amsfonts}

\let\saveamalg\amalg
\let\amalg\relax
\usepackage{mathabx}
\let\amalg\saveamalg

\usepackage{listings}

\usepackage{mathrsfs}
\usepackage{stmaryrd}
\usepackage{mathtools}
\usepackage{bbm}

\usepackage{url}
\usepackage{enumitem}
\usepackage[,all,cmtip]{xy}
\usepackage{wrapfig}
\usepackage{graphicx}
\usepackage{tabularx}
\usepackage{subfig}

\usepackage{pigpen}
\usepackage{tikz}
\usetikzlibrary{cd}
\usetikzlibrary{decorations.pathmorphing}
\usetikzlibrary{shapes,shapes.arrows,shapes.geometric}
\usetikzlibrary{calc}
\usetikzlibrary{positioning}
\pgfdeclarelayer{background}
\pgfdeclarelayer{overlay}
\pgfsetlayers{background,main,overlay}
\usepackage{listings}
\usepackage{booktabs}

\usepackage[normalem]{ulem}
\usepackage{xifthen}
\usepackage{enumitem}   
\usepackage{bigints}
\usepackage{bm}
\usepackage{booktabs}
\usepackage{diagbox}
\usepackage{multirow}
\usepackage{etoolbox,environ}

\let\saveamalg\amalg
\let\amalg\relax
\usepackage{mathabx}  
\let\amalg\saveamalg

\newcommand{\mor}{morphism}
\newcommand{\moric}{morphic}

\newenvironment{mygroup}{ { }{ } }

\newcommand{\pblemma}{PB lemma}

\newcommand{\imgarrow}{\enma{\cdot\,\rightarrow\,\cdot}}

\newcommand{\liftedX}[1]{\enma{\ovr{#1}}}

\newcommand{\opp}{\oppind}

\newcommand{\opX}[1]{\enma{#1^\oppind}}

\newcommand{\pullbackedX}[1]{\enma{\ovr{#1}}}
\newcommand{\pullbackedXindexedY}[2]{\enma{(\pullbackedX{#1})_{#2}}}

\newcommand{\mysubsubXY}[2]{\smallskip\noindent{\bf{#1\, #2}}}

\newcommand{\carrier}[1]{\enma{\lvert{#1}\rvert}}

\newcommand{\semmXUD}[3]{\enma{\semm{#1}^{#2}_{#3}}}

\newcommand{\Instcat}{\enma{\pmb{Inst}}}
\newcommand{\InstcatX}[1]{\enma{\Instcat(#1)}}

\newcommand{\ensureamth}[1]{\ensuremath{#1}}

\newcommand{\Ob}{\enma{\mathsf{Ob}}}
\newcommand{\Arr}{\enma{\mathsf{Arr}}}

\newcommand\dom{\enma{\mathsf{dom}}}
\newcommand\cod{\enma{\mathsf{cod}}}
\newcommand\domop{\enma{\mathsf{dom}^\oppind}}

\newcommand{\oppind}{\enma{\mathsf{op}}}

\newcommand{\righthookarrow}{\hookrightarrow}

\newcommand{\proj}[1]{\enma{\nfont{pr}_{#1}}}
\renewcommand{\proj}[1]{\enma{|_{{#1}}}}

\newcommand{\enma}[1]{\ensuremath{#1}}
\newcommand{\bfem}[1]{{\bf\em #1}}
\newcounter{defCounter}

\newcommand\tcat{\e{I\!\!I}}
\renewcommand\tcat{\enma{\mathbf{\large 1}}}

\newcommand\qqquad{\qquad\quad}

\newcommand\so{\ensuremath{\mathsf{src}}}
\newcommand\ta{\ensuremath{\mathsf{trg}}}

\newcommand\id[1]{\opname{id}_{{#1}}}
\renewcommand\id{\opname{id}}

\newcommand\figref[1]{Fig.~\ref{fig:#1}}

\newcommand\defref[1]{Def.~\ref{def:#1}}

\newcommand{\sectref}[1]{Sect.~\ref{sec:#1}}

\newcommand{\eqdef}{\ensuremath{\stackrel{\mathrm{def}}{=}}}

\newcommand\respace{\!}
\newcommand\lsemm{\left[ \respace\left[ }
\newcommand\rsemm{\right]\respace\right]}
\renewcommand\respace{\!\!} 
\renewcommand{\lsemm}{ [ \respace [ }
\renewcommand{\rsemm}{] \respace ]}

\newcommand{\semm}[1]{\ensuremath{\llbracket{}{#1}\rrbracket{}}}
\renewcommand{\semm}[1]{\ensuremath{\lsemm{#1}\rsemm}}

\newcommand\timm{{\times}}

\newcommand\inn{{\in}}

\newcommand\ovr[1]{\ensuremath{\overline{#1}}}
\newcommand\und[1]{\ensuremath{\underline{#1}}}

\newcommand\funimg[1]{\enma{\mathsf{Img}({#1})}}

\renewcommand\restriction[2]{#1\left\lceil_{#2}\mathstrut\right.}
\renewcommand\restriction[2]{\ensuremath{#1{\upharpoonright}_{#2}\mathstrut}}
\newcommand{\restr}[2]{\restriction{#1}{#2}}

\usepackage[capitalize]{cleveref}
\crefformat{equation}{(#2#1#3)}
\crefrangeformat{equation}{(#3#1#4) to~(#5#2#6)}

\theoremstyle{definition}
\newtheorem{proposition}{Proposition}  
\newtheorem{definition}{Definition}  
\newtheorem{corollary}{Corollary}    

\newtheorem{theorem}{Theorem}
\newtheorem{propo}{Proposition}
\numberwithin{propo}{section} 
\newtheorem{lemma}{Lemma}
\numberwithin{lemma}{section} 
\newtheorem{corol}[propo]{Corollary}

\newtheorem{fact}{Fact}

\newtheorem{defN}{Definition}
\numberwithin{defN}{section} 
\newtheorem{constrN}{Construction}
\numberwithin{constrN}{section}

\newtheorem{example}{{\em Example}}

\theoremstyle{remark}
\newtheorem{remark}{Remark}

\theoremstyle{remark}

\newenvironment{defin}[2][]{
	\refstepcounter{defCounter}
	\begin{trivlist}
		\item[\hskip \labelsep {\bfseries Definition \arabic{defCounter} } {(#1)}{\label{def:#2}}]
	}%
	{\end{trivlist}}

\newenvironment{mydef}[1][]
{
	\ifthenelse{\equal{#1}{}}{ \begin{definition}  }{ \begin{definition}[{#1}]  }%
		}
		{
		\end{definition} 
	}

	\newenvironment{myprop}[1][]
	{
		\ifthenelse{\equal{#1}{}}{ \begin{proposition} \ }{ \begin{proposition}[{#1}] }%
			}
			{\end{proposition}}
		
			\newenvironment{mylem}[1][]
			{
				\ifthenelse{\equal{#1}{}}{ \begin{lemma} \ }{ \begin{lemma}[{#1}] }%
					}
					{\end{lemma}
				}

				\newenvironment{mycor}[1][]
				{
					\ifthenelse{\equal{#1}{}}{ \begin{corollary} \ }{ \begin{corollary}[{#1}] }%
						}
						{\end{corollary}}

					\newenvironment{myexa}[1][]
					{
						\ifthenelse{\equal{#1}{}}{ \begin{example}[] \ }{ \begin{example}[{#1}] }%
							}
							{\end{example}}

						\newenvironment{myrem}[1][]
						{
							\ifthenelse{\equal{#1}{}}{ \begin{remark}[] \ }{ \begin{remark}[{#1}] }%
								}
								{\end{remark}}
							
							\newenvironment{myconstr}[1][]
							{
								\ifthenelse{\equal{#1}{}}{ \begin{construct}[] \ }{ \begin{construct}[{#1}] }%
									}
									{\end{construct}}

\newcommand\comprehension[2]{\ensuremath{\{{#1}|\;{#2}\}}}
\newcommand\comprehensionColon[2]{\ensuremath{\{{#1}{:} \;{#2}\}}}  
\newcommand\compr[2]{\comprehension{#1}{#2}}
\newcommand\comprcolon[2]{\comprehensionColon{#1}{#2}}

\newcommand\comprfam[2]{\ensuremath{\left(\mathstrut{#1}\left|\;{#2}\right.\right)}}

\newcommand\noteq{\ne}

\newcommand{\bicatnameXY}[2]{\enma{\pmb{\mathcal{#1}}\!\catname{#2}}}
\newcommand{\dcatnameXY}[2]{\enma{\pmb{\mathbb{#1}}\catname{#2}}}

\newcommand{\laxcat}{\catname{Lax}}

\newcommand{\spanbicat}{\bicatnameXY{S}{pan}}

\newcommand{\catname}[1]{\enma{\underline{\mathrm{#1}}}}  
\renewcommand{\catname}[1]{\enma{\mathsf{#1}}}

\newcommand{\spancat}{\catname{Span}}

\newcommand{\twocat}{\catname{2}}

\newcommand{\setcat}{\catname{Set}}

\newcommand{\XCatcatY}[2]{\enma{\mathsf{#1}}}
\newcommand{\Catcat}{\XCatcatY{CAT}{}}

\newcommand{\catcat}{\catname{Cat}}  

\newcommand\graphcat{\catname{Graph}}
\newcommand\gracat{\graphcat}
\newcommand\rgracat{\catname{rGraph}}

\newcommand{\spandcat}{\enma{\mathbb{S}\mathsf{pan}}(\catcat)} 
\renewcommand{\spandcat}{\dcatnameXY{S}{pan}}

\newcommand{\spandiag}[5]{\enma{
		#1\xleftarrow{#4}#2\xrightarrow{#5}#3
}}

\newcommand{\lenscat}{\enma{\catname{Lens}}}

\newcommand\spacename[1]{{\ensuremath{\mathbf{#1}}}}
\renewcommand\spacename[1]{{\ensuremath{\pmb{#1}}}}

\newcommand{\spX}[1]{\spacename{#1}}
\newcommand{\spA}{\spX{A}}
\newcommand{\spB}{\spX{B}}

\newcommand{\niceletter}[1]{\ensuremath{{\mathcal{#1}}}}
\renewcommand{\niceletter}[1]{\ensuremath{{\mathbb{#1}}}}
\newcommand{\nicex}[1]{\niceletter{#1}}
\newcommand{\nia}{\nicex{A}}

\newcommand{\CC}{\bbx{C}}

\newcommand{\nmf}[1]{\ensuremath{\mathsf{#1}}}

\newcommand{\typename}[1]{\enma{\mathsf{#1}}}

\newcommand{\length}{\typename{length}}

\newcommand{\has}{\nmf{has}}

\newcommand{\of}{\nmf{of}}

\newcommand{\covers}{\nmf{covers}}

\newcommand{\drives}{\nmf{drives}}

\newcommand{\Vehicle}{\nmf{Vehicle}}
\newcommand{\veh}{\nmf{veh}}
\newcommand{\Driver}{\nmf{Driver}}

\newcommand{\Wheel}{\nmf{\Wheel}}
\newcommand{\wh}{\nmf{wh}}

\newcommand{\bdate}{\nmf{bdate}}
\newcommand{\name}{\nmf{name}}

\newcommand{\Date}{\nmf{Date}}
\newcommand{\String}{\nmf{String}}
\newcommand{\Int}{\nmf{Int}}

\renewcommand{\spandiag}[5]{\enma{#1\xleftarrow{#2}#3\xrightarrow{#4}#5}}
\newcommand{\trirar}[3]{\ensuremath{#1{:}\;#2 \Rrightarrow #3}}

\newcommand{\frar}[3]{\ensuremath{{#1}{:}\;{#2}\rightarrow {#3}}}

\newcommand\frarxy[2]{\ensuremath{{#1}\rightarrow {#2}} }  
\newcommand\flarxy[2]{\ensuremath{{#1}\leftarrow {#2}}}

\newcommand\spanrar[3]{\enma{{#1}\!:{#2}\nrightarrow{#3}}}

\newcommand{\spanlar}[3]{\enma{{#1}\!:{#2}\nleftarrow{#3}}}

\newcommand\drar[3]{\mbox{$#1\!:#2\Rightarrow #3$}} 

\newcommand\flar[3]{\ensuremath{{#1}{:}~{#2\leftarrow {#3} }}}  

\usepackage{diagrams}

\newarrowhead{bt}\blacktriangleright\blacktriangleright\blacktriangleright\blacktriangleright
\newarrowtail{bt}\blacktriangleleft\blacktriangleleft\blacktriangleleft\blacktriangleleft
\newarrowmiddle{bar}\rtbar\ltbar\dtbar\utbar
\newarrowmiddle{|}\vert\vert--
\newarrowmiddle{||}\Vert\Vert==
\newarrowmiddle{=}==\Vert\Vert
\newarrowtail{=}==\Vert\Vert
\newarrowhead{l}\langle\langle\langle\langle 
\newarrowhead{r}\rangle\rangle\rangle\rangle 
\newarrowfiller{=}==\Vert\Vert
\newarrowfiller{d}\cdot\cdot\cdot\cdot
\newarrowmiddle{x}****
\newarrowmiddle{b}\bullet\bullet\bullet\bullet
\newarrowtail{b}\bullet\bullet\bullet\bullet
\newarrowmiddle{3}\equiv\equiv\vfthree\vfthree
\newarrowfiller{3}\equiv\equiv\vfthree\vfthree
\newarrowtail{3}\equiv\equiv\vfthree\vfthree
\newarrowtail{<=}\Leftarrow\Rightarrow{\@cmex7E}{\@cmex7F}
\newarrowfiller{bold}{\bf-}{\bf-}{\bf|}{\bf|}  
\newarrowfiller{o}{\hho}{\circ}{\circ}{\circ}  
\newarrowmiddle{>}\rtla\ltla\dtla\utla
\newarrowmiddle{d}\cdot\cdot\cdot\cdot

\newarrow{To}----{>}
\newarrow{Toang}----{->}
\newarrow{BMapsto}b---{>}
\newarrow{Mapsto}{|}---{>}
\newarrow{Mapsto}{b}---{>}
\newarrow{Dermapsto}{|}{dash}{}{dash}{>}
\newarrow{Dermapsto}{b}{dash}{}{dash}{>}
\newarrow{Dashmapsto}{b}{dash}{dash}{dash}{>}
\newarrow{Dotmapsto}{b}...{>}
\newarrow{Maps}b----
\newarrow{ID}33333

\newarrow{Dashes}{}{dash}{}{dash}{}
\newarrow{Dots}.....

\newarrow{EQ}=====
\newarrow{IDto}3333r  
\newarrow{NRelto}--+-{->}
\newarrow{Relto}--b-{->}
\newarrow{BSpanto}--b-{->} 
\newarrow{Embed}>---{>}
\newarrow{Embed}{blacktriangle}---{>}
\newarrow{EmbedDashed}{blacktriangle}{dash}{dash}{dash}{>}
\newarrow{Embdash}{blacktriangle}{dash}{dash}{dash}{>}
\newarrow{Embdot}{blacktriangle}...{>}
\newarrow{FEmb}{blacktriangle}---{>}
\newarrow{WEmb}{blacktriangle}---{>}
\newarrow{WEmbDot}{blacktriangle}...{>}
\newarrow{FEmbDot}{blacktriangle}...{>}
\newarrow{FEmb}{blacktriangle}---{blacktriangle}
\newarrow{WEmb}{blacktriangle}---{blacktriangle}
\newarrow{WEmbDot}{blacktriangle}...{blacktriangle}
\newarrow{FEmbDot}{blacktriangle}...{blacktriangle}
\newarrow{WEmbDash}{blacktriangle}{dash}{dash}{dash}{blacktriangle}
\newarrow{FEmbDash}{blacktriangle}{dash}{dash}{dash}{blacktriangle}

\newarrow{Dashto}{}{dash}{}{dash}{>} 
\newarrow{Dertoda}{}{dash}{}{dash}{>} 
\newarrow{Dotto}....{>} 
\newarrow{Dertodo}....{>} 

\newarrow{RDiagto}3333{r}
\newarrow{RDiagderto}{}{3}{}{3}{r}
\newarrow{LDiagto}3333{l}
\newarrow{LDiagderto}{}{3}{}{3}{l}
\newarrow{Mapsderto}{b}{dash}{}{dash}{>}
\newarrow{Into}C---{->}
\newarrow{Indashto}C{dash}{}{dash}{>}
\newarrow{Derinto}C{dash}{}{dash}{>}
\newarrow{Congruent}33333
\newarrow{Cover}----{triangle}
\newarrow{Monic}{>}---{>}
\newarrow{Isoto}{>}---{triangle}
\newarrow{ISA}===={=>}
\newarrow{Isato}===={=>}
\newarrow{Doubleto}===={=>}
\newarrow{ClassicMapsto}{|}--->
\newarrow{Entail}{|}----
\newarrow{PArrow}{o}--->
\newarrow{MArrow}----{>>}
\newarrow{MPArrow}{o}---{>>}
\newarrow{Ogogoto}oooo{->}
\newarrow{Metato}--3->
\newarrow{MetaMapto}{|}-3->
\newarrow{OneToMany}+---{o}
\newarrow{HalfDashTo}{}{dash}{}-{>>}
\newarrow{DLine}=====
\newarrow{Line}-----
\newarrow{Tline}33333
\newarrow{Dashline}{}{dash}{}{dash}{}
\newarrow{Dotline}{}{o}{}{o}{}
\newarrow{Curlyto}{curlyvee}--->           
\newarrow{Bito}<--->
\newarrow{Bito}{<}---{>}
\newarrow{Bidito}{<}==={>}
\newarrow{Bitrito}{bt}333{bt}
\newarrow{Corrto}<--->
\newarrow{Dercorrto}{<}{dash}{}{dash}{>}

\newarrow{Instofto}{curlyvee}...{>}
\newarrow{Instofderto}{curlyvee}{dash}{}{dash}{->}
\newarrow{Upd}===={=>}
\newarrow{Derupdto}{}{dash}{}{dash}{>}
\newarrow{Mchto}----{>}
\newarrow{Dermchto}{}{dash}{}{dash}{>}
\newarrow{Viewto}----{>->}
\newarrow{Derviewto}{}{dash}{}{dash}{>>}
\newarrow{Viewto}===={=>}
\newarrow{Hetmchto}===={=>}
\newarrow{Derviewto}{=}{}{=}{}{=>}
\newarrow{Idleto}===={=>}
\newarrow{Deridleto}{=}{}{=}{}{=>}

\newlength{\StatArrBody}
\newlength{\NodeFrameThickness}
\newcommand{\mynoteColored}[3]{
	\fbox{\bfseries\sffamily\scriptsize#1}
	{\small$\blacktriangleright$
		\textcolor{#3}{\textsf{\emph{#2}}}
		$\blacktriangleleft$}}

\newcommand{\zd}[1]{\mynoteColored{ZD}{#1}{red}}
\newcommand{\zdzd}[1]{\mynoteColored{ZD}{#1}{red}}

\newcommand{\PAST}[1]{\mynoteColored{PS}{#1}{red}}

\newcommand\mymarginnew[1]{%
                     \marginpar{\hspace{-5em}
                     	\textit{\footnotesize\textcolor{blue} {#1}}}
                     }

\newcommand\hlight[2]{%
\textcolor{blue} {#1} 
 {\mymarginnew{#2}} 
 }

\newcommand\zdhl[2]{\hlight{#1}{ZD:~#2}}

\newcommand\zdnewwno[3]{}

\newcommand\newno[2]{}

\newcommand{\myendinputX}[1]{
	\textcolor{red}{{\sc Stuff below Endinput:}} 
	\textcolor{blue}{#1} \newline ~~\rule{0cm}{0.5cm}
	 }

\newcommand{\END}{\zd{the END is inserted HERE}\end{document}}

\newcommand{\zddel}[1]{}  
\renewcommand{\PAST}[1]{}

\newtoggle{long}
\NewEnviron{includeLong}
{\iftoggle{long}{\BODY}{}}
\NewEnviron{includeShort}
{\iftoggle{long}{}{\BODY}}

\toggletrue{long}

\newcommand{\shortlong}[2]{#1}   

\newcommand{\ActTR}[2]{#1}   

\newcommand{\traceLong}[1]{\mynoteColored{Tracing incLong}{#1}{blue}}
\renewenvironment{includeLong}{\par \traceLong{includeLong Starts here}}{\par\traceLong{includeLong Ends here}}
\renewenvironment{includeLong}{}{}

\newcommand{\satty}{satisfiability}

\newcommand\eg{e.g.}
\newcommand\ie{i.e.}

\newcommand\wrt{w.r.t.}
\newcommand\etc{{\em etc}}
\newcommand\etal{{\em et al}}

\newcommand{\transf}{transformation}
\newcommand{\nattra}{natural \transf}

\newcommand\mty{multiplicity}
\newcommand\mties{multiplicities}

\newcommand\hical{hierarchical}
\newcommand\caty{category}
\newcommand{\caties}{categories}
\newcommand\hcaty{h-category}
\newcommand\cats{categories}
\newcommand\hcats{h-categories}

\newcommand\asson{association}

\newcommand\req{requirement}

\newcommand\corring{corresponding}

\newcommand\fwk{framework}

\newcommand{\dernode}[1]{\color{blue}\enma{#1}}

\newcommand{\odernode}[1]{\color{orange}\enma{#1}}
\newcommand{\orangenode}[1]{\color{orange}\enma{#1}}

\newcommand{\gapnamegap}[3]{\enma{\hspace{#1ex}{#2}\hspace{#3ex}}}

\usepackage{tikz-cd}
\usetikzlibrary{calc}
\usetikzlibrary{decorations.text, calc, arrows.meta}
\usetikzlibrary{decorations.markings, arrows.meta}
\usetikzlibrary{positioning,shapes,shadows}
\usetikzlibrary{decorations.pathmorphing}
\tikzcdset{arrow style=math font}

\tikzset{/tikz/inner sep=0.125ex}

\tikzset{
	marrow/.style={decoration={markings,mark=at position 0.5 with {\arrow{#1}}}, postaction=decorate}
}

\tikzset{commutative diagrams/.cd, 
	corr/.style = {negated, mapsto}, 
	Corr/.style = {leftrightarrow, "\bullet" description},
	Corrbi/.style = {leftrightarrow, "\circ" description},
	corr/.style = {mapsto, "\bullet" description},
	corr/.style = {mapsto, "\circ" description},
	corr/.style = {leftrightarrow, "\circ" description},
	corr/.style = {negated, mapsto},
	corrinst/.style = {negated, mapsto},
	Span/.style = {negated, rightarrow},
	span/.style = {negated, rightarrow},
	spanop/.style = {negated, leftarrow},
	Corr/.style = {negated, rightarrow},
	Spanop/.style = {negated, leftarrow},
	Corrop/.style = {negated, leftarrow},
	spaninst/.style = {negated, mapsto},
	spaninstop/.style = {negated, mapsfrom},
	corrop/.style = {negated, mapsfrom},
	 prop/.style = {rightarrow,  "\bullet" description},
	 Prop/.style = {rightarrow,  "\bullet" description, thick},
	 propop/.style = {leftarrow,  "\bullet" description},
	 Propop/.style = {leftarrow,  "\bullet" description, thick},
	biprop/.style ={leftrightarrow, "\bullet" description},
	corrbi/.style = {leftrightarrow},
	derived/.style = {dashed, blue},
	Derived/.style = {thick, dashed, blue},
	blackderived/.style={dashed, }
	bluederived/.style={dashed, blue}
	oderived/.style = {dashed, orange},
	Oderived/.style = {dashed, orange},
	gderived/.style = {dashed, darkgreen},
	gderived/.style = {dashed, caribbeangreen},
	oDerived/.style = {thick, dashed, orange},
	dotDerived/.style = {dotted, blue, thick },
	dotderived/.style = {dotted, blue},   
	dotoderived/.style = {dotted, orange, thick},
	dotOderived/.style = {dotted, orange, thick},
	oDotted/.style = {dotted, orange, thick},
	odotted/.style = {dotted, orange},
	oDashed/.style = {dashed, orange, thick},
	odashed/.style = {dashed, orange},
	s./.style = {s,.},
	notderived/.style = {blue},
}

\tikzset{negated/.style={
		decoration={markings,
			mark= at position 0.4 with {
				\node[transform shape] (tempnode) {{\footnotesize$\prime$}};
			}
		},
		postaction={decorate}
	}
}

\newcommand{\diagpar}[2]{[
	row sep = 2em, %
	column sep = 2em, %
	ampersand replacement=\&
	]}

\tikzstyle{Dot}=[fill=black, draw=black, shape=circle, scale=0.8]
\tikzstyle{Database}=[fill=white, draw=black, shape=cylinder, shape border rotate=90, aspect=0.25]
\tikzstyle{Class}=[fill=white, draw=black, shape=rectangle split, rectangle split parts=2, minimum width=1cm, minimum height=0.3cm, align=left]
\tikzstyle{ClassWithOps}=[fill=white, draw=black, shape=rectangle split, rectangle split parts=3, minimum width=1cm, minimum height=0.3cm, align=left]
\tikzstyle{Box}=[fill=white, draw=black, shape=rectangle]
\tikzstyle{ProcessStage}=[fill=white, draw=black, shape=signal, align=center, signal from=west, signal to=east]
\tikzstyle{Cloudy}=[fill=white, draw=black, shape=cloud, cloud puffs=10, cloud puff arc=120, aspect=2, inner ysep=1em]
\tikzstyle{BpmnTask}=[fill=white, draw=black, shape=rectangle, minimum width=4em, minimum height=2em, align=center, rounded corners]
\tikzstyle{BpmnGateway}=[fill=white, draw=black, shape=diamond, inner sep=0pt, minimum width=2em, minimum height=2em, thick]
\tikzstyle{BpmnEvent}=[fill=white, draw=black, shape=circle, minimum width=1.5em, minimum height=1.5em]
\tikzstyle{BpmnEndEvent}=[BpmnEvent, ultra thick]
\tikzstyle{BpmnIntermediateEvent}=[BpmnEvent, double]
\tikzstyle{BpmnMessage}=[o->, >=open triangle 45, dashed]
\tikzstyle{DiagramLabel}=[text=black!50]

\tikzstyle{Arrow}=[->]
\tikzstyle{DashArrow}=[->, dashed]
\tikzstyle{DotArrow}=[->, densely dotted]
\tikzstyle{DotLine}=[-, densely dotted]
\tikzstyle{Reference}=[->, >=angle 45]
\tikzstyle{Aggregation}=[->, >=open diamond, draw=black, fill=white]
\tikzstyle{Composition}=[->, >=diamond, draw=black, fill=black]
\tikzstyle{Inheritance}=[->, >=open triangle 45]
\tikzstyle{Realization}=[->, >=open triangle 45, dashed]
\tikzstyle{BpmnSequence}=[->, >=triangle 45]
\tikzstyle{BpmnMessageFlow}=[o->, >=open triangle 45, dashed]
\tikzstyle{BpmnAssociation}=[->, >=angle 45, densely dotted]
\tikzstyle{DiagramArity}=[-, dotted, draw=black!50]

\makeatletter

\newcommand{\pto}{}
\newcommand{\pgets}{}
\DeclareRobustCommand{\pto}{\mathrel{\mathpalette\p@to@gets\to}}
\DeclareRobustCommand{\pgets}{\mathrel{\mathpalette\p@to@gets\gets}}
\newcommand{\p@to@gets}[2]{%
	\ooalign{\hidewidth$\m@th#1\mapstochar\mkern5mu$\hidewidth\cr$\m@th#1\to$\cr}%
}
\makeatother

\newcommand{\semantics}[1]{\llbracket{}{#1}\rrbracket{}}

\newcommand{\arrow}[3]{{#2}{:}\;{#1} \to {#3}}

\newcommand{\partiallonginlinearrow}[3]{\xymatrix{{#1} \ar@{-^{`}}[r]|{#2} & {#3}}}

\newcommand{\relationlonginlinearrow}[3]{\xymatrix{{#1} \ar@{-|>}[r]|{#2} & {#3}}}

\newcommand{\longinlininclusionearrow}[3]{\xymatrix{{#1} \ar@{^{(}->}[r]|{#2} & {#3}}}

\newcommand{\longinlinemonoarrow}[3]{\xymatrix{{#1} \ar@{>->}[r]|{#2} & {#3}}}

\newcommand{\longinlineepiarrow}[3]{\xymatrix{{#1} \ar@{->>}[r]|{#2} & {#3}}}

\newcommand{\compose}[2]{{#1} \fatsemi {#2} }

\newcommand{\cat}[1]{\enma{\mathbb{#1}}}

\newcommand{\catArrows}[1]{\mathsf{Arr}_{#1}}
\newcommand{\catObj}[1]{\mathsf{Ob}_{#1}}
\renewcommand{\catArrows}[1]{\enma{\mathsf{Arr}(#1)}}
\renewcommand{\catObj}[1]{\enma{\mathsf{Ob}(#1)}}

\newcommand{\cosliceCat}[2]{{#1}{\setminus}{#2}}

\newcommand{\metaInclusion}{\enma{
		\xymatrix@C=2em{
			\\
			{} \ar@{^{(}->}[r] &{} 
			\\
		}
}}

\newcommand{\PPOO}{\enma{\PP{\star}\OO}}
\newcommand{\PPOE}{\enma{\PP{||}\OO_E}}	
\renewcommand{\PPOO}{\enma{\PP{||}\OO}}
\renewcommand{\PPOO}{\enma{\PP{||}\OO_E}}

	\newcommand{\etanota}{$\eta$-notation}

\newcommand{\PPstar}[1]{\enma{\PP^{\star}}} 

\newcommand{\Par}[1]{\enma{P^\arfun}}
\newcommand{\Pprimar}[1]{\enma{P"^\arfun}}

\newcommand{\OO}{\opsig}
\newcommand{\PP}{\predsig}

\newcommand{\skecat}[2]{\enma{{#1}||{#2}}}

\newcommand{\skecatgp}{\skecat{\cat{G}}{\predsig}}

\newcommand{\skecatgpo}{\skecat{\cat{G}}{\PPOO}}
\newcommand{\skecatgpmono}{\enma{\GG|\predsig}}

\newcommand{\skettografun}[1]{\enma{\widecheck{#1}}}

\newcommand{\stog}[1]{\enma{\skettografun{#1}}}
\renewcommand{\stog}[1]{\enma{G^S}}

\newcommand{\plabel}{\enma{\mathit{pred}}}
\renewcommand{\plabel}{\enma{{\mathit{\labell}}}}

\newcommand{\fancydom}[1]{\enma{\bigcirc#1}}
\newcommand{\fancycod}[1]{\enma{#1\bigcirc}}

\newcommand{\E}{\nfont{E}}

\newcommand{\monicsubcat}[1]{\enma{#1_{\mathrm{monic}}}}
\newcommand{\monicsubcatbr}[1]{\enma{(#1)_{\mathrm{monic}}}}

\newcommand{\NN}{\enma{\overrightarrow{N}}}
\renewcommand{\NN}{\enma{[0,N)}}
\newcommand{\NNop}{\opcat{\NN}}

\newcommand\forgetind{\enma{{-}\mathrm{indexing}}}

\newcommand{\opcat}[1]{\enma{#1^{\oppind}}}
\newcommand{\oppCat}[1]{\enma{{#1}^{\oppind}}}

\newcommand{\opCat}[1]{\oppCat}
\newcommand{\predsigop}{\opcat{\predsig}}

\newcommand\GGop{\opcat{\GG}}

\newcommand{\predsymbolX}[1]{\enma{\mathsf{[#1]}}}

\newcommand{\commsqpred}{\predsymbolX{=_{sqr}}}
\newcommand{\monicpred}{\predsymbolX{monic}}

\newcommand{\jmpred}{\predsymbolX{jm}}

\newcommand{\pbpred}{\predsymbolX{pb}}

\newcommand{\iistarp}{\enma{{\II{\star}_{\arfun}\predsig}}}

\newcommand{\HH}{\enma{\cat{H}}}
\renewcommand{\HH}{\enma{\mathcal{H}}}
\renewcommand{\HH}{\enma{\mathcal{G}}}
\renewcommand\SS{\enma{\mathbb{S}}}

\newcommand\II{\cat{H}}
\renewcommand\II{\enma{\mathcal{H}}}
\newcommand{\GG}{\enma{\mathbb{G}}}

\newcommand\op{\enma{\mathit{op}}}

\newcommand\comp{\,\compose{}{}}
\newcommand\carrcarr[1]{\enma{\carrier{\carrier{#1}}}}

\newcommand\depfun{\enma{\mathsf{hei}}}
\newcommand\heifun{\enma{\mathsf{hei}}}
\newcommand\Hei{\enma{\mathsf{Hei}}}

\newcommand\carr[1]{\carrier{#1}}

\newcommand{\predsig}{\enma{\mathcal{P}}}
\newcommand{\predicateSig}{\predsig}

\newcommand{\arfun}{\enma{\alpha}}

\newcommand{\arityOf}[1]{{#1}^\arfun}

\newcommand{\interpret}[2]{\semantics{#1}^{#2}}

\newcommand{\LDiagrcat}[1]{\enma{\mathbb{LD}\mathit{iagr}(#1)}}

\newcommand{\iQuery}[1]{\enma{\mathit{Index}(#1)}}

\newcommand{\iQueryEE}[1]{\enma{\mathit{Index}_{EE}}}
\newcommand{\iQueryEI}[1]{\enma{\mathit{Index}_{EI}}}

\newcommand{\labell}{\enma{\mathit{label}}}

\newcommand{\bind}{\enma{\mathit{bind}}}
\renewcommand{\bind}{\enma{\mathit{diagr}}}
\newcommand{\pbind}{\enma{\mathit{diagr}}}

\newcommand{\diagr}{\enma{\mathit{diagr}}}

\newcommand{\constrfun}[1]{\enma{{#1}_\mathrm{constr}}}

\newcommand{\opsig}{\enma{\mathcal{O}}}

\newcommand{\operationSig}{\opsig}

\newcommand{\discrete}[1]{\underline{#1}}

\newcommand\nfont[1]{\enma{\mathsf{#1}}}

\newcommand{\Graph}{\nfont{Graph}}
\newcommand{\graphind}{\nfont{graph}}

\newcommand{\Node}{\nfont{Node}}

\newcommand{\Arrow}{\nfont{Arrow}}

\newcommand{\opVar}{\enma{\phi}}
\newcommand\opvar{\opVar}

\newcommand{\opInterpret}[2]{\enma{\semm{{#1}}^{#2}}}

\newcommand{\dvar}{\enma{\delta}}
\newcommand{\dmorvar}{\enma{\Delta}}

\newcommand\arrcatC{\arrowCat{\CC}}
\newcommand\domC{\enma{\dom_{\CC}}}
\newcommand\codC{\enma{\cod_{\CC}}}
\renewcommand\domC{\dom} 
\renewcommand\codC{\cod} 

\renewcommand{\imgarrow}{\enma{\cdot\!\!\rightarrow\!\!\cdot}}

\newcommand{\imgmonic}{\enma{\cdot\!\!\rightarrowtail\!\!\cdot}}

\newcommand{\arrowCat}[1]{\enma{[\imgarrow, {#1}]}}

\newcommand{\monicarrowCat}[1]{\enma{[\imgmonic, \,{#1}]}}

\newcommand{\maCat}[1]{\monicarrowCat{#1}}

\newcommand{\ArrowDiag}{
\begin{tikzcd}[row sep = tiny, 
						ampersand replacement = \&]
A \ar[r, "r"] \& B 
\end{tikzcd}
}

\newcommand{\LoopDiag}{
	\begin{tikzcd}[row sep = small] 
		A\ar[loop right, "r",  loop right, distance=2em, %
		start anchor={[yshift=1ex]east}, end anchor={[yshift=-1ex]east}, 
		]
	\end{tikzcd}
}
\newcommand{\TwoParaArrowsDiag}{
\begin{tikzcd}[row sep = tiny, 
							ampersand replacement =\&]
A \ar[r,"r_1",yshift=0.25ex] \ar[r,"r_2" ',yshift=-0.25ex] \& B 
\end{tikzcd}
}
\newcommand{\e}{\enma{\mathsf{e}}}
\newcommand{\ecurry}{\enma{\e^\lambda}}

\newcommand{\get}{\enma{\mathsf{get}}}
\newcommand{\putl}{\enma{\mathsf{put}}}

\newcommand{\modelsdGS}{\enma{\modelsd_{G_S}}}
\newcommand{\modelsdCS}{\enma{\modelsd_{C_S}}}
\newcommand{\modelsdGSprim}{\enma{\modelsd_{G_{S'}}}}
\newcommand{\modelsdF}{\enma{\modelsd_f}}
\newcommand{\modelsdSigma}{\enma{\modelsd_{\Sigmaa}}}
\newcommand{\modelsdSigmaPrim}{\enma{\modelsd_{\Sigmaa'}}}

\newcommand{\carrsigop}{\opX{\carr{\csig}}}

\newcommand{\Satcat}{\nmf{Sat}}
\newcommand{\satascat}{\Satcat}
\newcommand{\Satcatphi}{\enma{\Satcat_\phi}}

\newcommand{\bmapc}{\enma{{\bmap_c}}}

\newcommand{\schemaind}{\enma{\mathsf{schema}}}
\newcommand{\schemaindgra}{\enma{\mathsf{schema}_{\mathrm{gr}}}}

\newcommand{\carrcdot}{\carr{\cdot}}

\newcommand{\triangleMorX}[1]{\enma{\apexX{#1}}}
\newcommand{\liftedTriangle}{\liftedX{\triangleMorX{d^\arfun}}}
\newcommand{\liftedTriangleDom}{\enma{d^{\arfun*}(x)}}
\renewcommand{\liftedTriangleDom}{\enma{x.\triangleMorX{d^\arfun}^*}}

\newcommand{\switchX}[1]{\enma{\kappa_{#1}}}

\newcommand{\morAsspanXdn}[1]{\enma{#1_{_\%}}}
\newcommand{\morAsspanXup}[1]{\enma{#1^{_\%}}}

\renewcommand{\id}{\enma{\mathsf{id}}}

\newcommand{\congg}{\enma{\cong}}
\newcommand{\gamm}{\enma{\alpha}}

\newcommand{\twip}{\enma{\propto}}
\renewcommand{\twip}{\enma{\alpha}}

\newcommand{\modelsd}{\enma{\vDash}}
\newcommand{\modesld}{\modelsd}
\newcommand{\modelsr}{\enma{\Dashv}}

\newcommand{\twimor}{\enma{\alpha}}

\newcommand{\CCsen}{\enma{\CC_\senind}}

\newcommand{\CCmod}{\enma{\CC_\modind}}

\newcommand{\twixsatXdn}[1]{\enma{\vDash^\twixSymbol_{#1}}}

\newcommand{\bangXdn}[1]{\enma{!_{#1}}}

\newcommand{\okTargetX}[1]{\PreinstcatX{#1}}
\newcommand{\okTargetsig}{\okTargetX{\csig}}
\newcommand{\OKobj}{\InstcatXdn{\csig}}
\newcommand{\ok}{\nmf{ok}}
\newcommand{\okArrow}{\ok}
\newcommand{\oksig}{\enma{\okArrow_\csig}}

\newcommand{\ellsig}{\ellX{\csig}}

\newcommand{\twimormor}{\enma{\twimor.\twimor}}

\newcommand{\twixdiagX}[1]{\enma{\mathbf{d}^\twixSymbol_#1}}
\newcommand{\spanfunX}[1]{\enma{\mathsf{Span}\,#1}}
\newcommand{\spanfuncongX}[1]{\enma{\mathsf{Span}_\cong\,#1}}

\newcommand{\SubprodOkDU}[2]{\enma{P_{#1}^{#2\ok}}}

\newcommand{\LDx}[1]{\enma{LD_{#1}}}

\newcommand{\LDG}{\LDx{G}}
\newcommand{\LDGprim}{\LDx{G'}}

\renewcommand{\LDx}[1]{\horcatX{#1}}
\newcommand{\satToPreinstX}[1]{\enma{p^{SM}_G}}

\newcommand{\XstardeltaUp}[1]{\enma{#1^*_\Delta}}
\newcommand{\fstardeltaUp}{\XstardeltaUp{f}}
\newcommand{\XstardeltaDn}[1]{\enma{#1_{*\Delta}}}
\newcommand{\fstardeltaDn}{\XstardeltaDn{f}}

\newcommand{\arrowcatdeltaX}[1]{\enma{#1^{\twocat\Delta}}}
\newcommand{\arrowcatdeltaGG}{\arrowcatdeltaX{\GG}}

\newcommand{\deltaarrowcatGG}{\arrowcatdeltaX{\GG}}

\newcommand{\arrowcatsigX}[1]{\enma{#1^{\twocat\csig}}}
\newcommand{\arrowcatsigGG}{\arrowcatsigX{\GG}}

\newcommand{\arrowcatdeltaGGcong}{\arrowcatdeltaX{\GG}}

\newcommand{\deltaslicecatGGG}{\deltaslicecatXY{\GG}{G}}
\newcommand{\deltaslicecatGGGprim}{\deltaslicecatXY{\GG}{G'}}

\newcommand{\slicecatdeltaXY}[2]{\enma{#1{/}_{\!\!{}_\Delta}#2}}
\newcommand{\slicecatdeltaGGG}{\slicecatdeltaXY{\GG}{G}}
\newcommand{\slicecatdeltaGGdot}{\slicecatdeltaXY{\GG}{\_}}
\newcommand{\slicecatdeltaGGGprim}{\slicecatdeltaXY{\GG}{G'}}
\newcommand{\slicecatdeltaGGGS}{\slicecatdeltaXY{\GG}{G_S}}
\newcommand{\slicecatdeltaGGGSprim}{\slicecatdeltaXY{\GG}{G_{S'}}}

\newcommand{\slicecatcsigXY}[2]{\enma{#1{/}_{\!\!{}_\csig}#2}}

\newcommand{\slicecatcsigGGG}{\slicecatcsigXY{\GG}{G}}

\newcommand{\slicecatsigGGdot}{\slicecatcsigXY{\GG}{\_}}

\newcommand{\slicecatsigGGGS}{\slicecatcsigXY{\GG}{G_S}}
\newcommand{\slicecatsigGGGSprim}{\slicecatcsigXY{\GG}{G_{S'}}}

\newcommand{\ssubset}{\;\subset\,}
\newcommand{\Obb}{\Ob\;}

\newcommand{\FCC}{FCC}
\renewcommand{\FCC}{CPB}

\newcommand{\hind}{\enma{\mathsf{h}}}
\newcommand{\vind}{\enma{\mathsf{v}}}
\renewcommand{\hind}{\enma{{S*}}}
\renewcommand{\vind}{\enma{{S'*}}}

\newcommand{\satsetFamilyXY}[2]{\comprfam{\models_#2\; \subset \Ob\; #1/#2\timm\ConstrsetXbra{#2}}{#2\in\Ob\, #1}}
\newcommand{\satsetFamilyGG}{\satsetFamilyXY{\GG}{G}}

\newcommand{\twistedSatcat}{\enma{\catname{TeSat}}}

\newcommand{\modelss}{\enma{\models\;}}
\newcommand{\driveVehOntologyXind}[1]{\enma{#1_\mathrm{DrvVeh}}}

\newcommand{\semmTofibTosemmX}[1]{\enma{\semm{#1}^{t_{\semm{.}}}}}

\newcommand{\satbaseXY}[2]{\enma{(#1, #1_\senind, #1_\modind, #2)}}
\newcommand{\satbaseCCE}{\satbaseXY{\CC}{E}}
\newcommand{\satbaseCC}{\satbaseX{\CC}}
\newcommand{\satbaseX}[1]{\enma{(#1, #1_\modind, #1^\e_\senind)}}

\newcommand{\satToSat}{\enma{\rho}}
\newcommand{\satToSatPrim}{\enma{\rho'}}

\newcommand{\Pobj}{\enma{P}}
\newcommand{\PobjPrim}{\enma{P'}}

\newcommand{\PobjX}[1]{\enma{P_{#1}}}
\newcommand{\PobjtwiX}[1]{\enma{P_{#1}^\twixSymbol}}

\newcommand{\okObj}{\enma{\catname{Ok}}}
\newcommand{\VerObj}{\enma{\mathit{V}}}
\newcommand{\ver}{\enma{\mathit{ver}}}
\newcommand{\twistedSatXdn}[1]{\enma{\vDash^\twixSymbol_{#1}}}

\newcommand{\modind}{\enma{\mathrm{mod}}}
\newcommand{\senind}{\enma{\mathrm{sen}}}
\newcommand{\conind}{\enma{\mathrm{con}}}

\newcommand{\projfunUD}[2]{\enma{p_{#2}^{#1}}}

\newcommand{\pEx}[1]{\projfunUD{E}{#1}}
\newcommand{\pMx}[1]{\projfunUD{M}{#1}}
\newcommand{\pSx}[1]{\projfunUD{S}{#1}}

\newcommand{\pMsig}{\projfunDU{\Sigmaa}{}}
\newcommand{\pSsig}{\projfunDUwithq{\Sigmaa}{}}
\newcommand{\pEsig}{\projfunDU{\Sigmaa}{E}}
\newcommand{\pMsigPrim}{\projfunDU{\Sigmaa'}{}}
\newcommand{\pSsigPrim}{\projfunDUwithq{\Sigmaa'}{}}
\newcommand{\pEsigPrim}{\projfunDU{\Sigmaa'}{E}}

\newcommand{\pEf}{\projfunDU{f_*}{E}}

\renewcommand{\pmod}{\projfunUD{}{\modind}}
\newcommand{\psen}{\projfunUD{}{\senind}}

\newcommand{\instfwk}{{\em inst}-\fwk}
\renewcommand{\instfwk}{institution \fwk}
\newcommand{\catnameSimpleX}[1]{\enma{\mathit{#1}}}
\newcommand{\Modcat}{\catnameSimpleX{Mod}}
\newcommand{\Sencat}{\catnameSimpleX{Sen}}
\newcommand{\eSencat}{\catnameSimpleX{eSen}}
\newcommand{\Evicat}{\catnameSimpleX{Evi}}
\newcommand{\Modcatsig}{\enma{\Modcat_\Sigmaa}}
\newcommand{\Sencatsig}{\enma{\Sencat_\Sigmaa}}
\newcommand{\eSencatsig}{\enma{\eSencat_\Sigmaa}}

\newcommand{\coddelta}{\enma{\cod^\Delta}}
\renewcommand{\cod}{\enma{\mathsf{cod}}}
\renewcommand{\dom}{\enma{\mathsf{dom}}}

\newcommand{\ModcatsigPrim}{\enma{\Modcat_{\Sigmaa'}}}
\newcommand{\SencatsigPrim}{\enma{\Sencat_{\Sigmaa'}}}
\newcommand{\eSencatsigPrim}{\enma{\eSencat_{\Sigmaa'}}}

\newcommand{\starsigX}[1]{\enma{#1_*^\csig}}

\newcommand{\stardeltaX}[1]{\enma{#1^*_\Delta}}
\newcommand{\fstarsig}{\starsigX{f}}

\newcommand{\fstardelta}{\stardeltaX{f}}

\newcommand{\Sigcat}{\catnameSimpleX{Sig}}
\newcommand{\Sigcatt}{\catname{Sig}}

\newcommand{\modelssig}{\enma{\vDash}}
\newcommand{\modelssigPrim}{\enma{\vDash'}}
\newcommand{\satrelsig}{\enma{\vDash_\Sigmaa}}
\newcommand{\satrelsigPrim}{\enma{\vDash_{\Sigmaa'}}}

\newcommand{\arrowcatX}[1]{\enma{#1^\twocat}}
\newcommand{\arrowcatCC}{\arrowcatX{\CC}}
\newcommand{\arrowcatCCcong}{\enma{\arrowcatX{\CC}_\cong}}
\newcommand{\arrowcatGGcong}{\enma{\arrowcatX{\GG}_\cong}}
\newcommand{\arrowcatGG}{\arrowcatX{\GG}}

\renewcommand{\HH}{\enma{\mathbb{H}}}
\renewcommand{\predsig}{\enma{\mathcal{C}}}
\renewcommand{\pblemma}{PB lemma}

\newcommand{\pbfunvX}[1]{\enma{\mathsf{pb}^\mathsf{v}_{#1}}}

\newcommand{\pbfunvG}{\pbfunvX{G}}

\newcommand{\pbfundeltavX}[1]{\enma{\mathsf{pb}^{\mathsf{v}\Delta}_{#1}}}

\newcommand{\pbfunvdeltaX}[1]{\enma{\mathsf{pb}^{\mathsf{v}\Delta}_{#1}}}

\newcommand{\key}{\nmf{key}}

\newcommand{\interopy}{interoperability}

\newcommand{\ovrrar}[1]{\enma{\overrightarrow{#1}}}

\newcommand{\revSatasarrowX}[1]{\ovrrar{\Dashv_{#1}}}
\newcommand{\revtwixSatasarrowX}[1]{\ovrrar{\Dashv^\twixSymbol_{#1}}}

\newcommand{\labfunSig}{\enma{\labfun^{\predsig}}}

\renewcommand{\arfun}{\enma{\mathsf{ar}}}

\newcommand{\XmapfunY}[2]{\enma{#1^#2_{\_}}}
\newcommand{\tmapfunY}[1]{\XmapfunY{\tmap}{#1}}

\newcommand{\tmapfunsig}{\tmapfunY{\predsig}}

\newcommand{\Skecat}{\catname{Ske}}

\newcommand{\sation}{satisfaction}

\newcommand{\vsfun}{\enma{\catname{es}}}

\renewcommand{\vsfun}{\enma{\vDash}}

\newcommand{\arrowlabXcolorY}[2]{\enma{\text{\textcolor{#2}{\footnotesize[#1]} } } }

\newcommand{\dfiblabelRed}{\arrowlabXcolorY{df}{red}}
\newcommand{\fiblabelRed}{\arrowlabXcolorY{fib}{red}}

\newcommand{\dfiblabelBlack}{\arrowlabXcolorY{df}{black}}
\newcommand{\fiblabelBlack}{\arrowlabXcolorY{fib}{black}}

\newcommand{\twixSymbol}{\enma{\between}}
\renewcommand{\twixSymbol}{\enma{\varpropto}}
\renewcommand{\twixSymbol}{\enma{\rtimes}}
\renewcommand{\twixSymbol}{\enma{\alpha}}

\newcommand{\twSubstX}[1]{\enma{\widetilde{#1}}}
\newcommand{\twSubstXsmall}[1]{\enma{\footnotesize\widetilde{#1}}}

\newcommand{\twix}[1]{\twistedX{#1}}

\newcommand{\twixU}[1]{\enma{#1^\twixSymbol}}
\newcommand{\twixD}[1]{\enma{#1_\twixSymbol}}
\renewcommand{\twix}[1]{\enma{\twixU{#1}}}

\newcommand{\satUD}[2]{\enma{\vDash^{#1}_{#2}}}

\newcommand{\specifin}{specification}

\newcommand{\setname}[1]{\enma{\mathit{#1}}}

\newcommand{\sysfunname}[1]{\enma{\mathit{#1}}}

\newcommand{\Sigmaa}{\enma{\Sigma}}

\newcommand{\ModsetX}[1]{\enma{\setname{Mod}_{#1}}}
\newcommand{\SensetX}[1]{\enma{\setname{Sen}_{#1}}}

\newcommand{\Modsetsig}{\ModsetX{\Sigmaa}}
\newcommand{\Sensetsig}{\SensetX{\Sigmaa}}
\newcommand{\ModsetsigPrim}{\ModsetX{\Sigmaa'}}
\newcommand{\SensetsigPrim}{\SensetX{\Sigmaa'}}

\renewcommand{\modelssig}{\enma{\vDash_\Sigmaa}}
\renewcommand{\modelssigPrim}{\enma{\vDash_{\Sigmaa'}}}

\newcommand{\senfun}{\sysfunname{sen}}
\renewcommand{\constrfun}{\sysfunname{constr}}

\newcommand{\modfun}{\sysfunname{mod}}

\newcommand{\satfun}{\funname{sat}}
\renewcommand{\satfun}{\enma{\vDash}}

\newcommand{\doublelift}[1]{\enma{{\bar {\bar #1}}}}

\renewcommand{\skecat}{\catname{Ske}}

\newcommand{\skecatsigGG}{\enma{\skecat(\GG,\predsig)}}

\newcommand{\globalbmap}{\bmap\bmap}
\renewcommand{\globalbmap}{\enma{\#_{\_}}}
\newcommand{\globaltmap}{\tmap\tmap}
\renewcommand{\globaltmap}{\tmapfun}

\newcommand{\bmapfunX}[1]{\enma{\bmap^{#1}_{\_}}}
\newcommand{\tmapfunX}[1]{\enma{\tmap^{#1}_{\_}}}
\newcommand{\bmapfun}{\enma{\bmap_{\_}}}
\newcommand{\tmapfun}{\enma{\tmap_{\_}}}

\newcommand{\carrpredsig}{\carr{\predsig}}

\newcommand{\clabelX}[1]{\enma{\widebar{#1}}}

\renewcommand{\Instcat}{\enma{\mathit{Inst}}}

\renewcommand{\InstcatX}[1]{\enma{\Instcat_{#1}}}

\newcommand{\preInstcatX}[1]{\enma{\Instcat^?_{#1}}}
\newcommand{\Instcatsig}{\InstcatX{\predsig}}

\newcommand{\preInstcatsig}{\preInstcatX{\predsig}}
\newcommand{\InstcatSig}{\InstcatX{\predsig}}

\newcommand{\Preinstcat}{\enma{\Instcat^?}}
\newcommand{\PreinstcatX}[1]{\enma{\Instcat^?_{#1}}}
\newcommand{\Preinstcatsig}{\enma{\Instcat^?_\predsig}}

\newcommand{\InstcatXdn}[1]{\enma{\Instcat_{#1}}}
\newcommand{\PreinstcatXupYdn}[2]{\enma{\Instcat^{{#1}?}_{#2}}}
\renewcommand{\PreinstcatXupYdn}[2]{\enma{\Instcat^{?}_{#2/#1}}}

\newcommand{\InstcatS}{\InstcatX{S}}

\newcommand{\ellX}[1]{\enma{\ell_{#1}}}
\newcommand{\ellGS}{\enma{\ell_{G_S}}}
\newcommand{\globalellGS}{\enma{\ellX{\_}}}

\newcommand{\globalecurry}{\enma{\e_{\_}^\lambda}}
\newcommand{\globalCS}{\enma{C_{\_}}}
\newcommand{\globalGS}{\enma{G_{\_}}}

\newcommand{\ConstrcatX}[1]{\enma{\mathit{Con}(#1)}}
\newcommand{\ConstrX}[1]{\ConstrcatX{#1}}
\renewcommand{\ConstrcatX}[1]{\apexX{C_{#1}}}
\renewcommand{\ConstrX}[1]{\ConstrcatX{#1}}

\newcommand{\ConstrcatXbra}[1]{\enma{\mathit{Con}(#1)}}
\newcommand{\ConstrsetXbra}[1]{\enma{\mathit{Con}(#1)}}

\newcommand{\labfun}{\enma{\mathit{lab}}}
\renewcommand{\labfun}{\enma{\ell}}

\newcommand{\labfunsig}{\enma{\labfun^\predsig}}
\renewcommand{\diagr}{\pbind}

\newcommand{\labfunSke}{\labfun}

\newcommand{\diagrfunSke}{\enma{\mathit{bmap}}}

\newcommand{\carrop}[1]{{\opX{\carr{#1}}}}

\newcommand{\semmD}[1]{\semmXUD{#1}{D}{}}

\newcommand{\datagra}{\enma{\mathbf{d}}}

\newcommand{\MapdatagraLab}{\enma{t_{\datagra}}}

\newcommand{\mlabell}{\enma{\mathit{mlab}}}
\renewcommand{\labell}{\enma{\mathit{lab}}}
\newcommand{\dlabX}[1]{\enma{\labell^{#1}_\datagra}}
\newcommand{\dmetalabX}[1]{\enma{\mlabell^{X}_\datagra}}
\newcommand{\dlab}{\dlabX{}} 
\newcommand{\dmetalab}{\dmetalabX{}}

\renewcommand{\dlab}[1]{\MapdatagraLab}
\renewcommand{\dmetalab}[1]{\MapdatagraLab'}

\newcommand{\semmX}[2]{\enma{\semm{#1}^{{#2}}}}
\newcommand{\semmt}[1]{\semmX{#1}{t}}

\renewcommand{\so}{\enma{\mathsf{s}}}
\renewcommand{\ta}{\enma{\mathsf{t}}}

\newcommand{\apexX}[1]{\enma{\widehat{#1}}}
\newcommand{\apexx}[1]{\enma{\hat{#1}}}

\renewcommand{\CC}{\bbx{C}}

\newcommand{\XtauY}[2]{\enma{#1_{\tauu_{#2}}}}
\renewcommand{\XtauY}[2]{\enma{#1_{}}}
\newcommand{\StauX}[1]{\XtauY{S}{#1}}
\renewcommand{\StauX}[1]{\XtauY{S}{}}

\newcommand{\Fset}{\enma{F}}

\newcommand{\tauu}{\diagImpl}
\newcommand{\implPart}{\enma{h}}
\newcommand{\arrowInj}[1]{\enma{\bar #1}}
\renewcommand{\implPart}[1]{\arrowInj{#1}}

\newcommand{\dedSys}{\enma{\mathsf{IL}}}  

\newcommand{\lambdaa}{\enma{\lambda}}

\newcommand{\lamm}{\enma{\lambdaa}}
\newcommand{\liftt}{\enma{y}}

\newcommand{\inj}{\enma{\mathsf{inj}}}
\newcommand{\lift}{\enma{\mathsf{lift}}}
\renewcommand{\diagrfunSke}{\enma{\bmap^S_{\_}}}
\newcommand{\modelsInjX}[1]{\enma{\models^\inj_{#1}}}
\newcommand{\modelsInj}{\modelsInjX{}}
\newcommand{\modelsLiftX}[1]{\enma{\models^\lift_{#1}}}

\newcommand{\InjMod}{\enma{\mathsf{Inj}}}
\newcommand{\InjTh}{\enma{\mathsf{Inj}^*}}

\newcommand{\DCL}{DCLog}
\renewcommand{\II}{\ensuremath{\mathcal{G}}}
\newcommand{\checkfun}{\ensuremath{\mathsf{check}}}

\newcommand{\arityX}[1]{\ensuremath{\arity_{#1}}}

\newcommand{\bmap}{\ensuremath{\mathsf{b}}}
\newcommand{\tmap}{\ensuremath{\mathsf{t}}}
\newcommand{\bmpa}{\bmap}
\newcommand{\csig}{\ensuremath{\pmb{C}_{\semm{}}}}
\newcommand{\csigsyn}{\ensuremath{\pmb{C}}}

\usepackage[capitalize]{cleveref}

\renewcommand{\zd}[1]{}
\renewcommand{\zdzd}[1]{}
\renewcommand{\nmf}[1]{\enma{\text{\small{\sf #1}}}}
\begin{document}

%

\maketitle
\shortlong{
\begin{abstract}
	Data constraints are fundamental for practical data modelling, and a verifiable conformance of a data instance to a  safety-critical constraint (satisfaction relation) is a corner-stone of safety assurance. Diagrammatic constraints are important as both a theoretical concepts and a practically convenient device. The paper shows that basic formal constraint management can well be developed within a finitely complete category  (hence the reference to Cartesianity  in the title). In the data modelling context, objects of such a category can be thought of as graphs, while their morphisms play two roles: of  data instances and (when being additionally labelled) of constraints. Specifically, a generalized sketch $S$ consists of a graph $G_S$ and a set of constraints $C_S$ declared over $G_S$, and appears as a pattern for typical data schemas (in databases, XML, and UML class diagrams). Interoperability of data modelling frameworks (and tools based on them) very much depends on the laws regulating the transformation of satisfaction relations between data instances and schemas when the schema graph changes: then constraints are translated co- whereas instances contra-variantly. Investigation of this transformation pattern is the main mathematical  subject of the paper.   
\end{abstract}
}%
{\zd{To be written afresh (and this is a long version)}
\begin{abstract}
	A diagrammatic operation takes a diagram of a specified shape as its input and completes it with new elements forming a bigger diagram  (e.g., arrow composition takes two consecutive arrows as its input and adds a third arrow resulting in a triangle). 
	Pre- and post-conditions are an important part of the story (e.g., monic preservation by pullbacks) so that diagram chasing smoothly integrates logic and algebra.  This technique transcends category theory and finds numerous applications in practice of software engineering, specifically in the multiview approach to system design, and in model management (in the sense of model-driven engineering). 
	These applications force us to consider a very general version of diagram chasing over generalized sketches (in the sense of Makkai) and give rise to a project (now in progress) reported in the present technical Report. 
	
We introduce the notions of diagrammatic signatures, terms and algebras over an underlying sketch category, and prove several basic results about them; specifically, that the category of views is symmetric monoidal. Our views are a modification of the standard notion of a Kleisli mapping, in which each view $v: V-> Q_v(S)$ carries its own term $Q_v$ for extending the target $S$ of the mapping  rather than using one unified but huge monad. Nevertheless, we prove that our localized Kleisli mappings generate a classical Kleisli triple and hence a monad.  
\end{abstract}
}
\section{Introduction}\label{sec:intro}

Constraints are fundamental for data modelling: they keep data integrity and ensure safety. If $D$ is a data instance over schema $S$, then adding a {\em constraint} $c$ to $S$ classifies instances into {\em valid} (and then we write $D\vDash c$) or (otherwise) invalid. Many constraint specification languages were developed, 
\eg, FOL and its fragments, database dependencies,  the OCL (Object Constraint Language) widely used in the UML/EML software development ecosystem \cite{oclBook-99}, or the diagrammatic language of  lifting constraints \cite{spivak-lifting14}  popular within the ACT community. 
These languages are successfully employed within their own ecosystems, but create severe interpretability 
problems when used in a heterogeneous environment \cite{antonio-edoc09,paige-intermod-models10}. 
Indeed, any systematic approach to model management would be centred around the notion of model mapping and traceability (cf. 
\cite{Bernstein07%
,spivak-algMMt-arxiv16%
,me-fase18
}), but what is a mapping between a full-fledged UML class diagram and a full-fledged relational schema? Unification via XML allows solving the problem for only simple constraints and does not help for complex constraints modelling complex requirements often appearing in system engineering. We need a unifying abstract \fwk\ for data and system modelling encompassing different schema specification languages including complex constraints. 
%
%

Developing such a \fwk\ can be traced back to so called {\em abstract model theory} \cite{barwise-axioms4amt-AML74} and, more recently, Goguen \& Burstall's {\em institutions}  \cite{GoguenBurstall-jofacm92}. 
%
%
The latter are practically a bare-bone \fwk\ for describing translation of a binary satisfaction relation (further called {\em Sat}),  $\satrelsig\subset \Modsetsig \timm \Sensetsig$,  between  sets   \Modsetsig\ of {\em \Sigmaa-models} and \Sensetsig\ of {\em \Sigmaa-sentences} (\ie, constraints),  when the logical signature \Sigmaa\ (of, say, relation and function symbols) 
changes. 
If such changes are specified by a signature \mor\ \frar{f}{\Sigmaa}{\Sigmaa'}, then the main and, in fact, the only substantial postulate of the institution theory states that Sat is preserved in a twisted way: for models with a function  \frar{f^*}{\ModsetsigPrim}{\Modsetsig} in the direction opposite to $f$, for sentences with \frar{f_*}{\Sensetsig}{\SensetsigPrim} in the same direction, and for all $m'\in\ModsetsigPrim$ and all $\phi\in\Sensetsig$, the {\em twisted Sat  
condition} 
\begin{equation}\label{eq:satAxiom-intheIntro}
	f^*(m')\satrelsig \phi \text{~iff~} m'\satrelsigPrim f_*(\phi)
\end{equation}
holds. 
Logicians refers to the Sat-axiom above as to `{The invariance of truth under change of notation}'; it may also be termed technologically as `{The main law of Sat-interoperability}'. 

There is a major deficiency of the  \instfwk\ wrt. its application to software interoperability, where models become many-sorted data instances, and sentences are constraints. To wit: institutions ignore the {\em local} nature of Sat  appearing in practice: 
to find if a model $m$ satisfies a constraint $c$, normally a small part of $m$ is to be examined and fixed if the constraint is violated. For industrial instances/models comprising thousands of elements over dozens of sorts, locality of constraints is their crucial feature (\eg, see \cite{me-ecmfa17} for how it can be used to make inter-model consistency checking more effective) and needs to be included into a principled \fwk.  It is here where the {\em Diagram Constraint Logic (DCL)} and  {\em generalized sketches} (further just {\em sketches}) presented in the paper come on the stage and place the constraints' locality at the very centre of the specification machinery, but otherwise strive to keep the \fwk\ as abstract as possible. We will briefly outline the main ideas. 

In DCL, both models (instances) and sentences (constraints) are arrows in a given \caty\ \GG\ with pullbacks, whose objects may be thought of as graphs to guide intuition (and we  will call them graphs), but existence of pullbacks is the only assumption about \GG\ we need. 
A {\em constraint} over \GG\ is a triple $(c, G_c, \semm{c})$ of a {\em constraint name} $c$, its {\em arity} graph $G_c\in \Ob\GG$, and a \caty\ $\semm{c} \hookrightarrow\GG/G_c$ of {\em valid} instances and their \mor s (typically, a full sub\caty\ but it's not a must); we require \semm{c} to be closed under iso\mor s. Then any graph  $G$ determines a {\em Cartesian Sat} $\modelsd_G \subset \GG/G\times\GG(G_c,G)$ by claiming  conformance $t \modelsd_G\bmap_c$ for an instance \frar{t}{\cdot}{G} and a constraint \frar{\bmap_c}{G_c}{G}  iff pulling $t$  back along $\bmap_c$ results in $\restr{t}{\bmapc}\in \Ob\semm{c}$. %
If $G$ changes by a \GG-\mor\ \frar{f}{G}{G'}, then instances/models change contravariantly via the pullback functor \frar{f^*}{\GG/G'}{\GG/G}, while  constraints change covariantly via the post-composition functor \frar{f_*}{\GG(G_c,G)}{\GG(G_c, G')}. Now the \pblemma\ implies the Sat-axiom \cref{eq:satAxiom-intheIntro} for Cartesian $\modelsd_{G}$ with  $G$ ranging over $\Ob\GG$.  
%
A major result of the paper (\cref{thm:mainTh} on \pageref{thm:mainTh}) states that any constraint signature (\ie, a collection of constraints with their arities and valid instances as described above) gives rise to an institution with sound logical inference. 
%
%
Thus, Cartesian Sat underlying the DCL is a well-defined logical \fwk, which is much more concrete than institutions but is much more abstract than constraint specification languages mentioned above. Indeed, in DCL, semantics of s constraint is postulated to exist as a category \semm{c} of valid $c$-instances but its implementation is not specified. 
This fine tuning has already allowed for several successful applications of DCL and sketches in diagrammatic domain and data modelling   \cite{
Diskin-diagrams00, me-dke03} and software engineering  \cite{rutleRLW-jlamp12,me-clafer16,rossini-versioning18}; see also applications of sketches to an important Model-Driven Engineering (MDE) problem of model consistency \cite{me-ecmfa17,patrick-faoc21}. In this sense, the present paper provides a new mathematical underpinning for a methodology being already employed in practice and, moreover, enrich the \fwk\ with several new features briefly outlined below. 

The paper makes special efforts to make DCL closer to system engineering (SE) practice. In the latter, a central role is played by the conformance relation between products and \req s. They can be modelled by, reps., instances and constraints/sentences, which reduces product-\req\ conformance to logical Sat discussed above. However, it is normal when at a given logical time moment,  different \req s are modelled by the same constraint, and different products by the same instance. It means that we need a theory of {\em indexed} Sat dealing with multisets rather than sets of logical constructs (\eg, a conjunctive theory/sketch becomes a multiset of atomic constraints if different \req s are modelled by the same constraint). Yet another new feature of DCL developed in the paper is considering instance \mor s as  {\em spans} in $\GG/G$, which model product updates, \eg, a span of monics
$\tmap\xleftarrow{i}\cdot\xrightarrow{j} \tmap'$ models a deletion of \tmap's part followed by addition of a new part resulting in instance $\tmap'$. 

Most importantly, the DCL of this paper transits from binary Sat to ternary {\em eSat}, whose third component refers to a piece of evidence supporting the claim $t\modelsd_G \bmpa_c$, and thus models a fundamental concept of safety assurance. In a bit more detail, as soon as we index valid $c$-instances by a mapping \frar{\tmapfun}{\semm{c}}{\GG/G_c}, for an  instance $t\in\GG/G$ and constraint $\bmapc\in \GG(G_c, G)$, we can (and should) write  $t\modelsd_G^e \bmapc$ when $\restr{t}{\bmapc}=\tmap_e$ for index $e\in\semm{c}$. We will also discuss and motivate the categorical setting for eSat, in which the latter is a ternary span of functors , whose apex and feet are \caties. This leads to the notion of an {\em institution with evidence (e-institution)}, which generalizes ordinary institutions for the eSat relation between models and sentences. Moreover, we will argue that our twisted eSat-axiom is closer to the intuition underlying Sat-interoperability than the usual binary Sat-axiom (see \cref{rema:eSat-vsSat-axiom} on \pageref{rema:eSat-vsSat-axiom}). Main \cref{thm:mainTh} mentioned above actually states that any constraint signature gives rise to a sound e-institution. 
\zd{A phrase about Th.2? }

Our plan for the rest of the paper is as follows. 
{
	Section 2 is a motivational background demonstrating the importance, and possibly intricate nature of constraints appearing in practice. Sect.3 provides a mild introduction to sketches to give the reader a basic intuition of basic concepts. Sect.4 presents a formal theory of twisted \mor s, including a novel \fwk\ of e-institutions, and Sect.5 exhibits the formal \fwk\ of Cartesian institutions. Sect.6 defines sketches and demonstrates that the sketch \fwk\ can be seen as a correct implementation of a very general specification \fwk\ for data and system modelling (\cref{thm:mainTh2} and its discussion). In Sect.7 related work is briefly discussed and Sect.6 concludes. Some auxiliary material is placed in Appendix.%
}

%
\tableofcontents
\begin{mygroup} 
	\renewcommand{\mysubsubXY}[2]{\subsection{#2}}
\section{Example:  Constraints are essential}
\label{sec:vehOntology}

We consider a detailed example of data modelling to demonstrate the importance, and possibly intricate nature of constraints appearing in practice. A special goal is to give the reader a touch and feel of a special world of data modelling, where one often needs to think out many fine details of data interconnection. 
%
%

\newcommand{\Sveh}{\ensureamth{S_\veh}}
\renewcommand{\Sveh}{\ensureamth{S}}
\newcommand{\lcdTo}{\nmf{lcdTo}}
\newcommand{\lcdBy}{\nmf{lcdBy}}
\newcommand{\hasprim}{\enma{\has_\nmf{dr}}}
\newcommand{\drivesprim}{\enma{\drives^*}}
\newcommand{\lcdprim}{\enma{\nmf{lcd}'}}
\newcommand{\toSat}{\nmf{toSat}}
\renewcommand{\implies}{\nmf{implies}}
\newcommand{\subjTo}{\nmf{subjTo}}
\newcommand{\subjToprim}{\nmf{subjTo'}}
\newcommand{\codee}{\nmf{code}}
\newcommand{\whData}{\nmf{whData}}

\mysubsubXY{2.1.1}{A sample database schema. }
Imagine a company providing a wide variety of courier and transportation services and thus using a range of vehicle from motorcycles to  heavy tracks. The inner grey box in \cref{fig:sample-schema} presents a schema for a fragment of the company's database, in which  rectangular boxes (\Vehicle, \etc) represent {\em classes} of objects of interest,  and arrows are directed {\em associations} between them. (We use here the UML parlance; database names for these notions would be entities and relationships.) 
Boxes consisting of two compartments, \eg, \Driver, encode also spans of {\em attribute} arrows as shown for two of them: names in the lower compartment of a class box (\eg, \bdate) are names  of arrows from the class node to a predefined (literal) value type specified after the colon (\eg, \Date). Dog-eared notes (also borrowed from the UML and pink with a colour display) denote {\em constraint declarations}, whose scopes are shown with (red) dotted edges. 
\begin{figure}[h]
	\centering
	\includegraphics[width=0.9\linewidth]{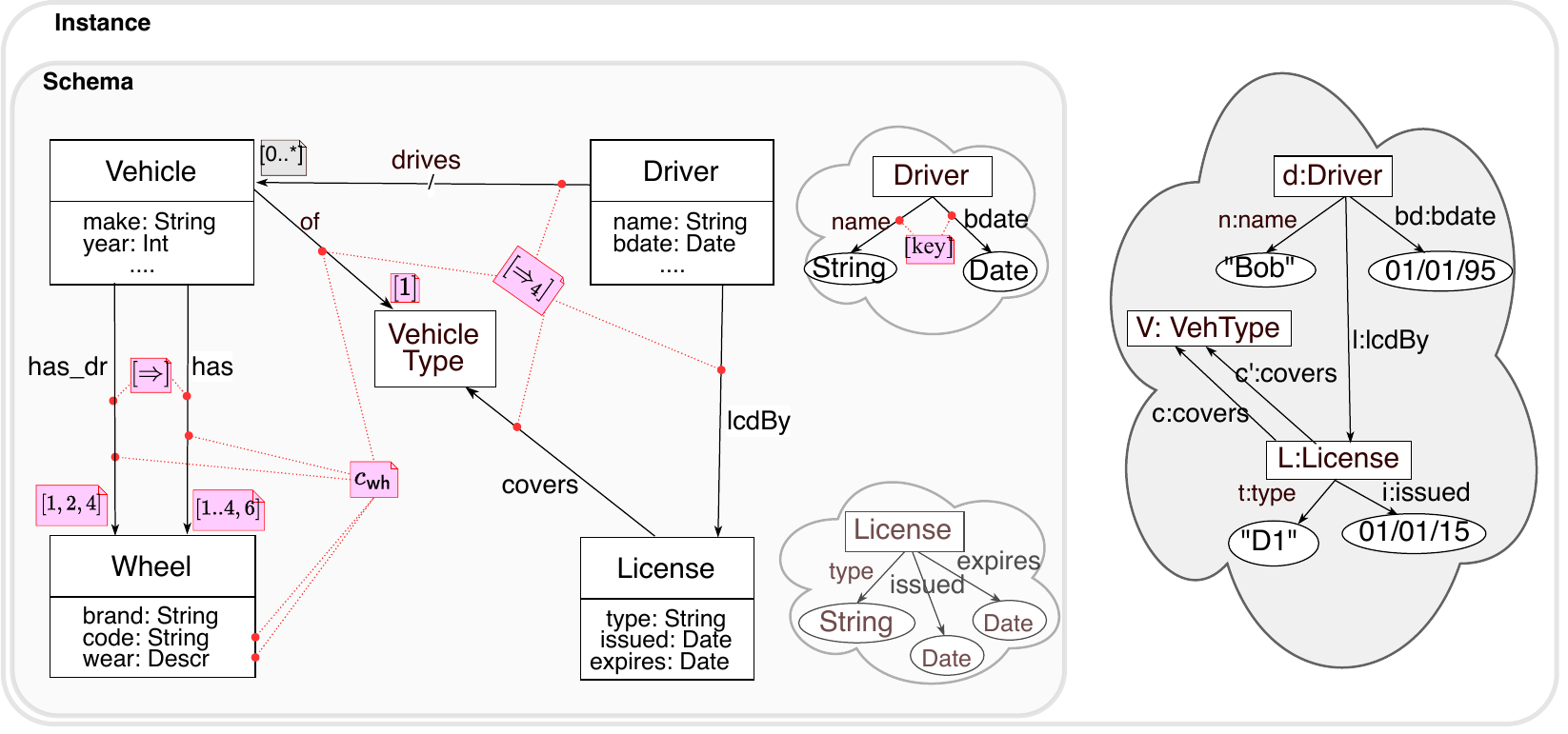}  
	\caption{Sample data schema $S$  and a fragment (in the rightmost cloud) of its instance
}
	\label{fig:sample-schema}
	\vspace{-2ex}
\end{figure}
Thus, the schema $S$ in \cref{fig:sample-schema} consists of two components: a) the carrier graph $G_S$ and b) a set of constraints $C_S$ over this graph, described below in detail.

a) Graph $G_S$ consists of six rectangular {class} nodes, 
several nodes referring to predefined value types (\String, \Date\ \etc) hidden in boxes, %
\footnote{graph $G_S$ is unlabelled and has only one node \String, one node \Date\ \etc; what is shown in \cref{fig:sample-schema} in cloud windows are diagrams in  $G_S$}
seven  {\em \asson} arrows between classes: \drives, \of, \lcdBy\ (read it as ``licencedBy''), \hasprim\ (``has driving'') \etc,  
and multiple {\em attribute} arrows from classes to predefined types: \name, \bdate, \etc.   

b) Set $C_S$ consists of multiple constraint declarations, 
each one has its scope in graph $G_S$ shown with red dotted edges. For example, the left constraint $\Rightarrow$ is declared for arrows \hasprim\ and \has. Numerical constraints (so called {\em \mties})  placed near arrow heads are declared for those arrows (\eg, [1] for arrow '\of' or [1..4,6] for arrow \has) and their scope is clear without scope lines (but, of course, they are recorded in the tool storing the schema). These constraints says that a vehicle may have 1, 2, or 4 driving wheels amongst (due to constraint $\Rightarrow$) its 1..4 or 6 wheels (\eg, a 3-wheel motorcycle may have 1 or 2 driving wheels).  
\label{page:wheel-attrs}
The scope of constraint $c_\wh$ includes three \asson s and two wheel attributes. 

There are several other constraints in set $C_S$, whose presence in the schema is not explicit in the visual diagram in \cref{fig:sample-schema} but can be restored based on syntactical conventions  --- we will discuss them later. 

\mysubsubXY{2.1.2}{Graph instances over the schema.}
Graph $G_S$ is a template for data populating the database, which we call {\em data instances} over $G_S$ as they change with time. A fragment of a possible instance is shown in a cloud window on the right side of the figure. It shows objects $d$, $V$, $L$ of types \Driver, \etc\ specified after colons, and values "Bob" \etc\ shown in ovals, whose type is implicit (\String\ for Bob, \Date\ for 01/01/95,  \etc). 

For an instance $D$,  the database should store sets $\semmD{\Vehicle}$, $\semmD{\Driver}$ \etc\ of {\em objects} for classes,  and sets \semmD{\drives}, \semmD{\of} \etc\ of directed {\em references} for \asson s, \eg, a \Driver-object $d\in\semmD{\Driver}$ may have multiple or none (\eg, if driver $d$ is on leave of absence) references to  \Vehicle-objects (note the grey multiplicity [0..1] attached to arrow \drives). 
References can be implemented in different ways, but after all, any \drives-reference in $D$ is a special  object that has exactly one object in \semmD{\Driver} as its {\em source} (which,  in the UML speak,  {\em owns} the reference), and exactly one object in \semmD{\Vehicle} as its {\em target}. We can thus model the set of \drives-references in $D$ by a span 
\[
\spandiag{\semmD{\Driver}}{\quad}{\semmD{\drives}}{\quad}{\semmD{\Vehicle}}
\]
whose unnamed legs are the source and the target functions. In general, reference spans may be not jointly monic (jm). Consider, \eg, an instance fragment in the cloud window on the right, which shows two reference links $c,c'$ of type \covers\ between the same pair of objects. It may happen because the  fact of covering a vehicle type by a license is regulated by normative documents, and it happened that two such documents establish that license $L$ covers the vehicle type $V$. In a finer model, we would add  to \covers-references references to such docs, but it would change the structure of both the schema and the instance graphs (we will need arrows from arrows to nodes), and it is often reasonable to keep graphs simple but admit multiple links of the same type between the same pair of objects. The UML does allow interpreting \asson s by non-jm spans, but in case when the span is required to be jm, the \asson\ is labelled by a keyword {\em unique} (see a detailed discussion in \cite{assonEnd-Milicev-tse07}).  For our schema in \cref{fig:sample-schema}, we may take a default assumption that if an \asson\ has a \mty\ declared \shortlong{for it}{(for at least one of its ends)}, then the span is assumed to be jm. 

In contrast to classes, sets of literal values \semm{\String}, \semm{\Date} \etc\ are defined by their types (\String, \Date, \etc) rather than stored in the database (\ie, are defined intensionally rather than extensionally). Attributes are thus implemented differently, but we can still model them by  spans, \eg, \spanrar{\semmD{\name}}{\semmD{\Driver}}{\semm{\String}} (and indeed, in a semi-structured data storage, \eg, XML-based, two different elements may have the same name and the same value
).  

\shortlong{Thus, an {instance} $D$ can be seen as a graph \mor\ \frar{\semmD{.}}{G_S}{|\spancat|} into the graph underlying the \caty\ of spans between sets. Equivalently,  an instance $D$ is a graph \mor\ \frar{t_D}{G_D}{G_S}, where $G_D$ is the graph of all objects and values involved in $D$ as nodes, and all references and attributes involved in  $D$ as arrows (below we will collectively refer to them as to {\em links}), and typing mapping $t_D$ maps each element in $G_D$ to its type in $G_S$. We will discus  this equivalence later in \cref{sec:skeIntro-instances}.)
}{
} 

\zdzd{
	Note \mty\ [1..2] near the source end of \asson\ drives, which is actually a \mty\ for span $\semm{\drives}^{-1}$.
}

\mysubsubXY{2.1.3}{Schema instances.} Not any $G_S$-instance as considered above is a {\em valid schema instance}: the latter must also satisfy all constraints in set $C_S$.  
\shortlong{
	We have already considered \mty\ constraints whose semantics is clear. We also assume that  the absence of an arrow's \mty\ means constraint [1..*] for that arrow by default, but then the absence of a \mty\ constraint is to be explicated by a grey pseudo-constraint note [0..*]. Similarly, we assume all attributes have \mty\ [1] by  default whereas any other \mty\ is to be shown including grey no-constraint note. 
	Constraint [\key] over \Driver's attributes states that each driver is uniquely identified by her name and birthdate, and an instance violating this condition is to be considered invalid.
}{
	For example,  if a $G_S$-arrow is not stroked, the corresponding span, \eg, \semm{\drivesprim} (here and below the script $D$ is omitted if the context is clear) is required to be jm (and thus represent a relation). Hence, a non-stroked arrow actually denotes a constraint of being jm imposed on the underlying arrow, but to ease notation, it is better to omit these constraints while stroke those few arrows, which are not constrained. Similarly, if we convert spans into multi-valued functions, then the majority of arrows in our  $G_S$ are instantiated by totally defined functions (\eg, set $\semm{\lcdBy}(d)$ is required to be non-empty for any $d\in\semm{\Driver}$), and it is reasonable to assume this constraint to be declared by default but mark few exclusions with a non-constraint (grey) label [0..*]. 
	\\
	We also assume that  the absence of \mty\ at the target end of an association means [1..*] (\ie, the span is totally defined) while the absence of \mty\ at the source means [0..*] (non-surjective span). 
	\\
	Constraint [1] near arrow `\of' states that span \semm{\of} is to be a totally-defined single-valued function. This is a simple case of a {\em multiplicity} constraints, while arrows \has\ and \hasprim\ (read them as `has wheels' and  `has driving wheels') are assigned with complex \mties, which constitute an important part of the Vehicle-has-Wheels ontology (\eg, there are three-wheel motorcycles with one or two driving wheels).
	\zd{??Write about the inverse \mty\ for \coverss$^{-1}$}
}
%
%
\shortlong{}{
	\zd{a big piece about commute is skipped -- see commented input}
}
%

Constraint $[\Rightarrow]$ formalizes an obvious property of the Vehicles-Wheels ontology  that driving wheels of a vehicle are to be  amongst its wheels, and as both spans are jm, we have subsetting $\semmD{\hasprim}\subseteq\semmD{\has}$ to hold for any instance $D$. \shortlong{}{
	\zd{ and thus there exists a function \frar{f^D_{[\Rightarrow]'}}{\semmD{\hasprim}}{\semmD{\has}} (necessarily injective as both spans are jm).}
} 
Similarly, constraint $[\Rightarrow]_4$, which  declares subsetting $\semmD{\drives}.\semmD{\of}\subseteq\semmD{\lcdBy}.\semmD{\covers}$, formalizes  an important safety \req\ that a driver can drive a vehicle of type $V$  only if  she is licenced to drive $V$-type vehicles.%
\footnote{Of course, it'd be better to declare binary  $[\Rightarrow]$ for composed \asson s introduced directly into the schema, but it needs having the query machinery in our schema formalism, which is beyond the paper's scope (but see \cite{myTR33-sketches}) and we thus hide composition inside the constraint $[\Rightarrow_4]$.  
} If the subsetting above holds for all $d\in\semmD{\Driver}$, we say that instance $D$ satisfies the constraint and write $D\modelsd [\Rightarrow_4]$. 
%
%

Finally, we discuss semantics of constraint $c_\wh$. Suppose there is a safety \req\ $R$ that demands all wheels of a vehicle to satisfy certain conditions depending on the vehicle type and the wheels' parameters encoded in their \codee-string. We can specify this by setting an admissible range $[R]_V$ (depending on the vehicle type $V$)  for some complex vehicle's attribute \whData\ so that a vehicle conforms to $R$ if the complex value of its \whData-attribute is within the range, \ie, $\semm{\whData}\in [R]_V$. More formally, if $D$ is an instance of our schema, then we consider $D\modelsd c_\wh$ iff for any vehicle $v\in\semmD{\Vehicle}$, we have $v.\semmD{\whData} \in [R]_V$ where $V=v.\semmD{\of}$. Clearly, we can rewrite this as $D\models c_\wh$ iff $v\modelsd c'_\wh$ for all $v\in \semmD{\Vehicle}$, and $v\modelsd c'_\wh$ iff the data collection 
$\left( (w.\semmD{\codee})_{{w\in v.\semmD{\has}}},\,  (w.\semmD{\codee})_{w\in v.\semmD{\hasprim}}, \, v.\semmD{\of}\right)
$ is within some predefined range \semm{c'_\wh}. The latter can be specified by a table in some normative document (or a complex formula if the description of wheel's wear is somehow formalized), having which the constraint monitoring system (CMS) can provide a definitive answer to whether vehicle $v$ satisfies $c'_\wh$ or not. In principle, it is possible that the CMS would ask a human expert (\eg, if the normative document suggests it for the $v$'s configuration of values in the table). Irrespectively to implementation, the CMS provides a certain Boolean value for the claim $D\modelsd c_\wh$, and optionally, a reference to a normative document or expert supporting the claim. In fact, for safety critical constraints such a reference is a must and usually called (a piece of) {\em evidence} supporting the claim. Then conformance becomes a ternary relation and we should write $D\modelsd^e c$, where $e$ refers to an object proving evidence.   
\zd{Constraint for $\covers^{-1}$ is not described. Maybe skip it? it's already too long section}

To summarize, each constraint $c\in C_S$ has its scope graph $G_c \subset G_S$ and checking of whether $D\models c$ only depends on the corresponding part $\restr{D}{G_c}$ of the instance over $G_c$: 
$$
D\modelsd_{G_S} c \text{~ iff (by definition)~} \restr{D}{G_c}\models_{G_c}c, 
\text{\ie, $\restr{D}{G_c}\in \semm{c}$}
$$
where the subindex near \modelsd\ points to the scope of conformance checking. 
These considerations will be made accurate in the next section.

\end{mygroup}

\renewcommand{\arityX}[1]{\enma{G_{#1}}}
\newcommand{\laxator}[2]{\enma{\ell^{#1}_{^{#2}}}}
\newcommand{\completion}[1]{\enma{\ovr{#1}}}
\newcommand{\classifyFun}{\enma{p}}

\section{Diagram constraint logic (DCL) and sketches:  
			An introduction } 
\label{sec:sketchesIntro}

The main ingredients of a semantically-oriented logical \fwk\ are {\em possibly valid} schema instances (or {\em pre}instances), constraints, and the \sation\ relation (Sat)  between them, which defines a subclass of {\em valid} (indeed) instances amongst preinstances. In this section, we will discuss and define these ingredients for our Diagram Constraint Logic (DC-logic or DCL). We begin with preinstances in \cref{sec:skeIntro-instances}, and then proceed to constraints \cref{sec:skeIntro-constraints}. Specifically, the diversity of possible implementations of constraints' semantics will be discussed in some detail:  understanding the difference between an abstract diagrammatic constraint and its concrete implementation in a given constraint specification language is, perhaps, a major stumbling block for  understanding the generalized sketches idea and its applications. Finally, in \cref{sec:skeIntro-Sat}, we define a pullback based Sat between preinstances and constraints, and give the notion of a sketch and its valid instances. 


\begin{mygroup}
\renewcommand{\semmD}[1]{\semmt{#1}}
\renewcommand{\completion}[1]{\enma{#1}}

\subsection{Possible instances in DCL}\label{sec:skeIntro-instances}

\mysubsubXY{3.1.1}{Fibrational vs. indexed.} 
Let $S=(G_S, C_S)$ be a schema as above. Its {\em possible instance}, or {\em preinstance}, or an {\em instance over graph} $G_S$, is a graph \mor\ 
\frar{t}{G_t}{G_S}, where $G_t$ is the graph of all objects and values involved in the instance as nodes, and all references and attributes involved in  the instance  as arrows, and {\em typing} $t$ maps each element in $G_t$ to its type in $G_S$. 
This setting will be referred to as  {\em fibred} (or {\em fibrational}) semantics for graph $G_S$.  As multiple $G_t$-arrows between the same two nodes $d,d'$ in $G_t$ can be typed by the same arrow \frar{a}{N}{N'} in $G_S$, inverting $t$ will map arrow $a$ to a span \spanrar{\semmD{a}}{\semmD{N}}{\semmD{N'}} between sets, and gives us a graph \mor\ \frar{\semm{.}}{G_S}{\carr{\spancat}} into the graph underlying the \caty\ of spans between sets. Having such a \mor s is referred to as an {\em indexed} semantics for $G_S$. The Grothendieck construction transforms  
an arbitrary graph \mor\ \frar{\semm{.}}{G_S}{\carr{\spancat}}  into a global graph of elements $\int\semm{.}$ supplied with projection \frar{t_{\semm{.}}}{\int \semm{.}}{G_S}. The two settings are equivalent: if we begin with $t$, then $\int (\semmD{.})\stackrel{i}{\cong} G_t$ and $i.t=t_{\semm{.}}$, and if we begin with \semm{\cdot}, then  $\semmTofibTosemmX{.}\cong \semm{.}$ in the sense of a canonical set iso\mor\ $\semmTofibTosemmX{N}\cong\semm{N}$ for any node $N$  in $G_S$, and a canonical  span iso\mor\ $\semmTofibTosemmX{a}\cong\semm{a}$ (in the vertical \caty\ of the double \caty\ \spandcat\ of functions and spans between sets) for any arrow $a$  in $G_S$. 

Including into this equivalence also morphisms between instances needs care and resorting to categories and functors rather than graphs and graph \mor s. Graphs can always be replaced by categories they freely generate, but the point is that in practice schemas are often non-free \caties, \ie, are presentations of non-free \caties\ given by their  underlying graphs and path equations, \ie, commutativity constraints declared over the \corring\ diagrams. For schemas as categories, we have the following general Grothendieck equivalence: $  \catcat/S\simeq \laxcat(S, \spanbicat)$, where  \spanbicat\ is the  bicategory of spans between sets and $S$ a \caty\ considered as a 2-discrete bi\caty. Indeed, laxity does matter in data modelling, \eg, consider adding \asson\  $\drives'$ from \Driver\ to \nmf{VehType} in the schema $S$ of \cref{fig:sample-schema}. In general, data links over this \asson\ can contain non-composed links, \ie, $\semmD{\drives'}\subseteq \semmD{\drives}.\semmD{\of}$ (composition is in \spancat),  if the database keeps information about driving vehicles of some type in the past but not currently.  

Below we will keep our graph-based fibrational setting as the main one: (pre)instances are graph \mor\ \frar{t}{G_t}{G_S} and instance \mor s are \mor s \frar{u}{G_t}{G_{t'}} such that $u.t'=t$. However, we will freely use inversion of $t$ into $\semmD{}$ for a given $t$; specifically, it is convenient for specifying semantics of many constraints. 

\mysubsubXY{3.1.2}{Attributes/values vs. References/Objects}
	The fundamental distinction between objects/references and values/attributes can be modelled by endowing graphs $G_S$ and \completion{G_t} with classifying  maps into the interval \caty\ \twocat\ (encompassing two objects 0 and 1, and the only non-identity arrow \frar{01}{0}{1}). Indeed, attribute arrows in $G_S$ go from classes to literal types but never in the opposite direction. (As for arrows between literal types, they are possible:  think, \eg, of arrow \frar{\length}{\String}{\Int}. ) However, constraint scopes' often cross-cut this distinction, \eg, suppose that each driver is uniquely identified by her name and her set of licenses; hence, in this paper we will ignore the distinction above and work with unclassified graphs. 
\shortlong{}{
	A pair $(X,\classifyFun)$ of a \caty\ $X$ and a functor \frar{\classifyFun}{X}{\twocat} is called a {\em barrel}. Thus, we refine the notion of a schema as a pair $S=(B_S, C_S)$ of a barrel $B_S$ and a set $C_S$ of constraints over it (formalized in the next section), and an unconstrained instance of $S$ is a barrel \mor\ \frar{t}{B_D}{B_S}, \ie, a functor \frar{t}{\carr{B_D}}{\carr{B_S}} such that
 $t.\classifyFun_S=\classifyFun_D$~. 
}
\end{mygroup}

\renewcommand{\tauu}{\enma{{f_c}}}
\newcommand{\tauuX}[1]{\enma{{f_{#1}}}}
\newcommand{\Xsubscrtauu}[1]{\enma{#1_{_\tauu}}}

\subsection{Abstract diagrammatic constraints and their implementation } 
\label{sec:skeIntro-constraints}

Let  \GG\ be a (small) \caty\ with pullbacks (PBC), whose objects are called {\em graphs}.
Given a graph  $G\in\Ob\GG$, we write $\GG(G,\_)$ and $\GG(\_,G)$ for the set of all arrows from $G$, resp, to $G$. Correspondingly, we have two slice categories: $\GG\setminus G$ and $\GG/G$, whose morphisms are commutative triangles. As \GG\ has pullbacks, all slice \caties\  $\GG/G$ have products.  
\begin{defN}[Abstract constraints]\label{def:abstrConstr}
	\shortlong{
		An {\em abstract  (diagrammatic) constraint} over \GG\ is a triple $(c, G_c, \semm{c})$ of a {\em constraint name} $c$, a graph $G_c$ called the {\em arity (shape)} of $c$, and a set  $\semm{c}\subset \Ob(\GG/G_c)$ of \GG-\mor s into $G_c$ called {\em valid instances} of $c$. We require set \semm{c} to be closed under iso\mor s in $\GG/G_c$.  
	}
	{  
		An {\em (abstract  diagrammatic) constraint} over \GG\ is given by its symbolic name/symbol $c$ and  
		a graph $\arityX{c}\in \Ob\GG$ 
		called the {\em arity shape} of $c$.  
		
		\zd{\sc to be rewritten! A part of the story below should be described already in Sect 2}
		A constraint $c$ is {\em (semantically) interpreted} if it is supplied with a Boolean-valued function \frar{\checkfun_c}{\GG(\_,\arityX{c})}{\{0,1\}} that classifies \GG-arrows \frar{t}{X}{\arityX{c}}. In applications, $X$ is a graph of {\em data values} while arrow $t$ provides their types and is often referred to as a {\em typing mapping} while the pair $(X,t)$ is a {\em instance} of $\arityX{c}$) 
		Instance $t$ is called {\em valid} if $\checkfun_c(t)=1$ and {\em invalid} if $\checkfun_c(t)=0$. If $t$ is valid, we write $t\models c$ and  say that $t$ {\em satisfies} $c$ (or {\em conforms to} $c$). Sometimes we will call mappings into \arityX{c}, whose validity or invalidity is not yet established,  {\em preinstances} of $c$. The set of all valid instances will be denoted by $\semm{c}$, and conversely, having a subset $\semm{c}\subset\GG/\arityX{c}$, we can define the corresponding \checkfun-function. Anyway, an interpreted constraint consists of a syntactical part $(c,\arityX{c})$ and a semantic part $\checkfun_c$ or, equivalently, \semm{c}. Moreover, set \semm{c} is required to be closed under isomorphisms in $\GG/\arityX{c}$: if there is an iso $t\stackrel{i}{\cong} t'$ in $\GG/\arityX{c}$, \ie, a \GG-iso\mor\ \frar{i}{X}{X'} commuting with typing, then $t\models c$ iff $t'\models c$. 
		
		A {\em (constraint) signature} \csigsyn\ is a set of constraints with their arities, \ie, a function \frar{S_{\_}}{\carr{\csigsyn}}{\Ob\GG} defined on some carrier set \carr{\csigsyn}; we will often denote the latter by \csigsyn\ to ease notation. n {\em interpreted} signature is a set \csig\ of interpreted constraints. 
	}  

\shortlong{		
An arbitrary arrow \frar{t}{X}{G_c} is called a {\em preinstance} of $c$, and if $t\in\semm{c}$, then we write $t\models c$.  (Sometime we will call preinstances just instances meaning ``instances whose validity nor invalidity  is not established''--- this is not perfect but follows the usual terminology.) Sometime we will write a preinstance as pair $(X,t)$ and call it a {\em typed} graph and $t$ its  {\em typing} mapping. 
}{} 
\qed\end{defN}
Constraints as defined above are called abstract because they do not specify how to implement the \checkfun\ function that classifies instances \frar{t}{X}{G_c} into valid and invalid -- it is only assumed that such an implementation exists and results in a subset of valid instances \semm{c}. \shortlong{}{\zd{this phrase should appear earlier: In practice, constraints are specified concretely based on some constraint implementation language, \eg, OCL \cite{OCLmanual00} or FOL, or with a diagrammatic implementation \zd{name for it, say, DCImpl?}as described in \cref{sec:diagImpl}}.}
\shortlong{}{We will mainly deal with interpreted constructs and thus often omit the adjective 'interpreted' in referring to them to ease terminology, but then we will mention 'uninterpreted' when  referring to their syntactical parts. Specifically, this is the case throughout the rest of \cref{sec:constr-and-sig}
} 
%
%

\begin{example}[FOL implementation of diagrammatic constraints]\label{ex:signature}
	Let \GG=\graphcat, \ie, {\em graphs} and {\em graph \mor} are ordinary directed multigraphs and  their morphisms.  
\cref{tab:sampleConstr} presents several typical constraints, whose  
	semantics is specified in the rightmost column in FOL: for a given instance   $t\in\GG(\_,G_c)$,  
	we consider $t\models c$ iff the corresponding FOL formula in the rightmost column is true for the graph \mor \ (\ie, a {\em model} in the logical sense) \frar{\semmt{.}}{G_c}{\carr{\spancat}}. To make formulas more compact,
	we identify a span \spanrar{\semmt{r}}{\semmt{A}}{\semmt{B}} with a discrete profunctor \frar{\semmt{r}}{\semmt{A}\times \semmt{B}}{\setcat} (think of sets \semmt{A}, \semmt{B} as discrete \caties). We also omit universal quantification if it is clear. 
%
	{\footnotesize 
\renewcommand{\shortlong}[2]{#1} 
\begin{table}[h]
 \vspace{-1.5mm}
 \caption{Sample Constraint\label{tab:sampleConstr}}
 \vspace{-1.5mm} 
 \centering
 \begin{tabular}{|c|c|c|}
 \hline
 Name, $c$ & Arity shape, $S_c$ & Semantics for $t\models c$ (in FOL)
 \\ \hline \hline 
[1..*]	& \ArrowDiag
 			&  
 			$\forall a\exists b\;  \semmXUD{r}{t}{}(a,b)\noteq\varnothing$
\\   \hline  
[0..1] & \ArrowDiag
 			&  
 			$\semmXUD{r}{t}{}(a,b_1){\noteq}\varnothing \wedge \semmXUD{r}{t}{}(a,b_2){\noteq}\varnothing\Rightarrow b_1=b_2$
\shortlong{}{\\ \hline 
[0..2] &  \ArrowDiag
 			&  
				$\begin{array}{l}
 					\semmXUD{r}{t}{}(a,h_1,b_1)\wedge  \semmXUD{r}{t}{}(a,h_2,b_2) \wedge \semmXUD{r}{t}{}(a,h_3,b_3) \\ 
					\Rightarrow  \quad b_1=b_2 \vee b_2=b_3\vee b_1=b_3
				\end{array}$  
} 
\shortlong{}{\\  \hline  
[=] & \CommTriangleDiag
&  $ 
			r_1^t.r_2^t \, \hookrightarrow r^t$ 
}  
\shortlong{}{\\  \hline  
[=$^!$] & \CommTriangleDiag
&  $
		\semmXUD{r_1}{t}{}.\semmXUD{r_2}{t}{}\, \xrightarrow{\cong} \semmXUD{r}{t}{}$ 
} 
 \\  \hline  
[$\Rightarrow$] & \TwoParaArrowsDiag
&  $\semmXUD{r_1}{t}{}(a,b) {\noteq} \varnothing \,\Rightarrow  \semmXUD{r_2}{t}{}(a,b){\noteq}\varnothing$ 
\shortlong{}{\\  \hline  
[trans] & \LoopDiag 
			& 
			$\semmXUD{r}{t}{}(a_1,a_2){\noteq}\varnothing\wedge \semmXUD{r}{t}{}(a_2,a_3){\noteq}\varnothing \Rightarrow \semmXUD{r}{t}{}(a_1,a_3){\noteq}\varnothing$
} 
\shortlong{}{\\  \hline  
[id] & \LoopDiag
		&  $r^t=\id{(A^t)}$ (that is, $\forall a\in A^t\exists!h\;\semmXUD{r}{t}{}(a,h,a)$)
} 
\\    \hline  
\end{tabular}
\vspace{-2.5mm}
\end{table}
} 
\end{example}
%
Semantics of constrains defined in  \cref{tab:sampleConstr} can be specified in another logical language, \eg, in a major for the EMF/UML  ecosystem constraint specification language OCL \cite{oclBook-99}   
Yet another implementation method is to employ a classical idea of injectivity: a categorical diagrammatic version of regular logic. 
\cref{sec:injectivityPrimer} provides some basics.

\begin{defN}[Regular (diagrammatic) constraints]\label{def:regularConstr}
	Let \GG\ be an arbitrary category (whose objects are called {\em graphs}).
	A  constraint $(c,G_c, \semm{c})$ over \GG\  is called {\em (diagrammatically) regular} if its semantics $\semm{c}\subset \Ob (\GG/G_c)$ can be specified as injectivity over a suitable $\GG/G_c$ arrow \tauu, \ie, $t\in\semm{c}$  iff $t\modelsInjX{\GG/G_c} \tauu$  as shown in diagram \cref{eqdiag:injectivity}(a) (so that arrow \tauu\ plays the role of a regular logic formula).    
	{

\newcommand{\lifting}{
	\xymatrix{ 
		P_\tauu \ar@{->}[r]^{x}
	\ar[d]_{\bar\tauu}
	& X_t \ar[d]^t
	\\
	Q_\tauu	
	\ar@{->}[r]_{q_\tauu}
	\ar@{-->}[ru]^y
	& S_\tauu
}}	

\newcommand{\injectivityUnpacked}{
		\xymatrix{
	&&X \ar[d]^t \ar@{<-}[lld]_{(\forall)x}
	\\
	P_\tauu 
	\ar[rr]_{p_\tauu} \ar[rd]_{\implPart{\tauu}}
	&&G_c
	\\
	&Q_\tauu \ar@{->}[ru]_{q_\tauu}
	\ar@{-->}[ruu]^{(\exists)y}
	&
}}

\renewcommand{\injectivityUnpacked}{
\begin{tikzcd}[row sep=3ex, column sep=5ex, ampersand replacement=\&]
\&\& X \\
P_\tauu \ar[rru, "(\forall)x"]  \ar[rr, "\Xsubscrtauu{p}", pos=0.55]
	\&\& G_c\ar[u, <-, "t" '] 
\\
\& Q_\tauu \ar[lu, <-, "\tauu"] \ar[ru, "\Xsubscrtauu{q}" '] 
				\ar[ruu, dashed, "\gapnamegap{1}{(\exists)y}{-1}", pos=0.1 ] 
\&
\end{tikzcd}
}

\newcommand{\injectivity}{
\xymatrix{
	 \Xsubscrtauu{p}  
	\ar[rr]^{(\forall)\hat x} 
	\ar[rd]_{\tauu}
	&&t
	\\
	&\Xsubscrtauu{q} \ar@{-->}[ru]_{(\exists) \hat \liftt}
	&
}}

\begin{equation}\label{eqdiag:injectivity}
	\begin{tabular}{c@{\qqquad}c@{\qqquad}c} 
\injectivity &\injectivityUnpacked & \lifting
\\ [25pt]
a)~ injectivity in $\GG/G_c$ 
& a')~ injectivity unpacked 
& b)~ lifting 
\end{tabular}
\end{equation}
} 
	
	\noindent 
	Diagram \cref{eqdiag:injectivity}a') unpacks this definition as a commutative diagram in \GG: arrows \tauu, $x$, $y$ in diagram a) become commutative triangles $(\tauu, \Xsubscrtauu{q}, \Xsubscrtauu{p})$, $(x,t,\Xsubscrtauu{p})$, $(y,t,\Xsubscrtauu{q})$ in a'). Ignore diagram (b)  until Def. \ref{constr:lifting}.
\end{defN}

\begin{figure}[h]
	\centering
	\begin{tabular}{c@{\qquad}c}
		\includegraphics[width=0.45\linewidth]{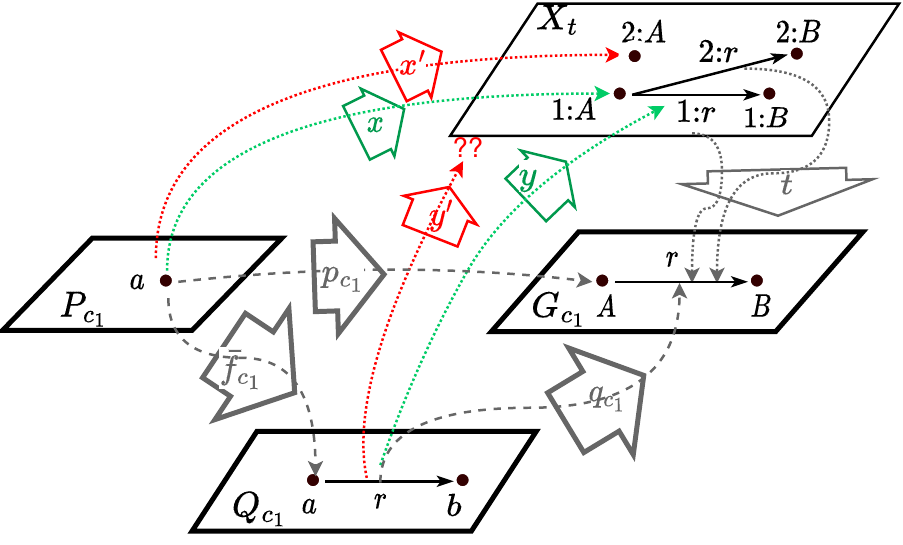}
		&
		\includegraphics[width=0.45\linewidth]{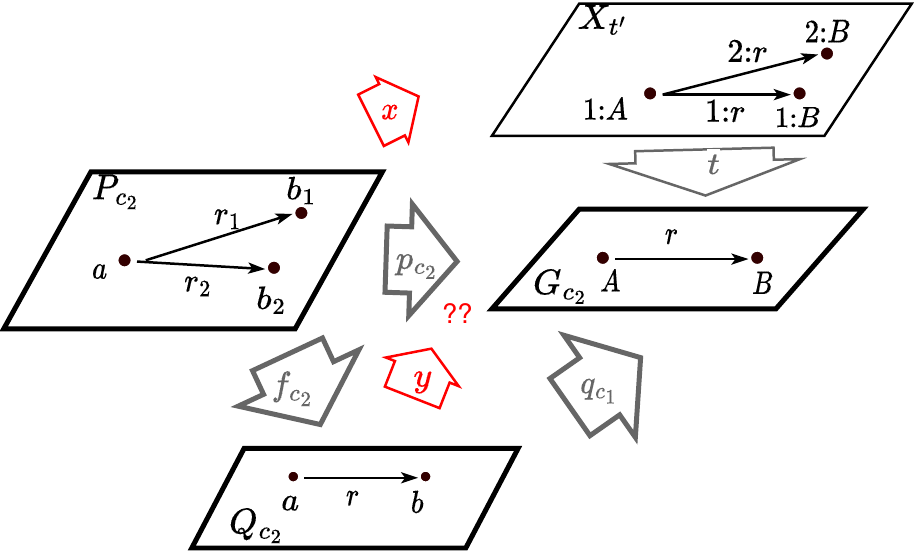}
		\\  [10pt]
		a) \parbox{0.42\linewidth}{Constraint [0..1] (left totality) via injectivity  (not all links are shown in mapping $t$)}
		&  b) Constraint [0..1]  (uniqueness)  
	\end{tabular}
	\caption{Two constraints from \cref{tab:sampleConstr}  implemented via injectivity}
	\label{fig:two-triangle-constraints}
\end{figure}

\begin{example}[Injectivity implementation of diagrammatic constraints]
	\label{ex:two-triangle-constr}
	\cref{fig:two-triangle-constraints}(a) specifies the top constraint in \cref{tab:sampleConstr} diagrammatically via injectivity. Green mapping \frar{x}{P_c}{X_t} can be factorized through $f_c$ via green mapping \frar{y}{Q_c}{X_t} whereas red \frar{x'}{P_c}{X_t} is not factorizable as node $2\in X_t$ of type $A$ does not have an arrow over $r$. Hence, injectivity wrt. triangle ``formula'' $(f_c, p_c, q_c)$ in \cref{fig:two-triangle-constraints}(a) guarantees the required semantics. Similarly, triangle formula  $(f_c, p_c, q_c)$ in subfigure (b) guarantees uniqueness, \ie, specifies semantics of the second row constraint in \cref{tab:sampleConstr}. (Indeed, an injective testing mapping \frar{x}{P_c}{X_t} cannot be factorized via $f_c$.)  
	\qed\end{example}

%
More examples can be found in paper \cite{spivak-lifting14},which exhibits  many examples of {\em lifting} constraints---yet another implementation of diagrammtic constraints. However, we now show that lifting and regular constraints are equivalent. 
\begin{defN}[Lifting constraints, Spivak \cite{spivak-lifting14}%
	]\label{constr:lifting}
	In a category \GG, take an object $S$ called a {\em schema}, and 
	
	\noindent 
	\begin{minipage}[h]{0.825\linewidth}
		then \mor s into $S$ are {\em possible instances (preinstances)}.  A {\em lifting constraint} over $S$  is a pair $\lamm=(m_ \lamm, n_\lamm)$ of  \mor s
		\frar{m_\lamm}{W_\lamm}{R_\lamm} and \frar{n_\lamm}{R_\lamm}{S} (see the inset diagram). An instance \frar{t}{X_t}{S} {\em satisfies} \lamm\ if for any \frar{x}{W}{X_t} making commutative square  in the diagram 
		there is a diagonal arrow \frar{\ell}{R}{X_t} (called a {\em lift}) such that both triangles commute. Then we write $t\modelsLiftX{\CC}\lamm$. \qed
	\end{minipage}
	\hspace{1ex}
	\begin{minipage}[h]{0.1\linewidth}
		\(\xymatrix{ 
			W_\lamm 
			\ar@{->}[r]^{x}
			\ar[d]_{m_\lamm}
			& X_t \ar[d]^t
			\\
			R_\lamm	
			\ar@{->}[r]^{n_\lamm}
			\ar@{-->}[ru]^\ell
			& S}  
	\)
\end{minipage}
\end{defN}
It is immediate to see that the diagram \cref{eqdiag:injectivity}(a'), and the inset diagram above reproduced with (a') names as diagram \cref{eqdiag:injectivity}(b), are the same. 
\begin{propo}[Equivalence] 
\label{propo:impl2injectivity} 
A regular constraint $c$ with formula $(f_c, p_c, q_c)$ gives rise to a lifting constraint $\lamm(c)$ over the schema $G_c$ with $m_{\lamm(c)}=f_c$,  $n_{\lamm(c)}=q_c$ such that 
$t\modelsLiftX{\GG}\lamm(c)$ ~iff~ $t\modelsInjX{\GG/G_c}\tauuX{c}$  for any preinstance $t$ over $S$ (note the change of the carrier category).
Conversely, any lifting constraint $\lamm=(m_\lamm, n_\lamm)$ over schema $S$ gives rise to a regular constraint $c(\lamm)$ with arity $G_c=S$ and  formula  $\tauuX{c(\lamm)}=(m_\lamm.n_\lamm, m_\lamm, n_\lamm)$ such that ~$t\modelsInjX{\GG/S}\tauuX{c(\lamm)}$  ~iff~ $t\modelsLiftX{\GG}\lamm$ for any $S$-preinstance $t$.  \shortlong{}{Obviously, 
	$\tauuX{\lamm(c)}=\tauuX{c}$ and $\lamm(\tauuX{c(\lamm)})=\lamm$. \zd{fix notation or remove at al!!}}
\qed\end{propo} 
\begin{corol}[Complete logic for the lifting constraint logic]
Four inference rules \cref{eq:injLogic-rules} \cref{sec:injectivityPrimer} for injectivity provide a sound and complete deductive calculus for the lifting constraint logic. 
\end{corol}

%
%

\renewcommand{\predsig}{\enma{\mathcal{C}}}
\subsection{DCL and sketches: an outline}\label{sec:skeIntro-Sat}
The  goal of the section is to define the main ingredients of the DC logic  in a quick manner to provide basic intuitions and to outline the \fwk\ without being too systematic---a detailed formal  presentation will follow in \cref{sec:skeFormal-heading}. The outline consists of  two parts: homogeneous  and heterogeneous DCL; the latter is important for interoperability. 

We begin with fixing an arbitrary \caty\ with pullbacks (PBC) \GG, whose objects will be called {\em graphs}. A {\em constraint signature} over \GG\ is a collection of constraints \predsig\  as defined in Def. \ref{def:abstrConstr}, \ie, a set of triples $ (c,G_c, \semm{c}) $ indexed by $c\in \carr{\predsig}\in\Ob\setcat$.

\mysubsubXY{3.3.1}{Homogeneous DCL}
Now we take an arbitrary graph $G\in\Ob\GG$ and consider it as a data schema. Possible instances (or preinstances) over this schema are just \mor s into $G$, \ie, graphs typed over $G$, and we define  $\Preinstcat_{G}=\GG/G$. Constraints for such instances are \predsig-labelled diagrams in $G$, \ie, pairs  $c=(\bar c,\bmap_c)$ of a constraint name $\bar c\in\carr{\predsig}$ and a \GG-arrow \frar{\bmap_c}{G_{\bar c}}{G_S} called a {\em binding mapping}. In an accurate terminology, such a pair should be called a {\em constraint declaration} but following an established imprecise software engineering jargon, we will often refer to both constraint declarations  and constraint names from $\predsig$ as to constraints delegating the distinction to the context of the phrase. Given $G$, we define set  $\ConstrsetXbra{G}=\comprcolon{(c,\bmap_c)}{c\in\carr{\predsig}, \bmap\in\GG(G_c, G)}$. 

Now for an instance $t\in\GG/G$ and a constraint $c\in \ConstrsetXbra{G}$, 
we define $t\models_G c$ iff pulling $t$ back along $\bmap_c$  results in a valid instance of $c$, \ie, iff $\restr{t}{\bmap_c}\in\semm{c}$, where \frar{\restr{t}{\bmap_c}}{\bmap_c^*(X)}{G_{c}} is a pullback image of $t$ (defined up to iso). This defines a  Sat-relation $\models_G\ssubset \Obb \GG/G \timm \ConstrsetXbra{G}$ and we can follow usual logical patterns in defining theories and closed classes of models. Specifically, the notion of a conjunctive theory becomes the notion of a {\em sketch}: the latter is a pair $S=(G_S, C_S)$ of a graph $G_S$ and a set $C_S\subset\ConstrsetXbra{G}$ of constraints over it.  A preinstance $t\in\Obb \GG/G$ is a {\em valid $S$-instance}, $t\modelss S$,  if $t\models_{G_S} c$ for all $c\in C_S$. This defines a class  $\Instcat(S)\subset\Preinstcat(S) := \GG/G_S$ of all valid $S$-instances. 

It is easy to see that 	constraints considered in the example of \cref{sec:vehOntology} can be seen as constituting a constraint signature \driveVehOntologyXind{\predsig} over the  \caty\ \graphcat\ as \GG. Furthermore, the database schema $S$ 
discussed in that section is nothing but a sketch in this signature, and validity of its instances is exactly the validity of instances over sketch $S$. 

\mysubsubXY{3.3.1}{Heterogeneous  DCL} Consideration above define a family 
\[ 
\satsetFamilyGG\ 
\]
of Sats. 

A major interoperability issue is how Sats over different schemas are related when a schema changes along \mor\  \frar{f}{G}{G'}:  
What can we say about interrelation between $\models_G$ and $\models_{G'}$? 
\begin{corol}[Sat-axiom for DCL]
	For any ``twisted'' pair of an instance $t\in \GG/G'$ and a constraint $c\in\ConstrcatXbra{G}$, the logical equivalence holds:  
	\begin{equation}\label{eq:twistedSat-cartesian} 
		f^*(t)\models_G c \text{ iff } t\models_{G'} f^\conind_*(c)
	\end{equation}
where functor  \frar{f^\conind_*}{\ConstrcatXbra{G}}{\ConstrcatXbra{G'}} is defined by setting $f^\conind_*(c,\bmap)=(c, f_*(\bmap))$. 
\qed \end{corol}
The above means that our Sat-family \satsetFamilyGG\ with twisted \mor s $(f^*,f_*^\conind)$ form an institution and hence conforms to normal Sat-interoperability patterns. 
%
%
In \cref{sec:skeFormal-heading}, we will generalize the above for more categorificated setting with arrows between constraints and, most importantly, for Sats supplied with evidence (e-institutions), and prove that any constraint signature gives rise to an e-institution again. 

%
%
\renewcommand{\Pobj}{\enma{\M\timm\S}}
\newcommand{\PobjTwix}{\enma{\M'\timm\S}}
\renewcommand{\PobjPrim}{\enma{\M'\timm\S'}}
\renewcommand{\satToSat}{\enma{\rho}}

\renewcommand{\id}{\enma{\mathsf{id}}}

\newcommand{\rr}{\enma{\mathbf{r}}}
\renewcommand{\ss}{\enma{\mathbf{s}}}

\newcommand{\TrispanToBispanMS}{
\begin{tikzcd}[ampersand replacement=\&,sep=small]
	\& \modelsd \&\& \E \\
	\M \& \textcolor{rgb,255:red,92;green,92;blue,214}{\M\timm\S} \& \S
	\arrow["\pM"', from=1-2, to=2-1]
	\arrow["\pS", from=1-2, to=2-3]
	\arrow[draw={rgb,255:red,92;green,92;blue,214}, dashed, from=2-2, to=2-3]
	\arrow["\pMS", color={rgb,255:red,92;green,92;blue,214}, dashed, from=1-2, to=2-2]
	\arrow[dashed, from=2-2, to=2-1]
	\arrow["\pE", from=1-2, to=1-4]
\end{tikzcd}
}

\newcommand{\TrispanToBispanSE}{
\begin{tikzcd}[ampersand replacement=\&, sep=scriptsize]
	\& \modelsd \&\& \E \\
	\M \&\& \S \& \textcolor{rgb,255:red,80;green,104;blue,226}{\S\timm\E}
	\arrow["\pE", from=1-2, to=1-4]
	\arrow["\pM"', from=1-2, to=2-1]
	\arrow["\pS"', from=1-2, to=2-3]
	\arrow["\pSE", color={rgb,255:red,80;green,104;blue,226}, dashed, from=1-2, to=2-4]
	\arrow[draw={rgb,255:red,80;green,104;blue,226}, dashed, from=2-4, to=1-4]
	\arrow[draw={rgb,255:red,80;green,104;blue,226}, dashed, from=2-4, to=2-3]
\end{tikzcd}
}

\newcommand{\switchDiag}{
\begin{tikzcd}[ampersand replacement=\&, row sep=scriptsize, column sep=small]
	{\M'\timm \S} \&\& {\modelsd_\twimor} \&\& \E \\
	\\
	{\M'} \&\& {\modelsr_\twimor} \&\& \S\timm\E
	\arrow["{\pi_{M'}}"', from=1-1, to=3-1]
	\arrow[from=1-3, to=1-1]
	\arrow[from=1-3, to=1-5]
	\arrow[from=3-3, to=3-1]
	\arrow[from=3-3, to=3-5]
	\arrow["{\pi_E}"', from=3-5, to=1-5]
	\arrow[dashed, from=1-3, to=3-1]
	\arrow[dashed, from=3-3, to=1-5]
	\arrow["{\switchX{\twimor}}"', from=1-3, to=3-3]
	\arrow["\cong"', curve={height=-12pt}, draw=none, from=1-3, to=3-3]
\end{tikzcd}
}

\newcommand{\switchDiagOddBig}{
\begin{tikzcd}[ampersand replacement=\&,sep=small]
	\&\& {\modelsd'_\twimor} \\
	{\M'\timm \S} \&\& {\modelsd_\twimor} \&\& \E \\
	\\
	{\M'} \&\& {\modelsr_\twimor} \&\& \S\timm\E \\
	\&\& {\modelsr'_\twimor}
	\arrow["{\pi_{M'}}"', from=2-1, to=4-1]
	\arrow[from=2-3, to=2-1]
	\arrow[from=2-3, to=2-5]
	\arrow[from=4-3, to=4-1]
	\arrow[from=4-3, to=4-5]
	\arrow["{\pi_E}"', from=4-5, to=2-5]
	\arrow[dashed, from=2-3, to=4-1]
	\arrow[dashed, from=4-3, to=2-5]
	\arrow["{\switchX{\twimor}}"', from=2-3, to=4-3]
	\arrow["{i_\twimor}", from=2-3, to=1-3]
	\arrow[from=1-3, to=2-1]
	\arrow[from=1-3, to=2-5]
	\arrow[from=5-3, to=4-1]
	\arrow[from=5-3, to=4-5]
	\arrow["{j_\twimor}", from=4-3, to=5-3]
	\arrow["\cong"', shift right=1, draw=none, from=2-3, to=1-3]
	\arrow["\cong"', shift right=1, draw=none, from=4-3, to=5-3]
	\arrow["\cong"', curve={height=-12pt}, draw=none, from=2-3, to=4-3]
\end{tikzcd}
}

\newcommand{\ternarySpanInCCinMtoSEform}{
\begin{tikzcd}[ampersand replacement=\&,column sep=small,row sep=scriptsize]
	\& \models \&\& \Evicat \\
	\Modcat \& \Sencat \&\& \textcolor{rgb,255:red,92;green,92;blue,214}{\Sencat\timm\Evicat}
	\arrow["{q_E}", from=1-2, to=1-4]
	\arrow["p"', from=1-2, to=2-1]
	\arrow["{q_S}"', from=1-2, to=2-2]
	\arrow[color={rgb,255:red,92;green,92;blue,214}, dashed, from=2-4, to=2-2]
	\arrow[color={rgb,255:red,92;green,92;blue,214}, dashed, from=2-4, to=1-4]
	\arrow["q"{description}, color={rgb,255:red,92;green,92;blue,214}, dashed, from=1-2, to=2-4]
\end{tikzcd}
}

\newcommand{\twistedMorphCompose}{
\begin{tikzcd}[ampersand replacement=\&,sep=scriptsize]
	\&\& {P_{\twimor.\twimor'}} \\
	\\
	\&\& {\#} \\
	P \& {P_\gamm} \& {P'} \& {P_{\gamm'}} \& {P''} \\
	\\
	\&\& \E
	\arrow["{1_S\timm r}", color={rgb,255:red,92;green,92;blue,214}, from=4-2, to=4-1]
	\arrow["{s\timm1_{M'}}"', color={rgb,255:red,92;green,92;blue,214}, from=4-2, to=4-3]
	\arrow["{1_{S'}\timm r'}", color={rgb,255:red,92;green,92;blue,214}, from=4-4, to=4-3]
	\arrow["{s'\timm1_{M''}}"', color={rgb,255:red,92;green,92;blue,214}, from=4-4, to=4-5]
	\arrow["{\rr:= 1_S\timm(r'.r)}"', curve={height=12pt}, from=1-3, to=4-1]
	\arrow["{1_S\timm r'}"{description, pos=0.7}, color={rgb,255:red,92;green,92;blue,214}, curve={height=6pt}, dashed, from=1-3, to=4-2]
	\arrow["{s\timm1_{M''}}"{description, pos=0.7}, color={rgb,255:red,92;green,92;blue,214}, curve={height=-12pt}, dashed, from=1-3, to=4-4]
	\arrow["{\ss:=(s.s')\timm1_{M''}}", curve={height=-12pt}, from=1-3, to=4-5]
	\arrow[dotted, from=3-3, to=4-2]
	\arrow[dotted, from=3-3, to=4-4]
	\arrow["{!_\#}", shift left=1, dotted, from=1-3, to=3-3]
	\arrow["{!_\timm}", shift left=1, color={rgb,255:red,92;green,92;blue,214}, dashed, from=3-3, to=1-3]
	\arrow["{\text{[pb]}}"{description}, color={rgb,255:red,150;green,150;blue,150}, draw=none, from=3-3, to=4-3]
	\arrow["\models"'{pos=0.3}, "\shortmid"{marking}, curve={height=12pt}, from=4-1, to=6-3]
	\arrow["{\models''}"{pos=0.3}, "\shortmid"{marking}, curve={height=-12pt}, from=4-5, to=6-3]
	\arrow["{\models'}", "\shortmid"{marking}, from=4-3, to=6-3]
	\arrow["\gapnamegap{1}{\models_\gamm}{-1}"'{pos=0.4}, "\shortmid"{marking}, from=4-2, to=6-3]
	\arrow["{\models_{\gamm'}}"{pos=0.4}, "\shortmid"{marking}, from=4-4, to=6-3]
\end{tikzcd}
}

\newcommand{\twistedMorphDefMainForAssurLevels}{ 
\begin{tikzcd}[ampersand replacement=\&,sep=small]
	\textcolor{rgb,255:red,210;green,86;blue,55}{M} \&\& \textcolor{rgb,255:red,210;green,86;blue,55}{\vDash} \&\& {S\timm_L\, E} \\
	\\
	\&\& {\twixU{\vDash}_f} \\
	\\
	\textcolor{rgb,255:red,40;green,176;blue,28}{M'} \&\& \textcolor{rgb,255:red,40;green,176;blue,28}{\vDash'} \&\& {S'\timm_L\, E}
	\arrow["{\satToSat_f}"{pos=0.7}, color={rgb,255:red,214;green,92;blue,92}, dashed, from=3-3, to=1-3]
	\arrow["{q'}", color={rgb,255:red,40;green,176;blue,28}, from=5-3, to=5-5]
	\arrow["{f^*}", from=5-1, to=1-1]
	\arrow["p"'{text={rgb,255:red,214;green,92;blue,92}}, draw={rgb,255:red,210;green,86;blue,55}, from=1-3, to=1-1]
	\arrow["q", color={rgb,255:red,210;green,86;blue,55}, from=1-3, to=1-5]
	\arrow["{p_f}"', from=3-3, to=5-1]
	\arrow["{q_f}"', from=3-3, to=1-5]
	\arrow["{p'}"', color={rgb,255:red,40;green,176;blue,28}, from=5-3, to=5-1]
	\arrow["{f_*\timm\id}", from=1-5, to=5-5]
	\arrow["{\text{[pb]}}"{description, pos=0.2}, color={rgb,255:red,210;green,86;blue,55}, draw=none, from=3-3, to=1-1]
	\arrow["{\satToSat'_f}"{pos=0.7}, color={rgb,255:red,40;green,176;blue,28}, dashed, from=3-3, to=5-3]
	\arrow["{\text{[pb]}}"{description, pos=0.2}, color={rgb,255:red,40;green,176;blue,28}, draw=none, from=3-3, to=5-5]
\end{tikzcd}
}

\newcommand{\twiMorphDefViaSpansSEtoM}{
\begin{tikzcd}[ampersand replacement=\&]
	\textcolor{rgb,255:red,214;green,92;blue,92}{M} \&\& \textcolor{rgb,255:red,214;green,92;blue,92}{\S\timm\E} \\
	\& \cong \\
	\textcolor{rgb,255:red,43;green,190;blue,30}{M'} \&\& \textcolor{rgb,255:red,40;green,176;blue,28}{\S'\timm \E}
	\arrow["{\morAsspanXup{r}}"', "\shortmid"{marking}, from=1-1, to=3-1]
	\arrow["\modelsr"', "\shortmid"{marking, text={rgb,255:red,214;green,92;blue,92}}, color={rgb,255:red,214;green,92;blue,92}, from=1-3, to=1-1]
	\arrow["{\modelsr'}", "\shortmid"{marking, text={rgb,255:red,40;green,176;blue,28}}, color={rgb,255:red,40;green,176;blue,28}, from=3-3, to=3-1]
	\arrow["{\morAsspanXdn{{(s\timm 1_\E)}}}", "\shortmid"{marking}, from=1-3, to=3-3]
	\arrow["{\modelsr_\twimor}"{description, pos=0.4}, color={rgb,255:red,214;green,92;blue,92}, curve={height=12pt}, from=1-3, to=3-1]
	\arrow["{\modelsr'_\twimor}"{description, pos=0.6}, color={rgb,255:red,40;green,176;blue,28}, curve={height=-12pt}, from=1-3, to=3-1]
\end{tikzcd}
}

\newcommand{ \twiMorphDefUnpackedSEtoM}{
\begin{tikzcd}[ampersand replacement=\&,sep=small]
	\textcolor{rgb,255:red,210;green,86;blue,55}{M} \&\& \textcolor{rgb,255:red,210;green,86;blue,55}{\modelsr} \&\& \S\timm\E \\
	\\
	\&\& {\modelsr_\twimor} \\
	\\
	\textcolor{rgb,255:red,40;green,176;blue,28}{M'} \&\& \textcolor{rgb,255:red,40;green,176;blue,28}{\modelsr'} \&\& {\S'\timm\E}
	\arrow["{\varrho_\twimor}"{pos=0.7}, color={rgb,255:red,214;green,92;blue,92}, dashed, from=3-3, to=1-3]
	\arrow["{q'}", color={rgb,255:red,40;green,176;blue,28}, from=5-3, to=5-5]
	\arrow["r", from=5-1, to=1-1]
	\arrow["w"'{text={rgb,255:red,214;green,92;blue,92}}, draw={rgb,255:red,210;green,86;blue,55}, from=1-3, to=1-1]
	\arrow["v", color={rgb,255:red,210;green,86;blue,55}, from=1-3, to=1-5]
	\arrow["{w_\twimor}"', dashed, from=3-3, to=5-1]
	\arrow["{v_\twimor}"', dashed, from=3-3, to=1-5]
	\arrow["{w'}"', color={rgb,255:red,40;green,176;blue,28}, from=5-3, to=5-1]
	\arrow["{s\timm 1_E}", from=1-5, to=5-5]
	\arrow["{\text{[pb]}}"{description, pos=0.2}, color={rgb,255:red,210;green,86;blue,55}, draw=none, from=3-3, to=1-1]
	\arrow["{\varrho'_f}"{pos=0.7}, color={rgb,255:red,40;green,176;blue,28}, dashed, from=3-3, to=5-3]
	\arrow["{\text{[pb]}}"{description, pos=0.2}, color={rgb,255:red,40;green,176;blue,28}, draw=none, from=3-3, to=5-5]
\end{tikzcd}
}

\newcommand{\twistedPBdiag}{
\begin{tikzcd}[ampersand replacement=\&,sep=scriptsize]
	\&\&\&\& \textcolor{rgb,255:red,210;green,86;blue,55}{\S} \\
	\&\& \textcolor{rgb,255:red,210;green,86;blue,55}{\M\timm\S} \&\& \textcolor{rgb,255:red,210;green,86;blue,55}{\vDash} \&\& \S\timm\E \\
	\\
	\M \&\& {\M'\timm S} \&\& \textcolor{rgb,255:red,214;green,92;blue,92}{\modelsd_\twimor} \& E \\
	\\
	\textcolor{rgb,255:red,40;green,176;blue,28}{\M'} \&\& \textcolor{rgb,255:red,40;green,176;blue,28}{\circ} \&\& \textcolor{rgb,255:red,153;green,92;blue,214}{\modelsr_\twimor}
	\arrow[draw={rgb,255:red,210;green,86;blue,55}, from=2-3, to=1-5]
	\arrow[""{name=0, anchor=center, inner sep=0}, "\pMS"', color={rgb,255:red,210;green,86;blue,55}, from=2-5, to=2-3]
	\arrow["\pMStwi"', color={rgb,255:red,214;green,92;blue,92}, dashed, from=4-5, to=4-3]
	\arrow[""{name=1, anchor=center, inner sep=0}, draw={rgb,255:red,153;green,92;blue,214}, dashed, from=6-5, to=6-3]
	\arrow[""{name=2, anchor=center, inner sep=0}, "{!_{\#1}}"{description}, from=4-3, to=6-3]
	\arrow["{r\timm 1_S}"'{pos=0.3}, color={rgb,255:red,214;green,92;blue,92}, from=4-3, to=2-3]
	\arrow[draw={rgb,255:red,214;green,92;blue,92}, dashed, from=4-5, to=2-5]
	\arrow["{\text{[pb]}_2}"{description, pos=0.6}, color={rgb,255:red,153;green,92;blue,214}, curve={height=12pt}, draw=none, from=4-3, to=6-5]
	\arrow[""{name=3, anchor=center, inner sep=0}, "\pE", color={rgb,255:red,210;green,86;blue,55}, from=2-5, to=4-6]
	\arrow["\pEtwi", color={rgb,255:red,214;green,92;blue,92}, from=4-5, to=4-6]
	\arrow[""{name=4, anchor=center, inner sep=0}, "{\text{[pb]}_0}"{description, pos=0.3}, "\lrcorner"{anchor=center, pos=0.125, rotate=180}, color={rgb,255:red,214;green,92;blue,92}, draw=none, from=4-5, to=2-3]
	\arrow[draw={rgb,255:red,214;green,92;blue,92}, from=2-3, to=1-5]
	\arrow[""{name=5, anchor=center, inner sep=0}, "\pSE"', color={rgb,255:red,153;green,92;blue,214}, from=2-5, to=2-7]
	\arrow[""{name=6, anchor=center, inner sep=0}, draw={rgb,255:red,153;green,92;blue,214}, dashed, from=6-3, to=6-1]
	\arrow[draw={rgb,255:red,214;green,92;blue,92}, from=4-3, to=6-1]
	\arrow["r", from=6-1, to=4-1]
	\arrow["{\pi_M}"', from=2-3, to=4-1]
	\arrow[draw={rgb,255:red,153;green,92;blue,214}, curve={height=-12pt}, dashed, from=6-3, to=2-3]
	\arrow[draw={rgb,255:red,153;green,92;blue,214}, curve={height=-24pt}, dashed, from=6-5, to=2-5]
	\arrow["{\text{[pb]}_1}"{description, pos=0.3}, "\lrcorner"{anchor=center, pos=0.125, rotate=180}, color={rgb,255:red,153;green,92;blue,214}, draw=none, from=6-3, to=4-1]
	\arrow["w"{description, pos=0.3}, curve={height=6pt}, from=2-5, to=4-1]
	\arrow[from=2-7, to=1-5]
	\arrow[from=2-7, to=4-6]
	\arrow["{!_{\timm3}}"{description, pos=0.7}, from=6-5, to=4-3]
	\arrow[""{name=7, anchor=center, inner sep=0}, "{!_{\#2}}"{description}, shift right=2, from=4-5, to=6-5]
	\arrow["{!_{\#4}}"{description}, shift right=3, from=6-5, to=4-5]
	\arrow["\pS", from=2-5, to=1-5]
	\arrow[""{name=8, anchor=center, inner sep=0}, draw=none, from=1-5, to=4-3]
	\arrow["\pSEtwi"', color={rgb,255:red,153;green,92;blue,214}, curve={height=24pt}, from=6-5, to=2-7]
	\arrow["\pMtwiPrim", curve={height=-12pt}, from=6-5, to=6-1]
	\arrow[draw={rgb,255:red,210;green,86;blue,55}, curve={height=-6pt}, squiggly, no head, from=3, to=0]
	\arrow[draw=none, from=1-5, to=4]
	\arrow[draw={rgb,255:red,153;green,92;blue,214}, shorten >=12pt, squiggly, no head, from=6, to=2]
	\arrow[draw={rgb,255:red,153;green,92;blue,214}, shorten >=11pt, squiggly, no head, from=1, to=7]
	\arrow[draw={rgb,255:red,153;green,92;blue,214}, curve={height=12pt}, shorten >=9pt, squiggly, no head, from=5, to=8]
\end{tikzcd}
}

\newcommand{\twistedMorphDefPackedForAssurLevel}{   
\begin{tikzcd}[ampersand replacement=\&,sep=small]
	\& \textcolor{rgb,255:red,214;green,92;blue,92}{M} \&\& \textcolor{rgb,255:red,214;green,92;blue,92}{S\timm_L\,E} \\
	\&\&\&\& \textcolor{rgb,255:red,172;green,166;blue,216}{S\timm_L\,E} \\
	\textcolor{rgb,255:red,172;green,166;blue,216}{M'} \\
	\& \textcolor{rgb,255:red,43;green,190;blue,30}{M'} \&\& \textcolor{rgb,255:red,40;green,176;blue,28}{S'\timm_L\,E}
	\arrow["{f^*}", draw={rgb,255:red,172;green,166;blue,216}, from=3-1, to=1-2]
	\arrow[draw={rgb,255:red,172;green,166;blue,216}, Rightarrow, no head, from=3-1, to=4-2]
	\arrow["{f^{**}}"', "\shortmid"{marking}, from=1-2, to=4-2]
	\arrow["\vDash"', "\shortmid"{marking, text={rgb,255:red,214;green,92;blue,92}}, color={rgb,255:red,214;green,92;blue,92}, from=1-4, to=1-2]
	\arrow["{\vDash'}", "\shortmid"{marking, text={rgb,255:red,40;green,176;blue,28}}, color={rgb,255:red,40;green,176;blue,28}, from=4-4, to=4-2]
	\arrow[draw={rgb,255:red,172;green,166;blue,216}, Rightarrow, no head, from=1-4, to=2-5]
	\arrow["{f_*\timm\id}", draw={rgb,255:red,172;green,166;blue,216}, from=2-5, to=4-4]
	\arrow["{(f_*\timm\id)_*}", "\shortmid"{marking}, from=1-4, to=4-4]
	\arrow["{\twixsatXdn{f}}"{description}, dashed, from=1-4, to=4-2]
\end{tikzcd}
}

\newcommand{\esatAxiomNewToOld}{
\begin{tikzcd}[ampersand replacement=\&,column sep=scriptsize,row sep=small]
	\Sigmaa \& \Modcatsig \& \Sencatsig\timm\Evicat \\
	\\
	{\Sigmaa'} \& \ModcatsigPrim \& \SencatsigPrim\timm\Evicat
	\arrow["f"{description}, from=1-1, to=3-1]
	\arrow["{f^*}", from=3-2, to=1-2]
	\arrow["{f^\ell_*\timm\, \mathsf{id}}"', from=1-3, to=3-3]
	\arrow["{\models_\Sigmaa}"{description}, draw=none, from=1-2, to=1-3]
	\arrow["{\models_{\Sigmaa'}}"{description}, draw=none, from=3-2, to=3-3]
\end{tikzcd}
}

\newcommand{\esatAxiomNew}{
\begin{tikzcd}[ampersand replacement=\&,sep=small]
	\Sigmaa \& \Modcatsig \& \eSencatsig \& {\Sencatsig\timm_L\,\Evicat} \\
	\\
	{\Sigmaa'} \& \ModcatsigPrim \& \eSencatsigPrim \& {\SencatsigPrim\timm_L\,\Evicat}
	\arrow["f"{description}, from=1-1, to=3-1]
	\arrow["{f^*}", from=3-2, to=1-2]
	\arrow["{f^\e_*}"', from=1-3, to=3-3]
	\arrow["{f^\ell_*\timm\, \mathsf{id}}"', from=1-4, to=3-4]
	\arrow["{=}"{description}, draw=none, from=1-3, to=1-4]
	\arrow["{=}"{description}, draw=none, from=3-3, to=3-4]
	\arrow["{\models_\Sigmaa}"{description}, dotted, tail reversed, from=1-2, to=1-3]
	\arrow["{\models_{\Sigmaa'}}"{description}, dotted, tail reversed, from=3-2, to=3-3]
\end{tikzcd}
}

\newcommand{\twiMorphDefViaSpansMStoE}{
\begin{tikzcd}[ampersand replacement=\&,column sep=scriptsize,row sep=small]
	\textcolor{rgb,255:red,210;green,86;blue,55}{\M} \&\& \textcolor{rgb,255:red,210;green,86;blue,55}{\Pobj} \\
	\textcolor{rgb,255:red,210;green,86;blue,55}{\S} \\
	\&\& \PobjTwix \&\&\& E \\
	\textcolor{rgb,255:red,40;green,176;blue,28}{\M'} \\
	\textcolor{rgb,255:red,43;green,190;blue,30}{\S'} \&\& \textcolor{rgb,255:red,40;green,176;blue,28}{\PobjPrim}
	\arrow[draw={rgb,255:red,210;green,86;blue,55}, from=1-3, to=2-1]
	\arrow[draw={rgb,255:red,210;green,86;blue,55}, from=1-3, to=1-1]
	\arrow[draw={rgb,255:red,40;green,176;blue,28}, from=5-3, to=4-1]
	\arrow[draw={rgb,255:red,43;green,190;blue,30}, from=5-3, to=5-1]
	\arrow["{\morAsspanXdn{{(1_{M'}\timm s)}}}"', "\shortmid"{marking, text={rgb,255:red,40;green,176;blue,28}}, color={rgb,255:red,40;green,176;blue,28}, from=3-3, to=5-3]
	\arrow["{\morAsspanXdn{{(r\timm 1_\S)}}}", "\shortmid"{marking, text={rgb,255:red,214;green,92;blue,92}}, color={rgb,255:red,214;green,92;blue,92}, from=3-3, to=1-3]
	\arrow["r", curve={height=-18pt}, from=4-1, to=1-1]
	\arrow["s"', curve={height=18pt}, from=2-1, to=5-1]
	\arrow[draw={rgb,255:red,214;green,92;blue,92}, from=3-3, to=2-1]
	\arrow[draw={rgb,255:red,40;green,176;blue,28}, from=3-3, to=4-1]
	\arrow["\modelsd", "\shortmid"{marking, text={rgb,255:red,210;green,86;blue,55}}, color={rgb,255:red,210;green,86;blue,55}, from=1-3, to=3-6]
	\arrow["{\modelsd'}"', "\shortmid"{marking, text={rgb,255:red,40;green,176;blue,28}}, color={rgb,255:red,40;green,176;blue,28}, from=5-3, to=3-6]
	\arrow["{\modelsd_\twimor}"{description}, color={rgb,255:red,210;green,86;blue,55}, curve={height=-12pt}, from=3-3, to=3-6]
	\arrow["{\modelsd'_\twimor}"{description}, color={rgb,255:red,40;green,176;blue,28}, curve={height=12pt}, from=3-3, to=3-6]
	\arrow["\cong"{description}, draw=none, from=3-3, to=3-6]
\end{tikzcd}
}

\newcommand{\twiMorphDefUnpackedMStoE}{
\begin{tikzcd}[ampersand replacement=\&,sep=scriptsize]
	\textcolor{rgb,255:red,210;green,86;blue,55}{\M\timm\S} \&\& \textcolor{rgb,255:red,210;green,86;blue,55}{\modelsd} \\
	\\
	{\M'\timm\S} \&\& {\modelsd_\twimor} \&\& E \\
	\\
	\textcolor{rgb,255:red,40;green,176;blue,28}{\M'\timm\S'} \&\& \textcolor{rgb,255:red,40;green,176;blue,28}{\modelsd'}
	\arrow["p"', color={rgb,255:red,210;green,86;blue,55}, from=1-3, to=1-1]
	\arrow["{p_\twimor}"', dashed, from=3-3, to=3-1]
	\arrow["{p'}"', color={rgb,255:red,40;green,176;blue,28}, from=5-3, to=5-1]
	\arrow["{1_{M'}\timm s}"{description}, color={rgb,255:red,40;green,176;blue,28}, from=3-1, to=5-1]
	\arrow["{r\timm 1_S}"{description}, color={rgb,255:red,214;green,92;blue,92}, from=3-1, to=1-1]
	\arrow["{\satToSat_\twimor}"{description}, color={rgb,255:red,214;green,92;blue,92}, dashed, from=3-3, to=1-3]
	\arrow["{\satToSat'_\twimor}"{description}, color={rgb,255:red,40;green,176;blue,28}, dashed, from=3-3, to=5-3]
	\arrow["{\text{[pb]}}"{description, pos=0.7}, color={rgb,255:red,40;green,176;blue,28}, draw=none, from=3-1, to=5-3]
	\arrow["q", color={rgb,255:red,210;green,86;blue,55}, from=1-3, to=3-5]
	\arrow["{q'}"', color={rgb,255:red,40;green,176;blue,28}, from=5-3, to=3-5]
	\arrow["{q_\twimor}", dashed, from=3-3, to=3-5]
	\arrow["{\text{[pb]}}"{description, pos=0.7}, color={rgb,255:red,214;green,92;blue,92}, draw=none, from=1-1, to=3-3]
\end{tikzcd}
}

\newcommand{\twistedMorphDefPackedMStoE}{
\begin{tikzcd}[ampersand replacement=\&,sep=scriptsize]
	\textcolor{rgb,255:red,210;green,86;blue,55}{\Pobj} \\
	\\
	{{P}_\twimor} \&\&\& E \\
	\\
	\textcolor{rgb,255:red,40;green,176;blue,28}{\PobjPrim}
	\arrow["{(1_S\timm r)_{_\%}}", "\shortmid"{marking, text={rgb,255:red,214;green,92;blue,92}}, color={rgb,255:red,214;green,92;blue,92}, from=3-1, to=1-1]
	\arrow["{(s\timm 1_{M'})_{\%}}"', "\shortmid"{marking, text={rgb,255:red,40;green,176;blue,28}}, color={rgb,255:red,40;green,176;blue,28}, from=3-1, to=5-1]
	\arrow["\vDash", "\shortmid"{marking, text={rgb,255:red,210;green,86;blue,55}}, color={rgb,255:red,210;green,86;blue,55}, from=1-1, to=3-4]
	\arrow["{\vDash'}"', color={rgb,255:red,40;green,176;blue,28}, from=5-1, to=3-4]
	\arrow["{\modelsd_\twimor}"{description}, color={rgb,255:red,214;green,92;blue,92}, curve={height=-12pt}, dashed, from=3-1, to=3-4]
	\arrow["{\modelsd'_\twimor}"{description}, color={rgb,255:red,40;green,176;blue,28}, curve={height=12pt}, from=3-1, to=3-4]
\end{tikzcd}
}

\newcommand{\satAxiomPackedMStoE}{
\begin{tikzcd}[ampersand replacement=\&,sep=small]
	\textcolor{rgb,255:red,214;green,92;blue,92}{M} \&\& \textcolor{rgb,255:red,214;green,92;blue,92}{S} \\
	\\
	\\
	\textcolor{rgb,255:red,43;green,190;blue,30}{M'} \&\& \textcolor{rgb,255:red,40;green,176;blue,28}{S'}
	\arrow["{f^{*^\%}}"', "\shortmid"{marking}, from=1-1, to=4-1]
	\arrow["{\vDash_{MS}}"', "\shortmid"{marking, text={rgb,255:red,214;green,92;blue,92}}, color={rgb,255:red,214;green,92;blue,92}, from=1-3, to=1-1]
	\arrow["{\vDash_{M'S'}}", "\shortmid"{marking, text={rgb,255:red,40;green,176;blue,28}}, color={rgb,255:red,40;green,176;blue,28}, from=4-3, to=4-1]
	\arrow["{f_{*_\%}}", "\shortmid"{marking}, from=1-3, to=4-3]
	\arrow["{\vDash_{M'S}^f}"{description}, dashed, from=1-3, to=4-1]
\end{tikzcd}
}
	
\newcommand{\morphComposeProof}{
\begin{tikzcd}[ampersand replacement=\&,sep=small]
	\textcolor{rgb,255:red,210;green,86;blue,55}{M} \& \textcolor{rgb,255:red,210;green,86;blue,55}{P} \& \textcolor{rgb,255:red,210;green,86;blue,55}{\vDash} \\
	\textcolor{rgb,255:red,210;green,86;blue,55}{S} \\
	\& {\twixU{P}_f} \& {\twixU{\vDash}_f} \\
	\\
	\& \textcolor{rgb,255:red,79;green,73;blue,233}{P_{f.g}} \& \textcolor{rgb,255:red,79;green,73;blue,233}{\vDash_{f.g}} \& \textcolor{rgb,255:red,40;green,176;blue,28}{P'} \&\& \textcolor{rgb,255:red,40;green,176;blue,28}{\vDash'} \\
	\textcolor{rgb,255:red,40;green,176;blue,28}{M'} \\
	\textcolor{rgb,255:red,43;green,190;blue,30}{S'} \& {\twixU{P}_g} \& {\twixU{\vDash}_g} \\
	{M''} \& {P''} \\
	{S''}
	\arrow[draw={rgb,255:red,210;green,86;blue,55}, dotted, from=1-2, to=2-1]
	\arrow[draw={rgb,255:red,210;green,86;blue,55}, dotted, from=1-2, to=1-1]
	\arrow["p"', color={rgb,255:red,210;green,86;blue,55}, from=1-3, to=1-2]
	\arrow[""{name=0, anchor=center, inner sep=0}, "{p_f}"', dashed, from=3-3, to=3-2]
	\arrow["{p_{fg}}"', color={rgb,255:red,79;green,73;blue,233}, from=5-3, to=5-2]
	\arrow["{1_S\timm f^*}"{description}, color={rgb,255:red,214;green,92;blue,92}, from=3-2, to=1-2]
	\arrow["{\satToSat_f}"{description}, color={rgb,255:red,214;green,92;blue,92}, dashed, from=3-3, to=1-3]
	\arrow["{\ell_f}"{description, pos=0.8}, color={rgb,255:red,79;green,73;blue,233}, dotted, from=5-3, to=3-3]
	\arrow["{\text{[pb]}}"{description, pos=0.8}, color={rgb,255:red,79;green,73;blue,233}, draw=none, from=3-2, to=5-3]
	\arrow["{f_*}"', curve={height=18pt}, from=2-1, to=7-1]
	\arrow[draw={rgb,255:red,214;green,92;blue,92}, dotted, from=3-2, to=2-1]
	\arrow["{\text{[pb]}}"{description, pos=0.8}, color={rgb,255:red,214;green,92;blue,92}, draw=none, from=1-2, to=3-3]
	\arrow[""{name=1, anchor=center, inner sep=0}, "{\satToSat'_f}", color={rgb,255:red,40;green,176;blue,28}, dashed, from=3-3, to=5-6]
	\arrow[color={rgb,255:red,40;green,176;blue,28}, from=3-2, to=5-4]
	\arrow[draw={rgb,255:red,210;green,86;blue,55}, curve={height=-6pt}, from=5-2, to=2-1]
	\arrow[color={rgb,255:red,92;green,92;blue,214}, dotted, from=5-2, to=8-1]
	\arrow["{g_*}"', curve={height=18pt}, from=7-1, to=9-1]
	\arrow["{1_S\timm g^*}"{description}, color={rgb,255:red,79;green,73;blue,233}, from=5-2, to=3-2]
	\arrow["{p'}"', color={rgb,255:red,40;green,176;blue,28}, from=5-6, to=5-4]
	\arrow[from=8-2, to=9-1]
	\arrow[from=7-2, to=7-1]
	\arrow[from=7-2, to=8-1]
	\arrow["{\#}", Rightarrow, from=5-3, to=5-4]
	\arrow["{\ell_g}"{description, pos=0.7}, dotted, from=5-3, to=7-3]
	\arrow["{\satToSat'_g}"{description}, color={rgb,255:red,79;green,73;blue,233}, dashed, from=7-3, to=5-6]
	\arrow[from=7-2, to=5-4]
	\arrow[from=5-2, to=7-2]
	\arrow[dashed, from=7-3, to=7-2]
	\arrow[from=8-2, to=8-1]
	\arrow["{f^*}", curve={height=-18pt}, from=6-1, to=1-1]
	\arrow[draw={rgb,255:red,40;green,176;blue,28}, dotted, from=3-2, to=6-1]
	\arrow["{g^*}", curve={height=-18pt}, from=8-1, to=6-1]
	\arrow["{1_S\timm g^*}"{description}, curve={height=24pt}, from=5-2, to=5-4]
	\arrow["{\text{[]}}"{description}, color={rgb,255:red,40;green,176;blue,28}, shorten <=16pt, shorten >=16pt, dotted, no head, from=0, to=1]
\end{tikzcd}
}


\newcommand{\satAxiomWithVerviaOk}{
	\begin{tikzcd}[ampersand replacement=\&,sep=small]
		\& \textcolor{rgb,255:red,214;green,92;blue,92}{M} \& \Pobj \& \textcolor{rgb,255:red,214;green,92;blue,92}{\Dashv} \\
		\& \textcolor{rgb,255:red,214;green,92;blue,92}{S} \\
		{} \&\& {\twix{P}} \&\&\& \VerObj \& \textcolor{rgb,255:red,214;green,92;blue,92}{\okObj} \\
		\& \textcolor{rgb,255:red,41;green,163;blue,41}{M'} \\
		\& \textcolor{rgb,255:red,41;green,163;blue,41}{S'} \& \textcolor{rgb,255:red,41;green,163;blue,41}{\PobjPrim} \& \textcolor{rgb,255:red,41;green,163;blue,41}{\Dashv'}
		\arrow[color={rgb,255:red,214;green,92;blue,92}, from=1-3, to=2-2]
		\arrow[color={rgb,255:red,214;green,92;blue,92}, from=1-3, to=1-2]
		\arrow[color={rgb,255:red,41;green,163;blue,41}, from=5-3, to=4-2]
		\arrow[color={rgb,255:red,41;green,163;blue,41}, from=5-3, to=5-2]
		\arrow["u"', color={rgb,255:red,214;green,92;blue,92}, dashed, from=1-4, to=1-3]
		\arrow["{u'}", color={rgb,255:red,41;green,163;blue,41}, dashed, from=5-4, to=5-3]
		\arrow["{!'}"', color={rgb,255:red,92;green,92;blue,214}, dashed, from=3-3, to=5-3]
		\arrow["{!}", color={rgb,255:red,96;green,71;blue,235}, dashed, from=3-3, to=1-3]
		\arrow["{f^*}"', color={rgb,255:red,245;green,80;blue,61}, curve={height=18pt}, from=1-2, to=4-2]
		\arrow["{f_*}", color={rgb,255:red,245;green,80;blue,61}, curve={height=-18pt}, from=5-2, to=2-2]
		\arrow[from=3-7, to=3-6]
		\arrow[from=3-3, to=4-2]
		\arrow[from=3-3, to=2-2]
		\arrow["\ver", color={rgb,255:red,214;green,92;blue,92}, dashed, from=1-4, to=3-7]
		\arrow["{\ver'}"', color={rgb,255:red,41;green,163;blue,41}, dashed, from=5-4, to=3-7]
		\arrow["{\twix{\ver}}", curve={height=-12pt}, from=3-3, to=3-6]
		\arrow[color={rgb,255:red,214;green,92;blue,92}, from=1-3, to=3-6]
		\arrow[color={rgb,255:red,41;green,163;blue,41}, from=5-3, to=3-6]
		\arrow["{\twistedSatXdn{f}}"{description}, curve={height=12pt}, tail reversed, from=3-3, to=3-6]
	\end{tikzcd}
}


\newcommand{\vsatDefTripleDiagram}{
\begin{tikzcd}[ampersand replacement=\&,sep=small]
	\& \textcolor{rgb,255:red,210;green,86;blue,55}{M} \& \textcolor{rgb,255:red,210;green,86;blue,55}{\Pobj} \& \textcolor{rgb,255:red,210;green,86;blue,55}{\Dashv} \&\&\&\& \textcolor{rgb,255:red,214;green,92;blue,92}{M} \&\& \textcolor{rgb,255:red,214;green,92;blue,92}{S} \&\& P \\
	\& \textcolor{rgb,255:red,210;green,86;blue,55}{S} \&\&\&\&\&\&\&\&\& \textcolor{rgb,255:red,172;green,166;blue,216}{S} \\
	{} \&\& {\twixU{P}_f} \& {\twistedSatXdn{f}} \&\& E \&\&\&\&\&\& {\twixU{P}_f} \&\& E \\
	\& \textcolor{rgb,255:red,40;green,176;blue,28}{M'} \&\&\&\&\& \textcolor{rgb,255:red,172;green,166;blue,216}{M'} \\
	\& \textcolor{rgb,255:red,43;green,190;blue,30}{S'} \& \textcolor{rgb,255:red,40;green,176;blue,28}{\PobjPrim} \& \textcolor{rgb,255:red,40;green,176;blue,28}{\Dashv'} \&\&\&\& \textcolor{rgb,255:red,43;green,190;blue,30}{M'} \&\& \textcolor{rgb,255:red,40;green,176;blue,28}{S'} \&\& {P'} \\
	\&\&\& {\text{*)} } \&\&\&\&\& {\text{a)}} \&\&\& {\text{b)}}
	\arrow[draw={rgb,255:red,210;green,86;blue,55}, from=1-3, to=2-2]
	\arrow[draw={rgb,255:red,210;green,86;blue,55}, from=1-3, to=1-2]
	\arrow[draw={rgb,255:red,40;green,176;blue,28}, from=5-3, to=4-2]
	\arrow[draw={rgb,255:red,43;green,190;blue,30}, from=5-3, to=5-2]
	\arrow["r"', color={rgb,255:red,210;green,86;blue,55}, from=1-4, to=1-3]
	\arrow["{r_f}"', from=3-4, to=3-3]
	\arrow["{r'}"', color={rgb,255:red,40;green,176;blue,28}, from=5-4, to=5-3]
	\arrow["{!^*_f}", color={rgb,255:red,214;green,92;blue,92}, dashed, from=3-3, to=1-3]
	\arrow["{\satToSat^*_f}"', color={rgb,255:red,214;green,92;blue,92}, dashed, from=3-4, to=1-4]
	\arrow["{\satToSat_*^f}", color={rgb,255:red,40;green,176;blue,28}, dashed, from=3-4, to=5-4]
	\arrow["{\text{[pb]}}"{description}, color={rgb,255:red,214;green,92;blue,92}, draw=none, from=3-3, to=1-4]
	\arrow["{\text{[pb]}}"{description}, color={rgb,255:red,40;green,176;blue,28}, draw=none, from=3-3, to=5-4]
	\arrow["e", color={rgb,255:red,210;green,86;blue,55}, from=1-4, to=3-6]
	\arrow["{e'}"', color={rgb,255:red,40;green,176;blue,28}, from=5-4, to=3-6]
	\arrow["{f^*}", curve={height=-18pt}, from=4-2, to=1-2]
	\arrow["{f_*}"', curve={height=18pt}, from=2-2, to=5-2]
	\arrow["{e_f}", from=3-4, to=3-6]
	\arrow[draw={rgb,255:red,214;green,92;blue,92}, from=3-3, to=2-2]
	\arrow[draw={rgb,255:red,40;green,176;blue,28}, from=3-3, to=4-2]
	\arrow["{f^*}", draw={rgb,255:red,172;green,166;blue,216}, from=4-7, to=1-8]
	\arrow[draw={rgb,255:red,172;green,166;blue,216}, Rightarrow, no head, from=4-7, to=5-8]
	\arrow["{f^{**}}"', "\shortmid"{marking}, from=1-8, to=5-8]
	\arrow["{\vDash_{MS}}"', "\shortmid"{marking, text={rgb,255:red,214;green,92;blue,92}}, color={rgb,255:red,214;green,92;blue,92}, from=1-10, to=1-8]
	\arrow["{\vDash_{M'S'}}", color={rgb,255:red,40;green,176;blue,28}, from=5-10, to=5-8]
	\arrow[draw={rgb,255:red,172;green,166;blue,216}, Rightarrow, no head, from=1-10, to=2-11]
	\arrow["{f_*}", draw={rgb,255:red,172;green,166;blue,216}, from=2-11, to=5-10]
	\arrow["{f_{**}}", "\shortmid"{marking}, from=1-10, to=5-10]
	\arrow["{\vDash_{M'S}^f}"{description}, dashed, from=1-10, to=5-8]
	\arrow["\vDash", "\shortmid"{marking}, from=1-12, to=3-14]
	\arrow["{\vDash'}"', "\shortmid"{marking}, from=5-12, to=3-14]
	\arrow["{!_*^f}", "\shortmid"{marking}, from=3-12, to=1-12]
	\arrow["{!_*^f}"', "\shortmid"{marking}, from=3-12, to=5-12]
	\arrow["{\twistedSatXdn{f}}", "\shortmid"{marking}, from=3-12, to=3-14]
	\arrow["{!_*^f}"', color={rgb,255:red,40;green,176;blue,28}, dashed, from=3-3, to=5-3]
\end{tikzcd}
}

\newcommand{\vsatDefMainDiag}{
\begin{tikzcd}[ampersand replacement=\&,sep=small]
	\& \textcolor{rgb,255:red,210;green,86;blue,55}{M} \& \textcolor{rgb,255:red,210;green,86;blue,55}{\Pobj} \\
	\& \textcolor{rgb,255:red,210;green,86;blue,55}{S} \\
	{} \&\& {\twixU{P}_f} \&\&\& E \\
	\& \textcolor{rgb,255:red,40;green,176;blue,28}{M'} \\
	\& \textcolor{rgb,255:red,43;green,190;blue,30}{S'} \& \textcolor{rgb,255:red,40;green,176;blue,28}{\PobjPrim}
	\arrow[draw={rgb,255:red,210;green,86;blue,55}, from=1-3, to=2-2]
	\arrow[draw={rgb,255:red,210;green,86;blue,55}, from=1-3, to=1-2]
	\arrow[draw={rgb,255:red,40;green,176;blue,28}, from=5-3, to=4-2]
	\arrow[draw={rgb,255:red,43;green,190;blue,30}, from=5-3, to=5-2]
	\arrow["{!^*_f}", color={rgb,255:red,214;green,92;blue,92}, from=3-3, to=1-3]
	\arrow["{f^*}", curve={height=-18pt}, from=4-2, to=1-2]
	\arrow["{f_*}"', curve={height=18pt}, from=2-2, to=5-2]
	\arrow[draw={rgb,255:red,214;green,92;blue,92}, from=3-3, to=2-2]
	\arrow[draw={rgb,255:red,40;green,176;blue,28}, from=3-3, to=4-2]
	\arrow["{!_*^f}"', color={rgb,255:red,40;green,176;blue,28}, from=3-3, to=5-3]
	\arrow["\vDash", "\shortmid"{marking, text={rgb,255:red,210;green,86;blue,55}}, color={rgb,255:red,210;green,86;blue,55}, from=1-3, to=3-6]
	\arrow["{\vDash'}"', color={rgb,255:red,40;green,176;blue,28}, from=5-3, to=3-6]
	\arrow["{\twistedSatXdn{f}}", "\shortmid"{marking}, from=3-3, to=3-6]
\end{tikzcd}
}

\newcommand{\vsatDefMainDiagUnpacked}{
\begin{tikzcd}[ampersand replacement=\&,sep=small]
	\&\& \textcolor{rgb,255:red,210;green,86;blue,55}{\Pobj} \& \textcolor{rgb,255:red,210;green,86;blue,55}{\Dashv} \&\&\&\& \textcolor{rgb,255:red,214;green,92;blue,92}{M} \&\& \textcolor{rgb,255:red,214;green,92;blue,92}{S} \\
	\&\&\&\&\&\&\&\&\&\& \textcolor{rgb,255:red,172;green,166;blue,216}{S} \\
	{} \&\& {\twixU{P}_f} \& {\twistedSatXdn{f}} \& E \\
	\&\&\&\&\&\& \textcolor{rgb,255:red,172;green,166;blue,216}{M'} \\
	\&\& \textcolor{rgb,255:red,40;green,176;blue,28}{\PobjPrim} \& \textcolor{rgb,255:red,40;green,176;blue,28}{\Dashv'} \&\&\&\& \textcolor{rgb,255:red,43;green,190;blue,30}{M'} \&\& \textcolor{rgb,255:red,40;green,176;blue,28}{S'} \\
	\&\&\& {\text{i)} } \&\&\&\&\& {\text{ii)}}
	\arrow["p"', color={rgb,255:red,210;green,86;blue,55}, from=1-4, to=1-3]
	\arrow["{\twixU{p}_f}"', dashed, from=3-4, to=3-3]
	\arrow["{p'}"', color={rgb,255:red,40;green,176;blue,28}, from=5-4, to=5-3]
	\arrow["{!_*^f}"', color={rgb,255:red,40;green,176;blue,28}, dashed, from=3-3, to=5-3]
	\arrow["{!^*_f}", color={rgb,255:red,214;green,92;blue,92}, from=3-3, to=1-3]
	\arrow["{\satToSat^*_f}"', color={rgb,255:red,214;green,92;blue,92}, dashed, from=3-4, to=1-4]
	\arrow["{\satToSat_*^f}", color={rgb,255:red,40;green,176;blue,28}, dashed, from=3-4, to=5-4]
	\arrow["{\text{[pb]}}"{description}, color={rgb,255:red,214;green,92;blue,92}, draw=none, from=3-3, to=1-4]
	\arrow["{\text{[pb]}}"{description}, color={rgb,255:red,40;green,176;blue,28}, draw=none, from=3-3, to=5-4]
	\arrow["e", color={rgb,255:red,210;green,86;blue,55}, from=1-4, to=3-5]
	\arrow["{e'}"', color={rgb,255:red,40;green,176;blue,28}, from=5-4, to=3-5]
	\arrow["{\twixU{e}_f}", from=3-4, to=3-5]
	\arrow["{f^*}", draw={rgb,255:red,172;green,166;blue,216}, from=4-7, to=1-8]
	\arrow[draw={rgb,255:red,172;green,166;blue,216}, Rightarrow, no head, from=4-7, to=5-8]
	\arrow["{f^{**}}"', "\shortmid"{marking}, from=1-8, to=5-8]
	\arrow["{\vDash_{MS}}"', "\shortmid"{marking, text={rgb,255:red,214;green,92;blue,92}}, color={rgb,255:red,214;green,92;blue,92}, from=1-10, to=1-8]
	\arrow["{\vDash_{M'S'}}", color={rgb,255:red,40;green,176;blue,28}, from=5-10, to=5-8]
	\arrow[draw={rgb,255:red,172;green,166;blue,216}, Rightarrow, no head, from=1-10, to=2-11]
	\arrow["{f_*}", draw={rgb,255:red,172;green,166;blue,216}, from=2-11, to=5-10]
	\arrow["{f_{**}}", "\shortmid"{marking}, from=1-10, to=5-10]
	\arrow["{\vDash_{M'S}^f}"{description}, dashed, from=1-10, to=5-8]
\end{tikzcd}
}

\newcommand{\satAxiomWithVer}{
	\begin{tikzcd}[ampersand replacement=\&,sep=small]
		\& \textcolor{rgb,255:red,210;green,86;blue,55}{M} \& \textcolor{rgb,255:red,210;green,86;blue,55}{\Pobj} \&\& \textcolor{rgb,255:red,210;green,86;blue,55}{\Dashv} \\
		\& \textcolor{rgb,255:red,210;green,86;blue,55}{S} \\
		{} \&\& {\twixU{P}_f} \&\& {\twistedSatXdn{f}} \&\& E \\
		\& \textcolor{rgb,255:red,40;green,176;blue,28}{M'} \\
		\& \textcolor{rgb,255:red,43;green,190;blue,30}{S'} \& \textcolor{rgb,255:red,40;green,176;blue,28}{\PobjPrim} \&\& \textcolor{rgb,255:red,40;green,176;blue,28}{\Dashv'}
		\arrow[color={rgb,255:red,210;green,86;blue,55}, from=1-3, to=2-2]
		\arrow[color={rgb,255:red,210;green,86;blue,55}, from=1-3, to=1-2]
		\arrow[color={rgb,255:red,40;green,176;blue,28}, from=5-3, to=4-2]
		\arrow[color={rgb,255:red,43;green,190;blue,30}, from=5-3, to=5-2]
		\arrow["p"', color={rgb,255:red,210;green,86;blue,55}, from=1-5, to=1-3]
		\arrow["{\twixU{p}_f}"', from=3-5, to=3-3]
		\arrow["{p'}"', color={rgb,255:red,40;green,176;blue,28}, from=5-5, to=5-3]
		\arrow["{!'}"', color={rgb,255:red,40;green,176;blue,28}, dashed, from=3-3, to=5-3]
		\arrow["{!}", color={rgb,255:red,214;green,92;blue,92}, dashed, from=3-3, to=1-3]
		\arrow["\satToSat"', color={rgb,255:red,214;green,92;blue,92}, dashed, from=3-5, to=1-5]
		\arrow["\satToSatPrim", color={rgb,255:red,40;green,176;blue,28}, dashed, from=3-5, to=5-5]
		\arrow["{\text{[pb]}}"{description}, color={rgb,255:red,214;green,92;blue,92}, draw=none, from=3-3, to=1-5]
		\arrow["{\text{[pb]}}"{description}, color={rgb,255:red,40;green,176;blue,28}, draw=none, from=3-3, to=5-5]
		\arrow["e", color={rgb,255:red,210;green,86;blue,55}, from=1-5, to=3-7]
		\arrow["{e'}"', color={rgb,255:red,40;green,176;blue,28}, from=5-5, to=3-7]
		\arrow["{f^*}", curve={height=-18pt}, from=4-2, to=1-2]
		\arrow["{f_*}"', curve={height=18pt}, from=2-2, to=5-2]
		\arrow["{\twixU{e}_f}", from=3-5, to=3-7]
		\arrow[color={rgb,255:red,214;green,92;blue,92}, from=3-3, to=2-2]
		\arrow[color={rgb,255:red,40;green,176;blue,28}, from=3-3, to=4-2]
	\end{tikzcd}
}

\renewcommand{\id}{\enma{1}}

\renewcommand{\Modsetsig}{\ModsetX{\Sigmaa}}
\renewcommand{\Sensetsig}{\SensetX{\Sigmaa}}

\newcommand{\V}{\enma{V}}
\renewcommand{\E}{\enma{E}}
\newcommand{\M}{\enma{M}}
\renewcommand{\S}{\enma{S}}

\newcommand{\vsatfunLifted}{\ovr{\vsfun}}
\newcommand{\vsfunLifted}{\vsatfunLifted}

\section{Institutions with evidence}

In the next section, we will begin with the classical definition, discuss its pros and cons, and how to fix the latter. In this discussion, we will use different symbols, \modelsd\ for a classical Sat, and \satascat\ for enriched Sat with evidence and a \caty\ structure. But afterwards, \ie, starting from \cref{sec:eInst-abstract} and on, we will use \modelsd\ for the enriched Sat too. In \cref{sec:eInst-abstract}, we consider the notion of eSat in an abstract setting relative to any finitely complete \caty\ (FCC) \CC, and give a \CC-relative definition of an institution with evidence. In \cref{sec:eInst-concrete} we briefly discuss categorical e-institution with elements. 

\subsection{Background: Why evidential institutions}\label{sec:ske4systems}
\begin{defN}[Institutions as indexed \caties\ and sets: Goguen-Burstall \cite{GoguenBurstall-jofacm92}]\label{def:instByGoguen}
	An {\em institution} is a quadruple 
	$$(\Sigcat, \senfun, \modfun, \models)$$ 
	of a small \caty\ \Sigcat\ of {\em signatures} (think, \eg, of signatures of operation and relation symbols),  functors 
	\frar{\senfun}{\Sigcat}{\setcat} and \frar{\modfun}{\Sigcat}{\opX{\catcat}} assigning to each signature $\Sigmaa\in \Ob\Sigcat$ (think of a set of predicate and operation symbols) its set of {\em (logical) sentences} $\Sensetsig=\senfun(\Sigmaa)$ and \caty\ $\Modsetsig=\modfun(\Sigmaa)$ of {\em models} and {\em model \mor s}, and a family of {\em \sation} relations (or just {\em Sats}) $\modelsd_\Sigmaa\subset\Ob\Modsetsig\timm\Sensetsig$ indexed by $\Sigmaa$ ranging over $\Ob\Sigcat$. These data are required to satisfy the following {\em (twisted) \sation\ axiom}: 
	for a signature \mor\ \frar{f}{\Sigmaa}{\Sigmaa'}  and any $m'\in\Ob\ModsetsigPrim$, $\phi\in\Sensetsig$,  $f^*(m')\modelssig\phi$ iff $m'\modelssigPrim f_*(\phi)$, where $f^*$ and $f_*$ are (standardly) denote functors $\modfun(f)$ and $\senfun(f)$ resp. as shown in diagram below
	%
{
\begin{equation}\label{eqdiag:satAxiom}
\begin{array}{c@{\quad}c} 
\begin{array}{cc} 
\begin{tikzcd}[row sep=4ex, column sep=2ex] 
\Sigmaa 
		\ar[d, "f"]
& \Modsetsig 
		\ar[d, <-, "f^*" ']
		\ar[r, phantom, "\modelssig"] & \Sensetsig  
\\
\Sigmaa'  
& \ModsetsigPrim 
		\ar[r, phantom, "\modelssigPrim"] 
& \SensetsigPrim 
		\ar[u, <-,  "f_*"]
\end{tikzcd}
&
\begin{array}{c} 
			f^*(m') \modelssig \phi'
			\\ [5pt]\text{iff}\rule{0ex}{1.ex}
			\\ [5pt] m' \modelssigPrim \ f_*(\phi)
\end{array}
\end{array} 
\end{array}	
\end{equation}
} 

\qed
\end{defN}\noindent 
The definition above is methodologically ``twisted'': \mor s between models are considered while \caties\ of sentences are discrete; moreover, even model \mor s alone allow us to think about Sat-\mor s but the definition above does not consider them either. Of course, adding arrows for the only sake of making the \fwk\ more categorical would not be satisfactory for the present paper's intended audience (the ACT community). 
However, below we will discuss the \fwk\ in the context of system engineering (SE) and see that fixing categorification gaps of \cref{def:instByGoguen} does make reasonable  SE-sense.  The discussion will go in two parts: homogeneous (\modelssig\ for a fixed \Sigmaa) and heterogeneous (\ie, the Sat-axiom \cref{eqdiag:satAxiom}).

\subsubsection{Homogeneous Sat: Model updates and evidence} 
\label{sec:instDiscussion--evidence-and -updates}
In the SE-context, models are products and sentences are \req s they must satisfy. If these \req s are safety-critical, then conformance $m\models \phi$ has to be explicitly supported by evidence, \eg, some verification procedure (simulation, model checking, a manual or automatic proof), or/and an reference to a normative document, or/and an expert opinion. If $e$ is such a piece of evidence, we write $m\models^e \phi$. Two pieces of evidence are better than one, and three are better than two, and for safety-critical Sats, having multiple pieces is a must. This makes Sat a ternary rather than a binary relation, \ie, we have a set \modelsd\ supplied with a (jointly-monic) triple of mappings $(p_M,p_S,p_E)$ into the  corresponding sets  \Ob\Modcat, \Sencat, and \Evicat\ of pieces of evidence. 

Now consider arrows. In the SE-context, model \mor s are product updates, and if we consider products as models in \Modcat, then an update \frar{u}{m}{m'} is a span of model \mor s: \spandiag{m}{i_u}{\apexx{u}}{j_u}{m'}, where $i_u$ and $j_u$ are typically injective. Then going from $m$ to $ \apexx{u}$ deletes some part of $m$ while going from $\apexx{u}$ to $m'$ adds new elements to  $m$. Model updates are related to Sat-updates in a complex way. If $m\modelsd^e_\phi$ and \frar{u}{m}{m'} is an update, then existence of some $e'$ with $m'\modelsd^{e'} \phi$ is uncertain. It is not a must but it may happen that indeed $m'\modelsd^{e'}\phi$ for some new piece of evidence $e'$, and we can consider the pair $(e, e')$ as an update \frar{u_E}{e}{e'}. Then it makes sense to make both sets, eSat-relation for a given $\phi$ and  \Evicat\  to be \caties\ (\Satcatphi\ and \Evicat\ resp.), and projections $p_{\phi M}$, $p_{\phi E}$ to be functors. 

However, we can model more than that.  
For example, suppose that sentence $\phi$ is existential and we have a non-deleting model update \frar{u}{m}{m'}, \ie, $u=(1_m,j)$. Then $m\models^e_\phi$ implies $m'\models'^e \phi$ with, importantly, the same $e$. Hence, update $u$ can be lifted to a Sat update  $\bar u \in \Arr\Satcatphi$
such that 
 $p_E(\bar u)=1_e$. Similarly, if sentence $\phi$ is universal while update $u$  is non-extending, \ie, $u=(i, 1_{m'})$, then $m\models^e\phi$ implies $m'\models'^e\phi$, update $u$ can be lifted, and $p_E(\bar u)=1_e$ again. 

%

\subsubsection{Homogeneous Sat: Sentence dependencies and evidence} 
On the sentence side, \mor s may also appear naturally, \eg, proofs-as-arrows is a classical example \cite{LambekScott-introduction}. In the context of DCL, another notion of \mor\ is important: we may consider {\em dependencies} between constraints, \eg, if a span is declared to be jointly monic, it is assumed by default that its legs are single-valued functions rather than general spans (see \cref{sec:skeFormal-constraints} for a detailed discussion of this example). Such dependencies naturally appear in the context of system engineering: considering conformance of a product to a \req\ often makes sense only if the product does meet another \req. In this context, a dependency of a \req\ $\phi$ on \req $\phi'$ may be seen as $\phi'$'s  contribution to $\phi$ and we draw an arrow \flar{d}{\phi}{\phi'}. Multiple contributors \flar{d}{\phi}{\phi'_i}, $i=1..n$ are a typical case, but it does not mean that $m\modelsd\phi'_i$ for all $i$  implies $m\modelsd \phi$. Thus, it makes sense to make \Sencat\ a category, whose arrows are (contributing) dependencies.  

Let us now consider how evidence should behave and thus make Sat a triple of functors from a \caty\ \satascat. Suppose $m\modelsd^e\phi$ and \flar{d}{\phi}{\phi'} is a valid dependency so that $m\modelsd^{e'} \phi'$. In the context outlined above, we may say that a piece of evidence $e'$ contributes to $e$, and hence their should be an evidence arrow \flarxy{e}{e'}. Thus, a dependency $d$ is lifted to an evidence arrow $\bar d\in \Satcat(\modelsd, \modelsd')$ such that $p_M(\bar d)=1_m$ while  \flar{p_E(\bar d)}{e}{e'} is a non-trivial and informative arrow. Of course, lifting should be required to work for any dependency into $\phi$ if $m\modesld^e\phi$, and be compositional: $\liftedX{d.d'}=\bar d. \bar d'$, if we want the dependency mechanism to be {\em sound}. Mathematically, it means that functor \frar{p_S}{\satascat}{\Sencat} is required to be a fibration whose lifts are $p_M$-vertical. 
%
%

%
%

\subsubsection{Heterogeneous Sat: a signature changes along a signature \mor\ \frar{f}{\Sigmaa}{\Sigmaa'}}
The Sat-axiom is a simple \req\ to functors $f^*$ and $f_*$ rather than anyhow substantial logical fact. Indeed, a property of model $m'$'s  reduct $f^*(m')$ is a property of  the model itself (\eg, a property of a country's capital is automatically the country's property). In this way any $\phi\in\Sencatsig$ can also be considered as an element $\phi'\in\SencatsigPrim$, and the Sat-axiom states that $f_*(\phi)$ is exactly this $\phi'$. This triviality of the Sat axiom can be accurately modelled in the eSat \fwk\ by requiring $f^*(m')\modelssig^e\phi$ iff $m'\modelssigPrim^e f_*(\phi)$, where both Sats refer to the same evidence $e$. It is this evidence preservation, which makes the condition a really simple technical \req\ that the Sat-axiom is intended to model.

In this setting, Sat-axiom becomes the following \req: 
\begin{equation}\label{eqdiag:eSatAxiom}
\begin{array}{c@{\quad}c} 
		\esatAxiomNewToOld
&
	\begin{array}{c} 
		f^*(m') \modelssig^e (\phi, l)
		\\ [5pt]\text{iff}\rule{0ex}{1.ex}
		\\ [5pt] m' \modelssigPrim^e \ (f_*(\phi), l)
	\end{array}
\end{array}
\end{equation}
Our goal in the rest of the section is to enrich the notion of institution with evidence and make it fully categorical with non-trivial arrows in all ingredients: spans/deltas for models and proofs/dependencies for sentences. We will begin with an abstract element-free \fwk, whose ingredients are objects and arrows of some \caty\ \CC\  (\cref{sec:eInst-abstract}) and then consider concrete evidential institutions, whose ingredients are \caties\ and functors (\cref{sec:eInst-concrete}).

\newcommand{\tcatCC}{\enma{\tcat_\CC}}
\renewcommand{\twix}[1]{\twixU{#1}}

\newcommand{\projXY}[2]{\enma{p_{#1#2}}}
\newcommand{\projXYtwi}[2]{\enma{p_{#1#2}^\twimor}}

\newcommand{\pM}{\projXY{M}{}}
\newcommand{\pS}{\projXY{S}{}}
\newcommand{\pE}{\projXY{E}{}}
\newcommand{\pMS}{\projXY{M}{S}}
\newcommand{\pSE}{\projXY{S}{E}}
\newcommand{\pME}{\projXY{M}{E}}

\newcommand{\pEtwi}{\projXYtwi{E}{}}
\newcommand{\pMStwi}{\projXYtwi{M}{S}}
\newcommand{\pSEtwi}{\projXYtwi{S}{E}}
\newcommand{\pMtwiPrim}{\projXYtwi{M'}{}}

\subsection{Abstract institutions without elements}\label{sec:eInst-abstract}
 The goal of this section is to consider Sat-axiom with evidence in an abstract setting. We begin  with a ternary span diagram (see \cref{eqdiag:trispans2bispans--twoDiags}  below)
 where nodes and arrows are objects and \mor s  of an arbitrary \caty\ \CC\ with finite limits (the diagram above uses products). An abstract setting will make them more transparent. Then we will return to categorical institutions with \CC\ being \catcat.  Although for our constructs, it is sufficient to require \CC\ to have pullbacks, we will assume \CC\ to be finitely complete \caty\ (FCC) as
 some of the constructs  can be written more compactly if \CC\ has products, and can usefully be illustrated with elements, \ie, mappings from a terminal object $\tcatCC$. Moreover, we need the abstract setting to clarify working with institutions in \catcat, and the latter does have products.   

Through this section (except \cref{sec:spans})  \CC\ is an arbitrary given FCC. 
We begin with recalling some basics of the notion of a span.
\subsubsection{Spans}\label{sec:spans}
Let \CC\ be an arbitrary \caty\ with pullbacks.
\begin{constrN}[Spans]\label{constr:spans}
	Given a pair of \CC-objects $A$ and $B$, a {\em span in \CC\ from $A$ to $B$} is a pair of arrows \spandiag{A}{p_r}{\hat r}{q_r}{B} with a common source called the {\em apex} of the span (and arrows $p_r$, $q_r$ are the {\em legs} of the span, and objects $A$, $B$ are its {\em feet}). We then write \spanrar{r}{A}{B}. To ease notation, we will normally write $p$, $q$, and $r$ for $p_r$, $q_r$, and $\hat r$ (thus denoting a span and its apex by the same symbol). 
	
	Two parallel spans 
	\spanrar{r}{A}{B}, \spanrar{r'}{A}{B} are {\em equivalent}, $r\cong r'$, if there is an iso\mor\ \frar{i}{\hat r}{\hat r'} such that $i.r'=r$ and $i.q'=q$. Pullbacks in \CC\ provide span composition, which is associative up to coherent iso\mor s and constitutes a bi\caty\ \spanfunX{\CC}. Considering spans up to equivalence \congg, gives us a \caty\ \spanfuncongX{\CC}, which we will use below and silently identify spans with their equivalence classes.  
\end{constrN}
\begin{constrN} \label{constr:mor2spans}
	There are two embedding: 
\begin{equation}\label{eq:mor2spans}
\frar{\morAsspanXdn{\_}}{\CC}{\spanfuncongX{\CC}}   \text{~~and~~ } 
\frar{\morAsspanXup{\_}}{\CC}{\spanfuncongX{\opX{\CC}}} 
\end{equation}
which map a \CC-\mor\ \frar{f}{A}{B} to spans 
\spanrar{\morAsspanXdn{f}}{A}{B}, where $\morAsspanXdn{f}= (\spandiag{A}{\id_A}{A}{f}{B} )$, and 
\spanlar{\morAsspanXup{f}}{A}{B}, where  $\morAsspanXup{f}= (\spandiag{B}{f}{A}{\id_A}{B} )$. 
 Composition is strictly preserved due to \congg-factorization, 
 %
 \end{constrN}
\begin{remark}
	Without \congg-factorization, \spanfunX{\CC} is a bi\caty\ and  hence supports the notion of adjointness. It can be proved that $\morAsspanXdn{f}\dashv \morAsspanXup{f}$ (see \cite{pare-spanConstr-tac10} for details). 
\end{remark}

%
\subsubsection{eSat spans}
\begin{defN}[Sat-base]\label{def:satBase}
	A {\em  
		base} is a quadruple  $(\CC, \CC_\senind, \CC_\modind, E)$ of a finitely complete \caty\ (FCC) \CC, two its FCC sub\caties\ (not necessarily disjoint) $\CC_\senind$ and $\CC_\modind$ closed under iso\mor s, whose objects are thought of as collections of {\em sentences} and {\em models} resp., and an  object  $E\in \Ob\CC$ thought of as a collection of {\em pieces of evidence}.  We will write $(\CC,E)$ for a base with $\CC_\senind=\CC =\CC_\modind$; such a base is just a pointed \caty. 
	We denote objects of $\CC_\modind$ and $\CC_\senind$ by $M, M', ..$ and $S, S',...$  resp. and omit explicit mentioning of \caties\ they belong.  
\end{defN}
\begin{defN}[ESats: Sat-spans with evidence]\label{def:esatSpan}
	A {\em  \satty\ span with evidence} ({\em eSat})  over base \satbaseXY{\CC}{E} is a ternary span in \CC, \ie, a triple of \CC-arrows $(\pM, \pS, \pE)$ 
	with a common source $\modelsd$ and feet (target objects)  $M$, $S$, $E$, resp. Using products, this ternary span can be written as a binary span in three different equivalent ways, two of which are shown in diagrams \cref{eqdiag:trispans2bispans--twoDiags}a),b)
\begin{equation}\label{eqdiag:trispans2bispans--twoDiags}
	\begin{array}{cc}
		\TrispanToBispanMS
		& \TrispanToBispanSE
		\\ a)  
		 & b)
	\end{array}
	\end{equation}
(which are two different views of the same ternary diagram, which can be restored from the either view). In this paper, representation a) will be the main  one as the object $\M\timm\S$ is especially  important in our context and models the notion of preinstances, \ie, all possible pairs (model, sentence). To ease notation, we will write $p$ for \pMS\ and $q$ for \pE. Thus, an eSat is 
 binary span  	$\M\timm\S\xleftarrow{p}\;\vDash\;\xrightarrow{q}\E$ or \spanrar{\modelsd}{\M\timm\S}{\E} (\ie, we denote the span and its apex by the same symbol). 
%
	
	If a triple of elements \frar{(m,\phi,e)}{\tcat}{\M\timm\S\timm\E} factors through \modelsd, then we say that  {\em model} $m$ satisfies {\em sentence} (or {\em specification}) $\phi$  as {evidenced} by a {\em (piece of) evidence} $e$ 
	and write $m\satUD{e}{}\phi$.  
	
	We require \satty\ to be invariant under iso\mor s: for any iso\mor s \frar{i_X}{X}{X} 	in the corresponding \caty\ (where $X\in\{\S, \M, \E\}$), there is a \CC-iso\mor\ \frar{i}{\vDash}{\vDash} such that $p_X.i_X =i.p_X$ with $X\in \{S, M, E\}$.  
\end{defN}
\subsubsection{Twisted \mor s of eSats} 
\begin{defN}
		A {\em twisted pair (of arrows)} is a pair $\twip=(s, r)$ of \mor s \frar{s}{S}{S'} in $\CCsen\subset\CC$ and  \frar{r}{M'}{M} in $\CCmod\subset\CC$, considered as an arrow  \frar{\twip}{(S,M)}{(S',M')} so that $r$ goes againt \twip\ (hence the name {\em twisted} and symbol \twip\ as a visual reminder).%
		\footnote{Symbol $r$ also reminds about 'reversing' and the 'model reduct' functor in the institution theory. Symbol $s$ is then appropriate for 'substitution' functors.} 
		Such pairs form category $\twixD{\CC}:=\CC_\senind\times\opX{\CC_\modind}$. 
\end{defN}
\begin{defN}[Twisted \mor s (MS-view)]\label{def:twistedMorph}	
 	Let  \spanrar{\modelsd}{\M\timm\S}{E} and  \spanrar{\modelsd'}{M'\timm S'}{E} be two eSat spans, and 
 	\frar{\twip=(s,r)}{(S,M)}{(S',M')} a twisted pair between their feet as shown in diagram \cref{eqdiag:twistedMorphDef--MStoE}a).  Then we can form two new parallel eSat spans, 
 	\spanrar{\modelsd_\twimor}{P_\twimor}{\E} and
 	\spanrar{\modelsd'_\twimor}{P_\twimor}{\E}, defined by span composition:
 as specified in diagram \cref{eqdiag:twistedMorphDef--MStoE}a). 
 	
	 A twisted pair \twimor\  is called a {\em (twisted) eSat \mor}, if spans $\modelsd_\twimor$ and $\modelsd'_\twimor$ are \congg-equivalent and can be identified in $\spanfuncongX{\CC}$. 	Then we write  \frar{\twimor}{\modelsd}{\modelsd'}. 
	 Diagram \cref{eqdiag:twistedMorphDef--MStoE}b shows this definition unpacked in \CC\ after the two equivalent spans are glued together.
%
\begin{equation}\label{eqdiag:twistedMorphDef--MStoE}
		\begin{tabular}{c@{~}cc} 
			\twiMorphDefViaSpansMStoE
			& \twiMorphDefUnpackedMStoE
&	$\begin{array}{c} 
	m'.r \models^e \phi
	\\ [5pt]\text{iff}\rule{0ex}{1.ex}
	\\ [5pt] m' \models'^e \phi.s
\end{array}$
			\\
			a)  &  b)  & c)
		\end{tabular}
	\end{equation}
%
\end{defN}
\begin{defN}[eSat axiom]
	We say  a twisted pair \frar{(s,r)}{(\S,\S')}{(\M,\M')} {\em satisfies eSat axiom} if for any triple of elements \frar{(m', \phi, e)}{\tcat}{\M'\timm\S\timm\E}, the logical equivalence  in diagram \cref{eqdiag:twistedMorphDef--MStoE}c) holds, \ie, $(m'.r, \phi, e)$ factors through \modelsd\ iff  $(m',\phi.s, e)$ factors through $\modelsd'$. \qed
\end{defN}
\begin{remark}[eSat- vs. Sat-axiom] \label{rema:eSat-vsSat-axiom} eSat axiom is stronger than the usual Sat axiom as it requires preservation of the evidence $e$ for both passaged (the top-down and the bottom-up) in implications c). We can probably construct a case where the Sat-axiom holds but the eSat one fails (when  $m'.r \models^e \phi$ for some \frar{e}{\tcat_\CC}{E} iff $m'\models'^{e'} \phi.s$ for some \frar{e'}{\tcat_\CC}{E}). We consider such cases as violating the ``pure notation'' nature of the Sat-axiom. 
\end{remark}
\begin{propo}
	If a twisted pair $(r,s)$ is a twisted \mor, then it satisfies the eSat axiom.
\end{propo}
{\em Proof.} The upper and lower pullbacks in diagram \cref{eqdiag:twistedMorphDef--MStoE}b) produce the upper and the lower statements in equivalence c). \qed 

%
\begin{propo}[Twisted \mor s compose]\label{prop:twistedMorph--Compose}
	Given a composable pair of twisted pairs,  \[
	(S,M) \xrightarrow{\twimor=(s,r)} (S',M') \xrightarrow{\twimor'=(s',r')} (S'', M'') 
	\] 
	if both pairs are twisted \mor s, then the composed pair $\twimor.\twimor'=(s.s', r'.r)$ is also a twisted \mor. 
\end{propo}
{\em Proof}. The proof is given by diagram \cref{eqdiag:twiMorph-compose} below. Arrows in the middle line $P-P''$ and below it specify the given data. and the two curved leftmost and rightmost arrows specify the twisted product $P_{\twimor.\twimor'}$  of the composed twisted pair. We will refer to them as to \rr\ and \ss\ to shorten notation. We need to prove that $\rr.\!\modelsd \;=\; \ss.\!\modelsd''$.  
\begin{equation}\label{eqdiag:twiMorph-compose}
	\twistedMorphCompose
\end{equation}
We take the pullback (in \CC) shown in the centre of the diagram triangularly, which gives us a span $(\#, \rr_\#, \ss_\#)$ whose legs from \#\ to $P$ and $P''$ are formed by composition but not shown in the diagram to avoid clutter.  It remains to prove there is an iso\mor\ \frar{i}{P_{\twimor.\twimor'}}{\#} such that $i.\rr_\#=\rr$ and $i.\ss_\#=\ss$. 

Two arrows from $P_\twimormor$ to the other two twisted products are due to the universality of products and specified in the diagram.  Then universality of pullbacks gives us a unique arrow $!_\#$ into \#. On the other hand, the universality of products gives us another unique arrow $!_\timm$ from \#.  These two are mutually inverse by the standard argument that otherwise the uniqueness of bang arrows would be violated.
\qed 

\zd{To be filled-in
\begin{remark}[On type-theoretic proofs] As mentioned, the result above can be obtained without products; having pullbacks would be sufficient. Moreover, according to Michael Shulman,.. \
\end{remark}
}
For the composition above, triples $(1_M, 1_S, 1_E)$ are identities, and we thus obtain a \caty\ \twistedSatcat\ of twisted \mor s between evidential  Sats. 
\begin{corol}[Category of twisted ESats] \label{corol:eSat-is-cat}
ESats and their twisted \mor s form a \caty\  (denoted 

\noindent $\twistedSatcat\satbaseCC$ or just \twistedSatcat\ if the base is clear) 
supplied with three projection functors: 
\begin{equation}\label{eq:twistedSpans-(mod-sen)}
\begin{tikzcd}[ampersand replacement=\&,sep=small]
	\twistedSatcat \\
	\&\& {\{\E\}} \\
	{\opX{\CCmod}} \& \CCsen
	\arrow["\pmod"', from=1-1, to=3-1]
	\arrow["\psen"', from=1-1, to=3-2]
	\arrow["{!_\E}", from=1-1, to=2-3]
\end{tikzcd}
\end{equation}
where $\{E\}$ is a singleton \caty\ whose the only object is $E$ and the only arrow is $1_E$ (and hence $!_E(\twip)=1_E$ for any \mor\ \frar{\twip}{\modelsd}{\modelsd'} in \twistedSatcat). 
\end{corol}
%
%
\begin{defN}[E-Institutions] 
An {\em evidential institution} ({\em e-institution})  over a base \satbaseCCE\  is a functor
\[ 
\frar{\satfun} {\Sigcatt} {\twistedSatcat\satbaseCCE},
\]
 where \Sigcatt\ is some \caty\ whose objects are called {\em signatures}. \cref{corol:eSat-is-cat} immediately gives us the usual presentation of an institution as a triple $(\modfun, \senfun, \modelsd)$ of functors \frar{\modfun=\;\modelsd.\pmod}{\Sigcatt}{\opX{\CCmod}} and  \frar{\senfun=\;\modelsd.\psen}{\Sigcatt}{\CCsen}, and a family of Sat-spans (now eSat-spans) of functors  indexed by signatures $\Sigmaa\in\Obb\Sigcatt$: 
 $$   
\Modcatsig\timm\Sencatsig\xleftarrow{~~\pMsig~~}\;\modelsd_\Sigmaa\; \xrightarrow{~~\pSsig~~}\E
 $$
 (where we write $\Modcatsig$ and $\Sencatsig$ for $ \modfun(\Sigmaa)$ and $\senfun(\Sigmaa)$) such that the eSat-axiom holds: if a signature changes by a \mor\ \frar{f}{\Sigmaa}{\Sigmaa'}, then the twisted pair $(\senfun(f), \modfun(f))$ is a twisted \mor\ \frarxy{\modelsd_\Sigmaa}{\modelsd_{\Sigmaa'}}
\qed\end{defN}

\subsection{Concrete institutions with elements}\label{sec:eInst-concrete}
\renewcommand{\pMsig}{\enma{p^M_\Sigmaa}}
\renewcommand{\pSsig}{\enma{p^S_\Sigmaa}}
\newcommand{\pMSsig}{\enma{(\pMsig,\pSsig)}}
\renewcommand{\pMSsig}{\enma{p_\Sigmaa}}
\newcommand{\pMSf}{\enma{p_f}}

\renewcommand{\pMsigPrim}{\enma{p^M_{\Sigmaa'}}}
\renewcommand{\pSsigPrim}{\enma{p^S_{\Sigmaa'}}}
\newcommand{\pMSsigPrim}{\enma{(\pMsigPrim,\pSsigPrim)}}
\renewcommand{\pMSsigPrim}{\enma{p_{\Sigmaa'}}}

\renewcommand{\pEsig}{\enma{q_\Sigmaa}}
\renewcommand{\pEsigPrim}{\enma{q_{\Sigmaa'}}}
\renewcommand{\pEf}{\enma{q_f}}

\newcommand{\eSataxiomForInst}{
\begin{tikzcd}[ampersand replacement=\&,sep=small]
	\Modcatsig\timm\Sencatsig \&\& \modelsdSigma \\
	\ModcatsigPrim\timm\Sencatsig \&\& \modelsdF \&\&\& \E \\
	\ModcatsigPrim\timm\SencatsigPrim \&\& \modelsdSigmaPrim
	\arrow[""{name=0, anchor=center, inner sep=0}, "\pMSsig"', from=1-3, to=1-1]
	\arrow["\pEsig", from=1-3, to=2-6]
	\arrow["\pEsigPrim"', from=3-3, to=2-6]
	\arrow["\pEf"{description}, dashed, from=2-3, to=2-6]
	\arrow[""{name=1, anchor=center, inner sep=0}, "\pMSsigPrim", from=3-3, to=3-1]
	\arrow["{\rho_f}"', dashed, from=2-3, to=1-3]
	\arrow["{\rho'_f}", dashed, from=2-3, to=3-3]
	\arrow["{f^*\timm 1_{\Sencatsig}}", from=2-1, to=1-1]
	\arrow["{1_{\ModcatsigPrim}\timm f_*}"', from=2-1, to=3-1]
	\arrow["\pMSf"', dashed, from=2-3, to=2-1]
	\arrow["\lrcorner"{anchor=center, pos=0.125, rotate=180}, draw=none, from=2-3, to=0]
	\arrow["\lrcorner"{anchor=center, pos=0.125, rotate=-90}, draw=none, from=2-3, to=1]
\end{tikzcd}
}

\begin{defN}[Categorical E-institutions]
	A {\em categorical e-institution} is an e-institution over base $(\catcat, \E)$, \ie, a functor \frar{\modelsd}{\Sigcat}{\twistedSatcat(\catcat, \E)} for some predefined \caty\ \E.  Unpacking this definition with \cref{def:esatSpan} and notation \cref{eq:twistedSpans-(mod-sen)},  gives us functors
	\[ 
	 \frar{\modfun=\;\modelsd.\pmod}{\Sigcat}{\opX{\catcat}} 
	 \text{~ and ~} 
	 \frar{\senfun=\;\modelsd.\psen}{\Sigcat}{\catcat},
	 \] and a 
	family of spans of functors indexed by $\Sigmaa\in\Ob\Sigcat$:
	\[
	\spandiag{\Modcatsig\timm\Sencatsig}{~\pMSsig~}{~\modelssig~}{~\pEsig~}{E}
	\]
	 such that for any signature \mor \frar{f}{\Sigmaa}{\Sigmaa'}, we have commutative diagram \cref{eqdiag:esatAxiom-concrete} a), in which both squares are pullbacks, and where we write $f^*$ for $\modfun(f)$ and $f_*$ for $\senfun(f)$. 
	 In terms of elements, it amounts to the {\em evidence preserving Sat-axiom} specified in \cref{eqdiag:esatAxiom-concrete}b) 
	%

\begin{equation}\label{eqdiag:esatAxiom-concrete}
\begin{array}{cc@{\quad}c} 
\begin{tikzcd}[row sep=6ex] 
\Sigmaa 
	\ar[d, "f"]
\\
\Sigmaa'  
\end{tikzcd}
&
\eSataxiomForInst
&
\begin{array}{c} 
	\\
			f^*(m') \models_{\Sigmaa}^e \phi
			\\ [10pt]\text{iff}\rule{0ex}{1.ex}
			\\ [10pt] m' \models_{\Sigmaa'}^e f_*(\phi)
\end{array}
\\ [10pt]
\multicolumn{2}{c}{\text{a)}} & \text{b)} 
\end{array}	
\vspace{-2ex}
\end{equation}

\medskip
\noindent

\qed\end{defN}
\begin{corol}
	Goguen-Burstall's institutions are e-institutions over base $(\catcat, \catcat, \setcat, \tcat_\catcat$) (\ie, all sentence \caties\ are discrete, and evidencing is trivial), and in which all apex categories \modelssig\ are discrete. Thus, these institutions can be seen as significantly decategorificited categorical e-institutions. 
	%
\end{corol}
\zd{... 
\begin{remark}[Sizes]
	 over base $(\Catcat, \Catcat, \setcat, \tcat)$ or $(\Catcat, \Catcat, \catcat, \tcat)$ for the fully categorical version.
\end{remark}
}
\begin{defN}[Sound institutions] 
	Let $$\frar{\modelsd}{\Sigcat}{\twistedSatcat(\catcat, \E)}$$ be a categorical e-institution with projections (the legs of the eSat span) $\pMsig$, $\pSsig$, $\pEsig$ for $\Sigmaa\in\Ob\Sigcat$. We call the institution {\em sound} if functor \pSsig\ is a fibrtaion whose lifts are \pMsig-vertical for any  signature \Sigmaa. 
	
This statements  unpacked means that given $m\modelssig^e \phi$,  for any arrow \flar{d}{\phi}{\phi'} in \Sencatsig\ (called a {\em contributing dependency}), there is a Cartesian lifting \flar{\liftedX{d}}{\modelssig}{\modelssig'} for which $\pMsig(\liftedX{d})=1_m$. (The domain of the lift is often denoted by $d^*(\modelssig)$ ). 
\end{defN}


		\newcommand{\arrocatcongX}[1]{\enma{\arrowcatX{#1}_\cong}}

\renewcommand{\twix}[1]{\twixD{#1}}
\renewcommand{\twix}[1]{\enma{#1_f}}
\renewcommand{\twixD}[1]{\enma{#1_f}}

\newcommand{\leftU}{\enma{\twix{h}, \twix{v}.f^*}}
\newcommand{\rightU}{\enma{\twix{h}.f_*, \twix{v}}}
\renewcommand{\leftU}{\bangXdn{f^*}}
\renewcommand{\leftU}{\enma{1\timm\,f^*}}
\renewcommand{\rightU}{\bangXdn{f_*}}
\renewcommand{\rightU}{\enma{f_*\timm\, 1}}

\newcommand{\leftDeltaU}{\bangXdn{f_\Delta^*}}
\newcommand{\rightDeltaU}{\bangXdn{f_*}}
\renewcommand{\leftDeltaU}{\enma{1\timm\, f_\Delta^*}}
\renewcommand{\rightDeltaU}{\enma{f_*\timm\, 1}}

\newcommand{\rightCsigU}{\bangXdn{f^\csig_*}}

\newcommand{\twixlemmaOneHHHtwistedMorphCompose}{
\begin{tikzcd}[ampersand replacement=\&,column sep=2.25em]
	\&\& {P_{f.f'}} \\
	\\
	\&\& {\#} \\
	P \& {P_f} \& {P'} \& {P_{f'}} \& {P''} \\
	\&\& \arrowcatGG_\cong
	\arrow["{1\timm f^*}", color={rgb,255:red,92;green,92;blue,214}, from=4-2, to=4-1]
	\arrow["{f_*\timm1}"', color={rgb,255:red,92;green,92;blue,214}, from=4-2, to=4-3]
	\arrow["{1\timm f'^*}", color={rgb,255:red,92;green,92;blue,214}, from=4-4, to=4-3]
	\arrow["{f'_*\timm1}"', color={rgb,255:red,92;green,92;blue,214}, from=4-4, to=4-5]
	\arrow["{1\timm(f'^*.f^*)}"', curve={height=12pt}, from=1-3, to=4-1]
	\arrow["{1\timm f'^*}"{description, pos=0.7}, color={rgb,255:red,92;green,92;blue,214}, curve={height=6pt}, dashed, from=1-3, to=4-2]
	\arrow["{f_*\timm1}"{description, pos=0.7}, color={rgb,255:red,92;green,92;blue,214}, curve={height=-12pt}, dashed, from=1-3, to=4-4]
	\arrow["{(f_*.f'_*)\timm1}", curve={height=-12pt}, from=1-3, to=4-5]
	\arrow[dotted, from=3-3, to=4-2]
	\arrow[dotted, from=3-3, to=4-4]
	\arrow["{!_\#}"{description}, curve={height=-12pt}, dotted, from=1-3, to=3-3]
	\arrow["{!_\timm}"{description}, color={rgb,255:red,92;green,92;blue,214}, curve={height=-12pt}, dashed, from=3-3, to=1-3]
	\arrow["{\text{[pb]}}"{description, pos=0.2}, color={rgb,255:red,150;green,150;blue,150}, draw=none, from=3-3, to=4-3]
	\arrow["\pbfunvX{G}"{description, pos=0.3}, curve={height=12pt}, from=4-1, to=5-3]
	\arrow["{\pbfunvX{G''}}"{description, pos=0.3}, curve={height=-12pt}, from=4-5, to=5-3]
	\arrow["{\pbfunvX{G'}}", from=4-3, to=5-3]
	\arrow["{\pbfunvX{f}}"{description, pos=0.3}, curve={height=12pt}, from=4-2, to=5-3]
	\arrow["{\pbfunvX{f'}}"{description, pos=0.3}, curve={height=-12pt}, from=4-4, to=5-3]
\end{tikzcd}
}

\newcommand{\arrowcatDeltaMorph}{
\begin{tikzcd}[ampersand replacement=\&,sep=small]
	\bullet \& \textcolor{rgb,255:red,40;green,159;blue,40}{\circ} \& \bullet \\
	\bullet \& \bullet \& \bullet
	\arrow["a"', from=1-1, to=2-1]
	\arrow["b", from=1-3, to=2-3]
	\arrow["{p_0}"', color={rgb,255:red,40;green,159;blue,40}, from=1-2, to=1-1]
	\arrow["{q_0}", color={rgb,255:red,40;green,159;blue,40}, from=1-2, to=1-3]
	\arrow["q"', color={rgb,255:red,40;green,159;blue,40}, from=2-2, to=2-3]
	\arrow["{\mathsf{id}}", Rightarrow, no head, from=2-2, to=2-1]
	\arrow["r"', color={rgb,255:red,40;green,159;blue,40}, from=1-2, to=2-2]
\end{tikzcd}
}

\newcommand{\arrowcatDeltaCompose}{
\begin{tikzcd}[ampersand replacement=\&,sep=small]
	\&\& \textcolor{rgb,255:red,92;green,92;blue,214}{\circ} \\
	\bullet \& \textcolor{rgb,255:red,40;green,159;blue,40}{\bullet} \& \bullet \& \textcolor{rgb,255:red,40;green,159;blue,40}{\bullet} \& \bullet \\
	\bullet \&\& \bullet \&\& \bullet
	\arrow["b"', from=2-3, to=3-3]
	\arrow["{p_0}"', color={rgb,255:red,40;green,159;blue,40}, from=2-2, to=2-1]
	\arrow["{q_0}", color={rgb,255:red,40;green,159;blue,40}, from=2-2, to=2-3]
	\arrow["{p'_0}"', color={rgb,255:red,40;green,159;blue,40}, from=2-4, to=2-3]
	\arrow["a"', from=2-1, to=3-1]
	\arrow["q"', color={rgb,255:red,40;green,159;blue,40}, from=3-1, to=3-3]
	\arrow["r"', color={rgb,255:red,40;green,159;blue,40}, from=2-2, to=3-1]
	\arrow[color={rgb,255:red,40;green,159;blue,40}, from=2-4, to=3-3]
	\arrow["{q'}"', from=3-3, to=3-5]
	\arrow["{q'_0}", color={rgb,255:red,40;green,159;blue,40}, from=2-4, to=2-5]
	\arrow["c"', from=2-5, to=3-5]
	\arrow[color={rgb,255:red,92;green,92;blue,214}, dashed, from=1-3, to=2-2]
	\arrow[color={rgb,255:red,92;green,92;blue,214}, dashed, from=1-3, to=2-4]
	\arrow["\lrcorner"{anchor=center, pos=0.125, rotate=-45}, draw=none, from=1-3, to=2-3]
\end{tikzcd}
}

\newcommand{\codfibrForDeltas}{
\begin{tikzcd}[ampersand replacement=\&,column sep=small,row sep=scriptsize]
	\bullet \& \textcolor{rgb,255:red,46;green,184;blue,46}{\bullet} \&\&\&\& a \\
	\&\& {B_0^q} \& {B_0} \&\&\&\& b \\
	\&\& A \& B \&\& {b_q}
	\arrow["q"', from=3-3, to=3-4]
	\arrow["b", from=2-4, to=3-4]
	\arrow["{q_b}"', dashed, from=2-3, to=2-4]
	\arrow["{b_q}", dashed, from=2-3, to=3-3]
	\arrow["\lrcorner"{anchor=center, pos=0.125}, draw=none, from=2-3, to=3-4]
	\arrow["a"', from=1-1, to=3-3]
	\arrow["{q_0}", color={rgb,255:red,46;green,184;blue,46}, from=1-2, to=2-4]
	\arrow["{p_0}"', color={rgb,255:red,46;green,184;blue,46}, from=1-2, to=1-1]
	\arrow["{\hat r}"{description, pos=0.4}, color={rgb,255:red,46;green,184;blue,46}, from=1-2, to=3-3]
	\arrow["{!_r}"{description}, color={rgb,255:red,70;green,53;blue,227}, dashed, from=1-2, to=2-3]
	\arrow["{r=(\hat r,p_0,q_0,q)}", from=1-6, to=2-8]
	\arrow["{\bar q = (q_b,q)}"', from=3-6, to=2-8]
	\arrow["{\#=(p_0,!_r, 1_A)}"{description}, color={rgb,255:red,70;green,53;blue,227}, dashed, from=1-6, to=3-6]
\end{tikzcd}
}

\newcommand{\composeForMianTh}{
\begin{tikzcd}[ampersand replacement=\&,sep=scriptsize]
	{P_{f.f'}} \\
	\\
	{P_G} \& {P_f} \& {P_{G'}} \& {P_{f'}} \& {P_{G''}} \\
	{\PreinstcatX{G}} \& {\PreinstcatX{f/\csig}} \& {\PreinstcatX{G'}} \& {\PreinstcatX{f'/\csig}} \& {\PreinstcatX{G''}} \\
	\\
	{\PreinstcatX{f.f'}}
	\arrow[from=3-2, to=3-1]
	\arrow[from=3-2, to=3-3]
	\arrow[from=3-4, to=3-3]
	\arrow[from=3-4, to=3-5]
	\arrow[from=1-1, to=3-1]
	\arrow[dashed, from=1-1, to=3-2]
	\arrow[from=1-1, to=3-3]
	\arrow[dashed, from=1-1, to=3-4]
	\arrow[from=1-1, to=3-5]
	\arrow[from=4-1, to=3-1]
	\arrow[dashed, from=4-2, to=3-2]
	\arrow[from=4-3, to=3-3]
	\arrow[dashed, from=4-4, to=3-4]
	\arrow[from=4-5, to=3-5]
	\arrow["{!_{f.f'/\csig}}", curve={height=-30pt}, from=6-1, to=1-1]
	\arrow[from=6-1, to=4-1]
	\arrow[dashed, from=6-1, to=4-2]
	\arrow[from=6-1, to=4-3]
	\arrow[dashed, from=6-1, to=4-4]
	\arrow[from=6-1, to=4-5]
	\arrow[from=4-2, to=4-1]
	\arrow[from=4-2, to=4-3]
	\arrow[from=4-4, to=4-3]
	\arrow[from=4-4, to=4-5]
\end{tikzcd}
}

\renewcommand{\odernode}[1]{\dernode{#1}}

\newcommand{\twixlemmaProofDefiningPBfunVDelta}{
\begin{tikzcd}[ampersand replacement=\&,column sep=scriptsize,row sep=2.25em]
	\&\& \textcolor{rgb,255:red,225;green,61;blue,55}{\bullet} \&\& \bullet \&\&\& \bullet \\
	\& \circ \&\& \circ \&\&\& \circ \\
	\bullet \&\& \textcolor{rgb,255:red,225;green,61;blue,55}{\bullet} \&\&\& \bullet \\
	\textcolor{rgb,255:red,225;green,61;blue,55}{\bullet} \&\& \textcolor{rgb,255:red,225;green,61;blue,55}{\bullet} \&\&\& S
	\arrow["{h'}"', from=4-3, to=4-6]
	\arrow["g", from=4-1, to=4-3]
	\arrow["h"', curve={height=18pt}, from=4-1, to=4-6]
	\arrow["{v'}"', from=3-6, to=4-6]
	\arrow["{u'}"', from=2-7, to=3-6]
	\arrow["{v'_{h'}}"'{pos=0.4}, color={rgb,255:red,225;green,61;blue,55}, from=3-3, to=4-3]
	\arrow["{u'_{h'}}"', color={rgb,255:red,225;green,61;blue,55}, dashed, from=2-4, to=3-3]
	\arrow["{v'_h}"', color={rgb,255:red,75;green,81;blue,231}, from=3-1, to=4-1]
	\arrow[curve={height=-12pt}, draw=none, from=2-4, to=4-3]
	\arrow["{\apexx{u}_{h'}}"{description}, curve={height=-6pt}, dashed, from=2-4, to=4-3]
	\arrow["u", from=2-7, to=1-8]
	\arrow["v", curve={height=-12pt}, from=1-8, to=4-6]
	\arrow["{\apexx{u}}"{description}, curve={height=-6pt}, dashed, from=2-7, to=4-6]
	\arrow["{u_{h}}", dashed, from=2-4, to=1-5]
	\arrow["{v_{h'}}", curve={height=-12pt}, from=1-5, to=4-3]
	\arrow["{\bar{h'}_v}", dotted, from=1-5, to=1-8]
	\arrow["{u'_{h}}"', draw={rgb,255:red,75;green,81;blue,231}, dashed, from=2-2, to=3-1]
	\arrow["{u_h}", color={rgb,255:red,225;green,61;blue,55}, dashed, from=2-2, to=1-3]
	\arrow["{\apexx{u}_{h}}"{description}, draw={rgb,255:red,75;green,81;blue,231}, curve={height=-6pt}, dashed, from=2-2, to=4-1]
	\arrow["{v_h}"{pos=0.4}, color={rgb,255:red,225;green,61;blue,55}, curve={height=-12pt}, from=1-3, to=4-1]
	\arrow["{\apexx{u}_g}"{description}, color={rgb,255:red,225;green,61;blue,55}, dashed, from=2-2, to=2-4]
	\arrow[dotted, from=2-4, to=2-7]
	\arrow["{\bar g_v}", dotted, from=1-3, to=1-5]
\end{tikzcd}
}

\newcommand{\twixlemmaProofDefiningPBfunV}{
	\begin{tikzcd}[ampersand replacement=\&,row sep=2.25em]
		\circ \&\& \circ \&\& \bullet \\
		\circ \&\& \circ \&\& \bullet \\
		\bullet \&\& \bullet \&\& \X
		\arrow["{h'}"', from=3-3, to=3-5]
		\arrow["g", from=3-1, to=3-3]
		\arrow["h"', curve={height=18pt}, from=3-1, to=3-5]
		\arrow["{v'}"', from=2-5, to=3-5]
		\arrow["u"', from=1-5, to=2-5]
		\arrow["v", curve={height=-12pt}, from=1-5, to=3-5]
		\arrow["{\bar{\bar g}}", color={rgb,255:red,75;green,81;blue,231}, dotted, from=1-1, to=1-3]
		\arrow[draw={rgb,255:red,75;green,81;blue,231}, dotted, from=1-3, to=1-5]
		\arrow["{\pbfunvX{G}(v',h')}"{pos=0.4}, color={rgb,255:red,75;green,81;blue,231}, dashed, from=2-3, to=3-3]
		\arrow[draw={rgb,255:red,75;green,81;blue,231}, dotted, from=2-3, to=2-5]
		\arrow["{\bar g}", color={rgb,255:red,75;green,81;blue,231}, dotted, from=2-1, to=2-3]
		\arrow[draw={rgb,255:red,75;green,81;blue,231}, dashed, from=1-1, to=2-1]
		\arrow["{\bar u}"', dotted, from=1-3, to=2-3]
		\arrow["{\pbfunvX{G}(v',h)}", color={rgb,255:red,75;green,81;blue,231}, dashed, from=2-1, to=3-1]
		\arrow["{\pbfunvX{G}(v,h)}"', color={rgb,255:red,75;green,81;blue,231}, curve={height=12pt}, dashed, from=1-1, to=3-1]
		\arrow[draw={rgb,255:red,75;green,81;blue,231}, dashed, from=1-1, to=2-3]
		\arrow[curve={height=-12pt}, draw=none, from=1-3, to=3-3]
	\end{tikzcd}
}
%
%

\newcommand{\twixlemmaCattydiagOne}{
\begin{tikzcd}[row sep=3ex, column sep=2.ex, ampersand replacement=\&]
\CC/\X \&
\& \CC/\X 
		\ar[rr, bend left=10, "f^*", <-] \&
\& \CC/\X' 
		\ar[from=1-1, to=1-7,  bend left =20, "f_*" ]
\&\& \CC/\X'
\\ [-2.5pt]
\&\dernode{P_G}
		\ar[lu, notderived, "h" '] 
		\ar[ru, notderived, "v"] 
\&\& \dernode{\twixU{P}_f}
		\ar[ru, dotted, "\gapnamegap{0}{\twix{v}}{-1}", pos=0.15, shorten =-1ex]
		\ar[lllu, dotted, "\gapnamegap{0}{\twix{h}}{-1}" ', pos=0.15, bend right=10]  
		\ar[ll, derived, "{\leftU}" '] 	
		\ar[rr, derived, "{\rightU}"  ]
		\ar[d, dotderived, "{\pbfunvX{f}}"]
\&\& \dernode{P_{G'}} 
		\ar[lu, notderived, "\gapnamegap{-1}{v'}{1}" ' , pos=0.15] 
		\ar[ru, notderived, "\gapnamegap{1}{h'}{-1.5} ", pos=0.25] 
\\ [2.5pt]
\&\&\& \arrowcatCC_\cong
		\ar[llu, <-, pos=0.35, "\pbfunvX{\X}"] 
		\ar[rru, <-, "\pbfunvX{\X'}" ', pos=0.35 ]  \&\&\&
		\\
\&\&\&\CC 
		\ar[uuulll, <-, bend left =25, "\dom_\X"] 
		\ar[uuurrr, <-, bend right =25, "\dom_{\X'}" '] 
		\ar[u, <-, "\cod"]
	\end{tikzcd}
}


%

\newcommand{\twixlemmaCattydiagTwo}{ 
\begin{tikzcd}[row sep=3ex, column sep=2.5ex, ampersand replacement=\&]
\GG/ G 
\&\& \deltaslicecatGGG
		\ar[rr, bend left=10, "{\fstardeltaUp}", <-] 
\&\& \deltaslicecatGGGprim
		\ar[from=1-1, to=1-7,  bend left =20, "{f_*}" ]
\&\& \GG/G' 
\\ [-2.5pt]  
\&\dernode{P_G}
		\ar[lu, notderived, "h" '] 
		\ar[ru, notderived, "v"]
\&\& \dernode{\twixU{P}_f}
		\ar[lllu, dotted, "\gapnamegap{-1.5}{\twix{h}}{1}" ', pos=0.15, bend right = 10]
		\ar[ru, dotted, "\gapnamegap{1}{\twix{v}}{-1.5}", pos=0.3, shorten=-1ex]  
		\ar[ll, derived, "{(1_\hind, \fstardeltaUp)}" '] 	
		\ar[rr, derived, "{(\fstardeltaDn, 1_\vind)} "  ]
		\ar[d, dotderived, "{\pbfundeltavX{f}}"]
\&\& \dernode{P_{G'}} 
		\ar[lu, notderived, "\gapnamegap{-1}{v'}{1}" ' , pos=0.15, shorten=-1ex] 
		\ar[ru, notderived, "\gapnamegap{1}{h'}{-1.5} ", pos=0.25] 
\\ [2.5pt]
\&\&\& \deltaarrowcatGG_\cong
		\ar[llu, <-, pos=0.35, "\pbfundeltavX{\X}"] 
		\ar[rru, <-, "\pbfundeltavX{\X'}" ', pos=0.35 ] 
		\\
		\&\&\&\CC 
		\ar[uuulll, <-, bend left =25, "\dom_\X"] 
		\ar[uuurrr, <-, bend right =25, "\dom_{\X'}" '] 
		\ar[u, <-, "\cod"]
	\end{tikzcd}
}


\newcommand{\twixlemmaCattydiagTwoWithOk}{
\begin{tikzcd}[row sep=4ex, column sep=2.5ex, ampersand replacement=\&]
\GG/ G 
\&\& \deltaslicecatGGG
		\ar[rr, bend left=10, "{\fstardeltaUp}", <-] 
\&\& \deltaslicecatGGGprim
		\ar[from=1-1, to=1-7,  bend left =20, "{f_*}" ]
\&\& \GG/G' 
\\ [-2.5pt]  
\&\dernode{P_G}
		\ar[lu, notderived, "h" '] 
		\ar[ru, notderived, "v"]
\&\& \dernode{\twixU{P}_f}
		\ar[lllu, dotted, "\gapnamegap{-1.5}{\twix{h}}{1}" ', pos=0.15, bend right = 10]
		\ar[ru, dotted, "\gapnamegap{1}{\twix{v}}{-1.5}", pos=0.3, shorten=-1ex]  
		\ar[ll, derived, "{\leftDeltaU}" '] 	
		\ar[rr, derived, "{\rightDeltaU} "  ]
		\ar[d, dotderived, "{\pbfundeltavX{f}}"]
\&\& \dernode{P_{G'}} 
		\ar[lu, notderived, "\gapnamegap{-1}{v'}{1}" ' , pos=0.15, shorten=-1ex] 
		\ar[ru, notderived, "\gapnamegap{1}{h'}{-1.5} ", pos=0.25] 
\&\& \odernode{\SubprodOkDU{G}{} }
\&\& \odernode{\SubprodOkDU{f}{\twixSymbol}} 
			\ar[ll, "\rho_f" ', dashed]
			\ar[rr, "\rho'_f", dashed]
\&\&    \odernode{\SubprodOkDU{G'}{} }
\\ [2.5pt]  
\&\&\& \deltaarrowcatGG_\cong
		\ar[llu, <-, pos=0.5, "\pbfundeltavX{\X}" ] 
		\ar[rru, <-, pos=0.5 , "\pbfundeltavX{\X'}" ',] 
\&\&\&\& \& \&    
\odernode{OK}  	\ar[from =3-10, to=3-4,  "\ok" ', pos=0.35]
			\ar[ull, <-, "q_G", derived]
			\ar[urr, <-, "q_{G'}" ', derived]
				\ar[u, <-, "q_f", derived, pos=0.65]
		\\   
\&\&\&\CC 
		\ar[uuulll, <-, bend left =25, "\dom_\X"] 
		\ar[uuurrr, <-, bend right =25, "\dom_{\X'}" ', pos=0.3] 
		\ar[u, <-, "\cod"]
\end{tikzcd}
}

\newcommand{\twixlemmaCattydiagOneWithok}{
	\begin{tikzcd}[row sep=4ex, column sep=2.5ex, ampersand replacement=\&]
		\GG/ G 
		\&\& \GG/G
		\ar[rr, bend left=10, "{f^*}", <-] 
		\&\& \GG/G'
		\ar[from=1-1, to=1-7,  bend left =20, "{f_*}" ]
		\&\& \GG/G' 
\\ [-2.5pt]  
\&\dernode{P_G}
		\ar[lu, notderived, "h" '] 
		\ar[ru, notderived, "v"]
\&\& \dernode{\twixU{P}_f}
		\ar[lllu, dotted, "\gapnamegap{-1.5}{\twix{h}}{1}" ', pos=0.15, bend right = 10]
		\ar[ru, dotted, "\gapnamegap{1}{\twix{v}}{-1.5}", pos=0.3, shorten=-1ex]  
		\ar[ll, derived, "{\leftU}" '] 	
		\ar[rr, derived, "{\rightU} "  ]
		\ar[d, dotderived, "{\pbfunvX{f}}"]
\&\& \dernode{P_{G'}} 
		\ar[lu, notderived, "\gapnamegap{-1}{v'}{1}" ' , pos=0.15, shorten=-1ex] 
		\ar[ru, notderived, "\gapnamegap{1}{h'}{-1.5} ", pos=0.25] 
\&\& \odernode{\SubprodOkDU{G}{} }
\&\& \odernode{\SubprodOkDU{f}{\twixSymbol}} 
		\ar[ll, "\rho_f" ', dashed]
		\ar[rr, "\rho'_f", dashed]
\&\&    \odernode{\SubprodOkDU{G'}{} }
		\ar[llllll, derived, bend right =25, "\liftedX{\ok}_{G'}" ']
		\\ [2.5pt]  
		\&\&\& \arrowcatGG_\cong
		\ar[llu, <-, pos=0.5, "\pbfunvX{\X}" ] 
		\ar[rru, <-, pos=0.5 , "\pbfunvX{\X'}" ',] 
		\&\&\&\& \& \&    
		\odernode{OK}  	\ar[from =3-10, to=3-4,  "\ok" ', pos=0.35]
		\ar[ull, <-, "q_G", derived]
		\ar[urr, <-, "q_{G'}" ', derived]
		\ar[u, <-, "q_f", derived, pos=0.65]
		\\   
		\&\&\&\CC 
		\ar[uuulll, <-, bend left =25, "\dom_\X"] 
		\ar[uuurrr, <-, bend right =25, "\dom_{\X'}" ', pos=0.3] 
		\ar[u, <-, "\cod"]
	\end{tikzcd}
}

\newcommand{\twixlemmaOneCompositional}{\twistedMorphCompose
}

\renewcommand{\csig}{\predsig}

\newcommand{\mainThChasingPiece}{
\begin{tikzcd}[ampersand replacement=\&,sep=scriptsize]
	\&\&\&\&\&\& \textcolor{rgb,255:red,205;green,100;blue,76}{\pbfundeltavX{G}} \\
	\&\&\&\&\&\& {d^\Delta_f} \\
	\deltaarrowcatGG \&\& \GG \&\& \textcolor{rgb,255:red,43;green,171;blue,43}{\GG/G'} \& \textcolor{rgb,255:red,43;green,171;blue,43}{\_{\times}\_} \& \textcolor{rgb,255:red,43;green,171;blue,43}{\deltaslicecatGGGprim} \\
	\& \textcolor{rgb,255:red,33;green,87;blue,212}{\preInstcatsig} \&\&\&\&\&\& \textcolor{rgb,255:red,43;green,171;blue,43}{\cdot} \\
	\Instcatsig \&\& {\carrop{\predsig}} \&\& \textcolor{rgb,255:red,43;green,171;blue,43}{\LDGprim} \& \textcolor{rgb,255:red,43;green,171;blue,43}{\preInstcatX{G'}} \&\& {\_\twixU{\times}\_} \\
	\&\&\& \textcolor{rgb,255:red,43;green,171;blue,43}{\Dashv_{G'}} \&\&\& \textcolor{rgb,255:red,205;green,100;blue,76}{\GG/G} \\
	\&\&\&\&\&\& \textcolor{rgb,255:red,209;green,140;blue,71}{\LDG} \& {\preInstcatX{\twixSymbol}} \\
	\&\&\&\&\&\& \textcolor{rgb,255:red,43;green,171;blue,43}{\twixU{\Dashv}_{f_*}}
	\arrow["\tmapfunsig", from=5-1, to=3-1]
	\arrow["\cod", from=3-1, to=3-3]
	\arrow["\arfun", from=5-3, to=3-3]
	\arrow["\labfunsig", from=5-1, to=5-3]
	\arrow["{\#_\predsig}", dashed, from=5-1, to=4-2]
	\arrow[dashed, from=4-2, to=3-1]
	\arrow[dashed, from=4-2, to=5-3]
	\arrow["{\text{[pb]}}"{description}, draw=none, from=4-2, to=3-3]
	\arrow["{\dom_{G'}}"', color={rgb,255:red,43;green,171;blue,43}, from=3-5, to=3-3]
	\arrow["{\ell_{G'}}"', color={rgb,255:red,43;green,171;blue,43}, dashed, from=5-5, to=5-3]
	\arrow["{\bmapfunX{G"}}", color={rgb,255:red,43;green,171;blue,43}, dashed, from=5-5, to=3-5]
	\arrow["{h'_0}"', color={rgb,255:red,43;green,171;blue,43}, dotted, from=3-6, to=3-5]
	\arrow["{v'}"{pos=0.6}, color={rgb,255:red,43;green,171;blue,43}, dashed, from=5-6, to=3-7]
	\arrow["{!_{G'}}"{pos=0.4}, color={rgb,255:red,43;green,171;blue,43}, dashed, from=5-6, to=3-6]
	\arrow["{\pbfundeltavX{G'}}"', color={rgb,255:red,43;green,171;blue,43}, curve={height=24pt}, dashed, from=3-6, to=3-1]
	\arrow["{\#_{G'}}"', color={rgb,255:red,43;green,171;blue,43}, dotted, from=5-6, to=4-2]
	\arrow["{\text{[pb]}_{1G'}}"{description, pos=0.6}, color={rgb,255:red,43;green,171;blue,43}, draw=none, from=5-5, to=3-3]
	\arrow["{\liftedX{\#_{G'}}}"', color={rgb,255:red,43;green,171;blue,43}, curve={height=-6pt}, dotted, from=6-4, to=5-1]
	\arrow["{v'_0}", color={rgb,255:red,43;green,171;blue,43}, dotted, from=3-6, to=3-7]
	\arrow["{\bmapfunX{G}}"', color={rgb,255:red,209;green,140;blue,71}, dashed, from=7-7, to=6-7]
	\arrow["{\ell_{G}}"'{pos=0.4}, color={rgb,255:red,209;green,140;blue,71}, dotted, from=7-7, to=5-3]
	\arrow["{\dom_{G}}"{description, pos=0.8}, color={rgb,255:red,209;green,140;blue,71}, curve={height=18pt}, dotted, from=6-7, to=3-3]
	\arrow["{\text{[pb]}_{1G}}"{description, pos=0.1}, color={rgb,255:red,209;green,140;blue,71}, curve={height=18pt}, draw=none, from=7-7, to=3-3]
	\arrow["{\twix{h}}", color={rgb,255:red,205;green,100;blue,76}, from=7-8, to=7-7]
	\arrow["{f_*}"'{pos=0.3}, curve={height=18pt}, from=6-7, to=3-5]
	\arrow["{\liftedX{\#_{f_*}}}"', color={rgb,255:red,43;green,171;blue,43}, dotted, from=8-7, to=6-4]
	\arrow["{f_*^{LD}}"', curve={height=6pt}, dashed, from=7-7, to=5-5]
	\arrow["{\text{[pb]}_{2f}}"{description, pos=0.3}, curve={height=-12pt}, draw=none, from=8-7, to=5-6]
	\arrow["{p^E_{f_*}}", color={rgb,255:red,43;green,171;blue,43}, curve={height=-12pt}, dotted, from=8-7, to=5-1]
	\arrow["{\text{[pb]}_{2G'}}"{description, pos=0.2}, color={rgb,255:red,43;green,171;blue,43}, draw=none, from=6-4, to=4-2]
	\arrow["{h'}"', color={rgb,255:red,33;green,87;blue,212}, dashed, from=5-6, to=5-5]
	\arrow["{\twixU{v_0}}"{description}, color={rgb,255:red,43;green,171;blue,43}, curve={height=6pt}, dotted, from=5-8, to=3-7]
	\arrow[draw={rgb,255:red,205;green,100;blue,76}, dotted, from=5-8, to=6-7]
	\arrow["{!_f}"', dashed, from=7-8, to=5-8]
	\arrow[draw={rgb,255:red,43;green,171;blue,43}, curve={height=6pt}, dashed, from=5-8, to=3-6]
	\arrow["{\#_{f_*}}"', color={rgb,255:red,43;green,171;blue,43}, curve={height=18pt}, dashed, from=7-8, to=5-6]
	\arrow[draw={rgb,255:red,205;green,100;blue,76}, curve={height=30pt}, dashed, no head, from=5-8, to=1-7]
	\arrow[draw={rgb,255:red,205;green,100;blue,76}, curve={height=24pt}, dashed, from=1-7, to=3-1]
	\arrow["{\twixU{v}}"', color={rgb,255:red,43;green,171;blue,43}, curve={height=30pt}, dashed, no head, from=7-8, to=4-8]
	\arrow[curve={height=18pt}, dashed, no head, from=5-8, to=2-7]
	\arrow[curve={height=30pt}, shorten >=37pt, dashed, from=2-7, to=3-1]
	\arrow[draw={rgb,255:red,43;green,171;blue,43}, curve={height=6pt}, dashed, from=4-8, to=3-7]
	\arrow["{\revSatasarrowX{G'}}", color={rgb,255:red,43;green,171;blue,43}, dotted, from=6-4, to=3-6]
	\arrow["{\liftedX{\#_\csig}}"', color={rgb,255:red,43;green,171;blue,43}, dashed, from=6-4, to=5-6]
	\arrow["{\revSatasarrowX{f_*}}", color={rgb,255:red,43;green,171;blue,43}, dotted, from=8-7, to=5-8]
	\arrow["{\liftedX{\liftedX{\#_\csig}}}"', color={rgb,255:red,43;green,171;blue,43}, from=8-7, to=7-8]
\end{tikzcd}
}

\newcommand{\mainThProofChasing}{
\begin{tikzcd}[ampersand replacement=\&,sep=scriptsize]
	\&\&\&\&\&\& \textcolor{rgb,255:red,205;green,100;blue,76}{\pbfundeltavX{G}} \\
	\&\&\&\&\&\& {\pbfundeltavX{f}} \\
	\deltaarrowcatGG \&\& \GG \&\& \textcolor{rgb,255:red,43;green,171;blue,43}{\GG/G'} \& \textcolor{rgb,255:red,43;green,171;blue,43}{\_{\times}\_} \& \textcolor{rgb,255:red,43;green,171;blue,43}{\deltaslicecatGGGprim} \\
	\& \textcolor{rgb,255:red,33;green,87;blue,212}{\preInstcatsig} \\
	\Instcatsig \&\& {\carrop{\predsig}} \&\& \textcolor{rgb,255:red,43;green,171;blue,43}{\LDGprim} \& \textcolor{rgb,255:red,43;green,171;blue,43}{\preInstcatX{G'}} \&\& {\_\twixU{\times}\_} \\
	\&\&\& \textcolor{rgb,255:red,43;green,171;blue,43}{\Dashv_{G'}} \&\&\& \textcolor{rgb,255:red,205;green,100;blue,76}{\GG/G} \\
	\&\&\&\&\&\& \textcolor{rgb,255:red,209;green,140;blue,71}{\LDG} \&\& {\preInstcatX{\twixSymbol}} \\
	\&\& \textcolor{rgb,255:red,205;green,100;blue,76}{\Dashv_{G}} \&\&\&\& \textcolor{rgb,255:red,43;green,171;blue,43}{\twixU{\Dashv}_{f_*}}
	\arrow["\tmapfunsig", from=5-1, to=3-1]
	\arrow["\cod", from=3-1, to=3-3]
	\arrow["\arfun", from=5-3, to=3-3]
	\arrow["\labfunsig", from=5-1, to=5-3]
	\arrow["{\#_\predsig}", dashed, from=5-1, to=4-2]
	\arrow[dashed, from=4-2, to=3-1]
	\arrow[dashed, from=4-2, to=5-3]
	\arrow["{\text{[pb]}}"{description}, draw=none, from=4-2, to=3-3]
	\arrow["{\dom_{G'}}"', color={rgb,255:red,43;green,171;blue,43}, from=3-5, to=3-3]
	\arrow["{\ell_{G'}}"', color={rgb,255:red,43;green,171;blue,43}, dashed, from=5-5, to=5-3]
	\arrow["{\bmapfunX{G"}}", color={rgb,255:red,43;green,171;blue,43}, dashed, from=5-5, to=3-5]
	\arrow["{h'_0}"', color={rgb,255:red,43;green,171;blue,43}, dotted, from=3-6, to=3-5]
	\arrow["{v'}"{pos=0.6}, color={rgb,255:red,43;green,171;blue,43}, dashed, from=5-6, to=3-7]
	\arrow["{!_{G'}}"'{pos=0.8}, color={rgb,255:red,43;green,171;blue,43}, dashed, from=5-6, to=3-6]
	\arrow["{\pbfundeltavX{G'}}"', color={rgb,255:red,43;green,171;blue,43}, curve={height=24pt}, dashed, from=3-6, to=3-1]
	\arrow["{\#_{G'}}"', color={rgb,255:red,43;green,171;blue,43}, dotted, from=5-6, to=4-2]
	\arrow["{\text{[pb]}_{1G'}}"{description, pos=0.6}, color={rgb,255:red,43;green,171;blue,43}, draw=none, from=5-5, to=3-3]
	\arrow["{\liftedX{\#_{G'}}}"', color={rgb,255:red,43;green,171;blue,43}, curve={height=-6pt}, dotted, from=6-4, to=5-1]
	\arrow["{v'_0}", color={rgb,255:red,43;green,171;blue,43}, dotted, from=3-6, to=3-7]
	\arrow["{\bmapfunX{G}}"', color={rgb,255:red,209;green,140;blue,71}, dashed, from=7-7, to=6-7]
	\arrow["{\ell_{G}}"'{pos=0.4}, color={rgb,255:red,209;green,140;blue,71}, dotted, from=7-7, to=5-3]
	\arrow["{\dom_{G}}"{description, pos=0.8}, color={rgb,255:red,209;green,140;blue,71}, curve={height=18pt}, dotted, from=6-7, to=3-3]
	\arrow["{\text{[pb]}_{1G}}"{description, pos=0.1}, color={rgb,255:red,209;green,140;blue,71}, curve={height=18pt}, draw=none, from=7-7, to=3-3]
	\arrow["{\twix{h}}", color={rgb,255:red,205;green,100;blue,76}, from=7-9, to=7-7]
	\arrow["{f_*}"'{pos=0.3}, curve={height=18pt}, from=6-7, to=3-5]
	\arrow["{\liftedX{f_*^?}}"', color={rgb,255:red,43;green,171;blue,43}, dotted, from=8-7, to=6-4]
	\arrow["{f_*^{LD}}"', curve={height=6pt}, dashed, from=7-7, to=5-5]
	\arrow["{\text{[pb]}_{2f}}"{description, pos=0.3}, curve={height=-12pt}, draw=none, from=8-7, to=5-6]
	\arrow["{\liftedX{\twixdiagX{f}}}"{description}, curve={height=-12pt}, dashed, from=8-7, to=5-1]
	\arrow["{\text{[pb]}_{2G'}}"{description, pos=0.2}, color={rgb,255:red,43;green,171;blue,43}, draw=none, from=6-4, to=4-2]
	\arrow["{h'}"', color={rgb,255:red,33;green,87;blue,212}, dashed, from=5-6, to=5-5]
	\arrow["{\twixU{v_0}}"{description}, color={rgb,255:red,43;green,171;blue,43}, curve={height=6pt}, dashed, from=5-8, to=3-7]
	\arrow[draw={rgb,255:red,205;green,100;blue,76}, dotted, from=5-8, to=6-7]
	\arrow["{!_f}"', dashed, from=7-9, to=5-8]
	\arrow["{(f_*,\, 1)}"{description}, color={rgb,255:red,43;green,171;blue,43}, curve={height=6pt}, dashed, from=5-8, to=3-6]
	\arrow["{f_*^?}"'{pos=0.7}, color={rgb,255:red,43;green,171;blue,43}, curve={height=18pt}, dashed, from=7-9, to=5-6]
	\arrow[draw={rgb,255:red,205;green,100;blue,76}, curve={height=30pt}, dashed, no head, from=5-8, to=1-7]
	\arrow[draw={rgb,255:red,205;green,100;blue,76}, curve={height=24pt}, dashed, from=1-7, to=3-1]
	\arrow[curve={height=18pt}, dashed, no head, from=5-8, to=2-7]
	\arrow[curve={height=30pt}, shorten >=35pt, dashed, from=2-7, to=3-1]
	\arrow["{\revSatasarrowX{G'}}"'{pos=0.6}, color={rgb,255:red,214;green,92;blue,214}, dotted, from=6-4, to=3-6]
	\arrow["{\liftedX{\#_\csig}}"', color={rgb,255:red,43;green,171;blue,43}, dashed, from=6-4, to=5-6]
	\arrow["{\revSatasarrowX{f_*}}"'{pos=0.6}, color={rgb,255:red,153;green,92;blue,214}, curve={height=6pt}, dashed, from=8-7, to=5-8]
	\arrow["{\liftedX{\liftedX{\#_\csig}}}"', color={rgb,255:red,43;green,171;blue,43}, from=8-7, to=7-9]
	\arrow["{\twixU{v}}"', color={rgb,255:red,43;green,171;blue,43}, curve={height=30pt}, dashed, from=7-9, to=3-7]
	\arrow["{\liftedX{f^{*\Delta?}}}", color={rgb,255:red,205;green,100;blue,76}, dotted, from=8-7, to=8-3]
	\arrow["{\liftedX{\#_G}}", color={rgb,255:red,205;green,100;blue,76}, dotted, from=8-3, to=5-1]
	\arrow["{\twixdiagX{f}}"{description, pos=0.1}, curve={height=18pt}, dashed, from=5-8, to=4-2]
\end{tikzcd}
}

\newcommand{\idtoidX}[1]{\enma{(\pullbackedXindexedY{\tmapfun^\predsig}{#1}.v,\,
		\pullbackedXindexedY{\tmapfun^\predsig}{#1}.v)}}

\newcommand{\sattosatX}[1]{
	\pullbackedX{!^\twixSymbol_{f_*}}
}

\newcommand{\XNode}{\orangenode{\id_{\Dashv^{\twix{}}_f}}}
\renewcommand{\XNode}{\revtwixSatasarrowX{f}}

\newcommand{\curvedArr}{(\sattosatX{}, \sattosatX{})}
\renewcommand{\curvedArr}{}

\newcommand{\mainThDiamondSpansInside}{
	\begin{tikzcd}[ampersand replacement=\&, row sep = small]
		\&\&\& \textcolor{rgb,255:red,209;green,140;blue,71}{\bmapfun^G} \& \textcolor{rgb,255:red,209;green,140;blue,71}{\#_G} \& \textcolor{rgb,255:red,209;green,140;blue,71}{1_{\deltaslicecatGGG}} \\
		\&\&\&\& \textcolor{rgb,255:red,209;green,140;blue,71}{1_{\Dashv_G}} \\
		{\#^{\csig}} \&\& {\arfun^\csig} \&\& \XNode \\
		\&\&\&\& \textcolor{rgb,255:red,50;green,200;blue,50}{1_{\Dashv_{G'}}} \\
		\&\&\& \textcolor{rgb,255:red,50;green,200;blue,50}{\bmapfun^{G'}} \& \textcolor{rgb,255:red,50;green,200;blue,50}{\#_{G'}} \& \textcolor{rgb,255:red,50;green,200;blue,50}{1_{\deltaslicecatGGGprim}}
		\arrow["{(v, v_0)}", color={rgb,255:red,209;green,140;blue,71}, dashed, from=1-5, to=1-6]
		\arrow["{(h,h_0)}"', color={rgb,255:red,209;green,140;blue,71}, dashed, from=1-5, to=1-4]
		\arrow["{(\ell_G,\,\dom_G)}"', color={rgb,255:red,209;green,140;blue,71}, from=1-4, to=3-3]
		\arrow["{(\ell^\csig, \cod)}", from=3-1, to=3-3]
		\arrow[color={rgb,255:red,209;green,140;blue,71}, curve={height=12pt}, dashed, from=2-5, to=3-1]
		\arrow[draw={rgb,255:red,209;green,140;blue,71}, dashed, from=2-5, to=1-4]
		\arrow[draw={rgb,255:red,50;green,200;blue,50}, from=5-4, to=3-3]
		\arrow[draw={rgb,255:red,50;green,200;blue,50}, dashed, from=5-5, to=5-6]
		\arrow[draw={rgb,255:red,50;green,200;blue,50}, curve={height=-12pt}, dashed, from=4-5, to=3-1]
		\arrow[draw={rgb,255:red,50;green,200;blue,50}, dashed, from=4-5, to=5-4]
		\arrow["{(f_*^{LD}, f_*)}", curve={height=6pt}, dashed, from=5-4, to=1-4]
		\arrow["{(\fstardeltaUp,\, \fstardeltaUp)}"', curve={height=-6pt}, dashed, from=5-6, to=1-6]
		\arrow[color={rgb,255:red,209;green,140;blue,71}, from=2-5, to=1-5]
		\arrow[draw={rgb,255:red,209;green,140;blue,71}, dashed, from=2-5, to=1-6]
		\arrow[draw={rgb,255:red,50;green,200;blue,50}, from=4-5, to=5-5]
		\arrow[draw={rgb,255:red,50;green,200;blue,50}, dashed, from=4-5, to=5-6]
		\arrow[dashed, from=5-5, to=5-4]
		\arrow[dotted, from=3-5, to=1-6]
		\arrow[dotted, from=3-5, to=5-4]
		\arrow[from=3-5, to=2-5]
		\arrow[from=3-5, to=4-5]
	\end{tikzcd}
}

\newcommand{\mainThDiamondSpansOutside}{
\begin{tikzcd}[ampersand replacement=\&,sep=scriptsize]
	\&\&\&\&\& \textcolor{rgb,255:red,46;green,184;blue,46}{\revSatasarrowX{G'}} \\
	\&\& {} \\
	\& {} \&\&\&\&\& \textcolor{rgb,255:red,46;green,184;blue,46}{\bmapfun^{G'}} \& \textcolor{rgb,255:red,46;green,184;blue,46}{!_{G'}} \& \textcolor{rgb,255:red,46;green,184;blue,46}{1_{\deltaslicecatGGGprim}} \\
	{\revSatasarrowX{f}} \&\&\& {\tmapfun^{\csig}} \&\& {\arfun^\csig} \\
	\&\& {} \&\&\&\& \textcolor{rgb,255:red,225;green,114;blue,81}{\bmapfun^{G}} \& \textcolor{rgb,255:red,225;green,114;blue,81}{!_{G}} \& \textcolor{rgb,255:red,225;green,114;blue,81}{1_{\deltaslicecatGGG}} \\
	\& {} \\
	\&\&\&\&\& \textcolor{rgb,255:red,225;green,114;blue,81}{\revSatasarrowX{G}}
	\arrow["{(v', v'_0)}", color={rgb,255:red,46;green,184;blue,46}, dashed, from=3-8, to=3-9]
	\arrow["{(h',\,h'_0)}"', color={rgb,255:red,46;green,184;blue,46}, dashed, from=3-8, to=3-7]
	\arrow["{(\ell_G,\,\dom_G)}"'{pos=0.6}, color={rgb,255:red,46;green,184;blue,46}, from=3-7, to=4-6]
	\arrow["{(\ell^\csig, \cod)}", from=4-4, to=4-6]
	\arrow["{(\liftedX{\#_{G'}},\,\pbfundeltavX{G'})}"{pos=0.3}, color={rgb,255:red,46;green,184;blue,46}, from=1-6, to=4-4]
	\arrow[color={rgb,255:red,46;green,184;blue,46}, dashed, from=1-6, to=3-7]
	\arrow[color={rgb,255:red,225;green,114;blue,81}, from=5-7, to=4-6]
	\arrow[color={rgb,255:red,225;green,114;blue,81}, dashed, from=5-8, to=5-9]
	\arrow[color={rgb,255:red,225;green,114;blue,81}, from=7-6, to=4-4]
	\arrow[color={rgb,255:red,225;green,114;blue,81}, dashed, from=7-6, to=5-7]
	\arrow["{(f_*^{LD}, f_*)}"', curve={height=6pt}, from=5-7, to=3-7]
	\arrow["{(\fstardeltaUp,\, \fstardeltaUp)}", curve={height=6pt}, from=3-9, to=5-9]
	\arrow["{(\liftedX{\#_\csig}, 1)}"{pos=0.8}, color={rgb,255:red,46;green,184;blue,46}, curve={height=-12pt}, from=1-6, to=3-8]
	\arrow[color={rgb,255:red,46;green,184;blue,46}, curve={height=-24pt}, dashed, from=1-6, to=3-9]
	\arrow[color={rgb,255:red,225;green,114;blue,81}, curve={height=12pt}, from=7-6, to=5-8]
	\arrow[color={rgb,255:red,225;green,114;blue,81}, curve={height=24pt}, dashed, from=7-6, to=5-9]
	\arrow[color={rgb,255:red,225;green,114;blue,81}, dashed, from=5-8, to=5-7]
	\arrow["{(\liftedX{f_*^?}, (f_*,1))}", color={rgb,255:red,46;green,184;blue,46}, curve={height=-12pt}, dashed, from=4-1, to=1-6]
	\arrow["{(\liftedX{f^{*?}},\, (1,f^*))}"'{pos=0.8}, color={rgb,255:red,225;green,114;blue,81}, curve={height=18pt}, dashed, from=4-1, to=7-6]
	\arrow[color={rgb,255:red,46;green,184;blue,46}, curve={height=18pt}, dashed, from=4-1, to=5-7]
	\arrow["{(\twixdiagX{f},\, \pbfundeltavX{f})}"{pos=0.6}, dotted, from=4-1, to=4-4]
	\arrow["{\text{[pb]}}"{description, pos=0.3}, color={rgb,255:red,46;green,184;blue,46}, shorten <=21pt, shorten >=16pt, dotted, no head, from=2-3, to=5-3]
\end{tikzcd}
}

\section{Diagram constraint logic (DCL) as a Cartesian institution
	}\label{sec:skeFormal-heading}
Throughout this section,  \GG\ is an arbitrary  \caty\ with pullbacks and \frar{f}{G}{G'} is a \GG-arrow. We will begin with basic results about interoperability between slice \caties\ over \GG. Our applications need two extensions for this basic setting. First, considering \mor s between instances as spans in slice \caties; this is done in Sect. 5.2. Second, constraints are labelled diagrams, \ie, we need to consider labelling of slice \caties\ objects; this is done in Sect.5.4, after in Sect.5.3 the notion of a constraint signature is defined.  
%

 
\zd{It seems this blah-blah is not needed}
%

\renewcommand{\CC}{\GG}
\newcommand{\X}{G}

\renewcommand{\vind}{\enma{\mathsf{v}}}
\renewcommand{\hind}{\enma{\mathsf{h}}}

\renewcommand{\twixSymbol}{{}}


\subsection{Cartesian Interoperability 1: Basics} 
The following results are basic for slice \caties\ and their interrelations along $f$. 
\begin{lemma}[Substitution functors, P. Freyd \cite{Freyd-aspects}]\label{lemma:dirSubst}
There is an adjunction between slice \caties:
	$f_*{:}\, \GG/G \rightleftarrows\GG/G'\,{:}f^* $, $f_* \dashv f^*$,
	where functor $f_*$ (or $\GG/f$) is defined by postcomposition with $f$, and functor $f^*$ is defined by pulling back over $f$. 
	\qed
\end{lemma}
\begin{constrN}[in which the \pblemma\ is heavily employed by default.]\label{constr:PBfunVconstr}
	We take an object $G\in\GG$  and the product of \caties\ $\GG/G\timm\GG/G$,  in which one copy of $\GG/G$ is considered {\em horizontal} (h) and the other is {\em vertical} (v). Having an arrow $v$ in the v-copy of $\GG/G$ and an arrow $h$ in the h-copy of $\GG/G$, pulling $v$ back along $h$ gives rise to an operation \frar{\pbfunvX{G}}{\GG/G\timm\GG/G}{\arrowcatX{\GG}} (where \twocat\ is the interval \caty\ and  \arrowcatGG\ is the arrow \caty\ of \GG) as specified in diagram 
	\cref{eqdiag:PBfunVconstruction} below. In this diagram, all squares are PB, and pair $(\bar{\bar g}.\bar{u},\, g )$ is a \arrowcatCC-\mor\ \frar{\pbfunvX{G}(u,g)}{\pbfunvX{G}(v,h)}{\pbfunvX{G}(v',h')}).
	In applications, objects of \arrowcatGG\ are data instances, which are defined up to their domain  iso\mor\ (software object identities are not visible). Hence, it makes sense to identify objects of \arrowcatX{\GG}, which have the same codomain and iso\moric\ domains (but objects with iso\moric\ codomain are still diffeent), which makes \pbfunvX{G} a single-valued operation. 
	Composition and identity are strictly preserved after $\cong$-factorization. Thus, we have a   
	functor \frar{\pbfunvX{G}}{\GG/G\timm\GG/G}{\arrowcatX{\GG}_\cong}, where$\cong$ denotes the equivalence just specified.
	\begin{equation}\label{eqdiag:PBfunVconstruction}
		\twixlemmaProofDefiningPBfunV		
	\end{equation}
\end{constrN}
Now yet another application of the \pblemma\ gives us the following result.
\begin{mygroup}
\begin{lemma}[Twisted substitution 1: Basic]\label{lemma:twistedSubst} 
Diagram \cref{eqdiag:cattyDiag1} of \caties\ and functors  (without its OK-structure, \ie, arrows going into and from node OK, their domain and codomain objects) commutes for any \frar{f}{G}{G'}:
\renewcommand{\arrowcatCCcong}{\arrowcatX{\CC}}
\begin{equation}\label{eqdiag:cattyDiag1}
\begin{tabular}{c} 
		\twixlemmaCattydiagOneWithok
\end{tabular}
\end{equation}
\noindent In this diagram, 
\caties\ $P_G$, $\twixU{P}_f$, and $P_{G'}$ are products, whose projections are referred to as  `horizontal' (h) and 'vertical' (v),%
\footnote{twisted product $\twixU{P}_f$ depends on $f$ in that its feet are determined by $f$'s \dom\ and \cod}
and functors $f^*$ and $f_*$ are as in \cref{lemma:dirSubst}. 
Category \arrocatcongX{\GG} and functors $(\pbfunvX{X})_{X\in\Ob\GG}$ are  described above.  
%
As product $P_f$ is twisted, there is no a direct arrow from \twixU{P} to \arrowcatGGcong, but there are two indirect paths as shown. The lemma states that they commute, and the diagonal is denoted by \pbfunvX{f}.  \qed 
\end{lemma}
\end{mygroup}
\begin{mygroup}
	\renewcommand{\arrowcatdeltaX}[1]{\arrowcatX{#1}}
	\renewcommand{\deltaslicecatGGG}{\enma{\GG/G}}
	\renewcommand{\deltaslicecatGGGprim}{\enma{\GG/G'}}
	 \renewcommand{\slicecatdeltaGGG}{\enma{\GG/G}}
Now we consider the OK-part of \cref{eqdiag:cattyDiag1}
\begin{constrN}[Cartesian eSat-axiom]\label{constr:okStory4twistedlemma}
	Let \flar{\ok}{\arrowcatGGcong}{OK} be a functor thought of as a classifier (now we consider the  OK-part of diagram \cref{eqdiag:cattyDiag2--withOk}), \ie, selecting ``good'' elements of $\arrowcatGG$; it may be monic but not necessary. Pulling back functors \pbfunvX{G} and  \pbfunvX{G' } along \ok\ 
	gives us two ``okayed'' \caties\ with corresponding evidential functors $q_G$, $q_{G'}$ into $OK$ (horizontal lifted arrows into the \corring\ products are not shown to avoid clutter). If functor \ok\ is injective, then \SubprodOkDU{G}{} and \SubprodOkDU{G"}{} are subproduct \caties\ as PBs preserve monics. We will inaccurately refer to these \caties\ as to ok-subproducts even if \ok\ is not injective. 
	
	As for the ok-subproduct of the twisted product $P_f$, there are two ways to obtain them: via the left and the right arrows from $P_f$ to $P$ and $P'$ resp. But  \cref{lemma:twistedSubst} states that the two paths commute, then the \pblemma\ implies that the two \ok-subproduct are the same up to iso\mor, and we thus have a commutative square in the right part of \cref{eqdiag:cattyDiag1}, whose diagonal ( denoted $q_f$) can be considered as the ok-subproduct of the twisted product.  
	Comparing this diagram with the diagram defining twisted \mor s of eSpans \cref{def:twistedMorph} shows that the tuple  $(f^*, f_*, \; (\SubprodOkDU{f}{\twixSymbol}, \rho_f, \rho'_f, q_f))$ is a twisted \mor s of spans $(\SubprodOkDU{G}{},\liftedX{\ok}_G, q_G)$ and $(\SubprodOkDU{G'}{}, \liftedX{\ok}_{G'}, q_{G'})$. 
\end{constrN}
\begin{propo}[Cartesian institutions 1]\label{prop:cartInst-1}
	For a \caty\ $\GG$ with pullbacks, any functor \frar{\ok}{OK}{\arrowcatdeltaX{\GG}} (\ie, a classifier for \GG-arrows) gives rise to a e-institution 
	\frar{\models^\ok}{\GG}{\twistedSatcat(\catcat, OK)}. 
\end{propo}
{\em Proof.}  The two institution's ingredients are defined as follows. 

a) Functors \frar{\modfun}{\GG}{\opX{\catcat}} and \frar{\senfun}{\GG}{\catcat} are given by  $\modfun(G)=\deltaslicecatGGG$ with $\modfun(f)=f^*_\Delta$ and $\senfun(G)=\GG/G$ with $\senfun(f)=f_*$. 

b) The family of eSats is given by triples 
$(\liftedX{\ok}_G.v, \liftedX{\ok}_G.h, q_G)_{G\in\Ob\GG}$ (in diagram \cref{eqdiag:cattyDiag2--withOk}  the lifted arrows 
\flar{ \liftedX{\ok} }{P}{ \SubprodOkDU{G}{}} and  
\flar{ \liftedX{\ok} }{f}{ \SubprodOkDU{f}{}} 
are not shown to avoid clutter). Moreover, the triple $(f_*, f^*, \mathrm{span\,}\SubprodOkDU{f}{})$ is a twisted \mor.  In terms of elements, for any object $e\in \Obb OK$, we have a span
\spanrar{\models_G^e}{\deltaslicecatGGG}{\GG/G}, whose apex \caty\ is the $q_G$-fiber over $e$, and $v \models_G^{e} h$ iff $\pbfunvX{G}(h,v)=\ok(e)$. 

Finally, we see that mapping $f\mapsto (f^*, f_*,\SubprodOkDU{f}{\twixSymbol})$ is functorial:  
given 
$G\xrightarrow{~f~} G' \xrightarrow{~f'~}G''$, we have $(f.f')^*=f'^*.f^*$ and $(f.f')_*=f_*.f'_*$, and the proof of \cref{prop:twistedMorph--Compose} shows that $\SubprodOkDU{f.f'}{}$ is given by span composition of \SubprodOkDU{f}{} and \SubprodOkDU{f'}{} (\ie, by taking PB of cospan $(\rho'_f, \rho_{f'})$). 
\qed

%
%
\end{mygroup}

 
 \subsection{Cartesian Interoperability 2: Slice \caties\ with delta \mor s}
 Our discussion in \cref{sec:vInst-motivation} shows that SE applications need instance \mor\ to be spans (also called {\em deltas}) rather that ordinary arrows. 
\begin{defN}[Delta slice  \caties]
	Category \deltaslicecatGGG\ has objects and arrows of \caty\  $\spanfuncongX{(\GG/G)}$ but arrow composition is defined differently as span composition in \GG, \ie, via pullbacks in \GG\ rather than in $\GG/G$ --- the latter may not have PBs at all. 
\end{defN}
\begin{defN}[Delta arrow \caties]
Category  $\arrowcatdeltaGG$ has objects of \arrowcatGG\ as objects and {\em special} spans in \spanfuncongX{\arrowcatGG} as \mor s, namely, those spans  \spanrar{r}{a}{b} in \arrowcatGG, whose codomain component  $\cod(r)$ is $q_\%=(1,q)$ for some \GG-arrow $q$ 
(see \cref{constr:mor2spans} for $q\mapsto q_\%$). Commutative diagram \cref{eqdiag:arrowcatMorphAsDeltas}(upper a) unpacks this definition and shows a (representative of a \congg-class) \mor\ \frarxy{a}{b} in \arrowcatdeltaGG. 
\begin{equation}\label{eqdiag:arrowcatMorphAsDeltas}
\begin{array}{c@{\quad}c}
	\begin{array}{c}
		\arrowcatDeltaMorph
		\\ 	\arrowcatDeltaCompose
	\end{array}  
& \codfibrForDeltas
\\ a) &  b)
\end{array}
\end{equation}
Composition is defined via pullbacks in \GG\  for the upper components and arrow composition in \GG\ for the lower components: diagram \cref{eqdiag:arrowcatMorphAsDeltas} (lower a)g shows how it works (the span of dashed arrows is produced by PB). Due to \congg-factorization, so defined delta composition is strongly associative, and \mor s whose all horizontal arrows are \GG-identities are identities in \arrowcatdeltaGG.  
To ease notation, we will write 

i) \spanrar{(p_0, q_0, q)}{a}{b} if arrow $\hat r$ is clear from the context, and 

ii)  \frar{(q_0,q)}{a}{b}  if the leg $p_0$ is the identity of $a$'s domain. 
\end{defN}
\begin{lemma}[Codomain fibration for deltas]\label{lemma:codfibr4deltas}
	Functor \frar{\coddelta}{\arrowcatdeltaGG}{\GG} is a fibration, whose lifting is provided by pullbacks.
\end{lemma} 
{\em Proof.}  Given $\frar{b}{B_0}{B}\in\Ob\arrowcatdeltaGG$ and \frar{q}{A}{B} in \GG, 
the lifted $q$ is \frar{
	 (q_b,q)
	}{b_q}{b}, where \GG-arrows $b_q$ and $q_b$ are produced by PB of $(q,b)$ as shown in diagram b)  above (its the standard codomain lifting). Now if \spanrar{r}{a}{b} is any \mor\ in \arrowcatdeltaGG\ with $\coddelta(r)=q$, the universality of the pullback square in b) provides a unique arrow $!_r$ and then  \spanrar{\#=(p_0, !_r, 1_A)}{a}{b_q} is a unique arrow over $A$ in \arrowcatdeltaGG\ that makes triangle specified in diagram b) on the right commuting. This proves that lifting $\bar q$ is weakly Cartesian over $q$. To prove that lifting is Cartesian, it remains to show compositionality: $\liftedX{q.q'}=\bar q.\bar q'$, which is indeed demonstrated by a routine diagram chasing.  \qed
\begin{lemma}[Twisted substitution 2: Vertical arrow \mor s are  deltas] \label{lemma:twistedSubst2-Deltas} 
Any \GG-arrow \frar{f}{G}{G'} gives rise to a functor \frar{f^*_\Delta}{\slicecatdeltaGGGprim}{\slicecatdeltaGGG} and  commuting diagram \cref{eqdiag:cattyDiag2--withOk} of \caties\ and functors  (ignore its $OK$-part for a moment). 
\begin{figure}[h]
\begin{equation}\label{eqdiag:cattyDiag2--withOk}
	\twixlemmaCattydiagTwo
\end{equation}
\end{figure}
\end{lemma}
{\em Proof.} 
Mapping 
\fstardeltaUp\ is obtained by postcomposition of spans (\ie, span-shape diagrams) in \deltaslicecatGGG\ 
with $f^*$. 
It preserves span composition (recall that it is defined via PB in \GG) as $f^*$ is the right adjoint to $f_*$ and hence preserves pullbacks. Identity preservation is obvious. On objects, \pbfundeltavX{G} acts as \pbfunvX{G}.
For a pair of \mor s, horizontal \frar{g}{h}{h'} in $\GG/G$ and vertical \spanrar{u=(\apexX{u}, u,u')}{v}{v'} in  \deltaslicecatGGG, we define arrow 
\[ \frar{\pbfundeltavX{G}(u,g)}
	{\pbfundeltavX{G}(v,h) }
	{\pbfundeltavX{G} (v', h')}
	\]
in \arrowcatdeltaGGcong\ as the following tuple $(\hat r, p_0, q_0, q)$ (see diagram \cref{eqdiag:arrowcatMorphAsDeltas} above)
specified by diagram \cref{eqdiag:twixlemmaProof-DefPBfunV-Delta}:
\begin{equation}\label{eqdiag:twixlemmaProof-DefPBfunV-Delta}
	\twixlemmaProofDefiningPBfunVDelta
\end{equation}
In this diagram, we write $x_h$ for $\pbfundeltavX{G}(x, h)$ for an element $x\in\GG/G$, the source and the target arrows of $\pbfundeltavX{G}(u,g)$ are shown red and inter-connecting arrows needed for defining the \mor\ are dashed red. Thus, 
$\dom(\hat r)=\dom(\apexx{u}_h)$, 
$p_0=u_h$, 
$q_0= \apexx{u}_g.u'_{h'}$, 
and $q=g$. 
This definition is compositional as operations $\pbfunvG(\_,h)$ and $\pbfunvG(\_, h')$ are nothing but   $h^*$ and $h'^*$ and hence preserve pullbacks as being right adjoints; then 
$\pbfundeltavX{G}(g, u_1.u_2)$ is defined as above with $(\apexX{u_1.u_2})$ playing the role of $\apexx{u}_g$. 
Identity preservation is obvious and we obtain a functor \frar{\pbfundeltavX{G}} {\slicecatdeltaGGG\timm\GG/G} {\arrowcatdeltaGG_\cong} such that $\pbfundeltavX{G}.\cod=h.\dom_G$. 
Now similarly to the proof of \cref{lemma:twistedSubst}, the \pblemma\ implies commutativity $\leftDeltaU.\pbfundeltavX{G}=\rightDeltaU.\pbfundeltavX{G'}$ and hence the entire diagram \cref{eqdiag:cattyDiag2--withOk}. 
\qed
%
\renewcommand{\arfun}{\enma{\mathsf{ar}}}

\newcommand{\cartInstHomogenDiag}{
\begin{tikzcd}[ampersand replacement=\&,sep=scriptsize]
	\arrowcatdeltaGG \&\& \GG \&\& {\GG/G} \& \textcolor{rgb,255:red,5;green,20;blue,5}{\_{\times}\_} \&\& \textcolor{rgb,255:red,1;green,4;blue,1}{\slicecatdeltaGGG} \\
	\& \textcolor{rgb,255:red,33;green,87;blue,212}{\okTargetsig} \\
	\Instcatsig \&\& {\carrop{\csig}} \&\& \textcolor{rgb,255:red,33;green,87;blue,212}{\LDx{G}} \& \textcolor{rgb,255:red,43;green,171;blue,43}{\preInstcatX{G}} \\
	\\
	\&\&\&\& {\modelsd_G}
	\arrow["\tmapfunsig", from=3-1, to=1-1]
	\arrow["\coddelta", from=1-1, to=1-3]
	\arrow["\arfun"'{pos=0.7}, from=3-3, to=1-3]
	\arrow["{\ellX{\csig}}", from=3-1, to=3-3]
	\arrow["{\ok_\csig}", color={rgb,255:red,33;green,87;blue,212}, dashed, from=3-1, to=2-2]
	\arrow[draw={rgb,255:red,33;green,87;blue,212}, dashed, from=2-2, to=1-1]
	\arrow["{\text{[fib]}}"{description}, draw={rgb,255:red,33;green,87;blue,212}, dashed, from=2-2, to=3-3]
	\arrow["{\text{[pb]}_\csig}"{description}, color={rgb,255:red,32;green,83;blue,203}, draw=none, from=2-2, to=1-3]
	\arrow["{\dom_G}"', from=1-5, to=1-3]
	\arrow["{\ellX{G}}"', color={rgb,255:red,33;green,87;blue,212}, dashed, from=3-5, to=3-3]
	\arrow["{\bmapfunX{G}}"', color={rgb,255:red,33;green,87;blue,212}, dashed, from=3-5, to=1-5]
	\arrow["{h^0_G}"{text={rgb,255:red,3;green,12;blue,3}}, from=1-6, to=1-5]
	\arrow["{h_G}", color={rgb,255:red,33;green,87;blue,212}, from=3-6, to=3-5]
	\arrow["{v_G}"{description, text={rgb,255:red,3;green,12;blue,3}}, from=3-6, to=1-8]
	\arrow["{!_G}"{description}, dashed, from=3-6, to=1-6]
	\arrow["{\pbfundeltavX{G}}"'{text={rgb,255:red,32;green,83;blue,203}}, curve={height=24pt}, dashed, from=1-6, to=1-1]
	\arrow["{\#_G}"{description, pos=0.6}, draw={rgb,255:red,33;green,87;blue,212}, dashed, from=3-6, to=2-2]
	\arrow["{\text{[pb]}_{G}}"{description, pos=0.6}, color={rgb,255:red,32;green,83;blue,203}, draw=none, from=3-5, to=1-3]
	\arrow[""{name=0, anchor=center, inner sep=0}, "{\pEx{G}}", color={rgb,255:red,33;green,87;blue,212}, curve={height=-12pt}, dashed, from=5-5, to=3-1]
	\arrow[""{name=1, anchor=center, inner sep=0}, "{\liftedX{\oksig}}"{description}, color={rgb,255:red,33;green,87;blue,212}, dashed, from=5-5, to=3-6]
	\arrow["{v^0_G}"'{text={rgb,255:red,3;green,12;blue,3}}, from=1-6, to=1-8]
	\arrow["{\text{[fib]}}"{description}, shift right=3, color={rgb,255:red,230;green,76;blue,127}, draw=none, from=3-1, to=3-3]
	\arrow["{\text{[dfib]}}"{description}, shift right=3, color={rgb,255:red,32;green,83;blue,203}, draw=none, from=3-3, to=3-5]
	\arrow["{\pSx{G}}"{pos=0.8}, dashed, from=5-5, to=3-5]
	\arrow["{\pMx{G}}"{description}, curve={height=30pt}, dashed, from=5-5, to=1-8]
	\arrow["{\text{[dfib]}}"', shift right=1, draw=none, from=1-3, to=1-5]
	\arrow["{\text{[fib]}}"', shift right=1, draw=none, from=1-1, to=1-3]
	\arrow["{\text{[pb]}}"{description}, color={rgb,255:red,32;green,83;blue,203}, dotted, no head, from=0, to=1]
\end{tikzcd}
}

\newcommand{\predsigDefDiag}{
\begin{tikzcd}[ampersand replacement = \&, row sep=2ex, column sep = 4ex]
	\Instcat_\predsig
	\&\& \deltaarrowcatGG 
	\\ \& \dernode{\Instcat^?_\predsig}	\&
	\\
	\carrop{\predsig} \ar[rr, "\arfun"]  \&\& \GG
	\ar[from=1-1, to=1-3, "\tmapfunX{\predsig}"] 
	\ar[from=1-1, to=3-1, "\labfunSig" ', thick ] 
	\ar[from=1-1, to=3-1, phantom, "\gapnamegap{2.5}{\fiblabelRed}{-2.} ",  thick ]
	\ar[from=1-1, to=2-2, derived, "\oksig"]
	\ar[from=1-3, to=3-3, "\coddelta" ', thick ]
	\ar[from=1-3, to=3-3, phantom, "\gapnamegap{2.5}{\fiblabelBlack}{-2.} ",  thick ]
	\ar[from=2-2, to=3-1, derived]  
	\ar[from=2-2, to=1-3, derived]
	\ar[from=2-2, to=3-3, phantom, gray,  "\text{[pb]}", pos=0.25]
	\end{tikzcd}
}


\newcommand{\fibrMorphDiagGen}{
\begin{tikzcd}[ampersand replacement=\&,sep=scriptsize]
	\bullet \& \circ \& \bullet \\
	\bullet \& \bullet
	\arrow["{\tmap_i}"', from=1-1, to=2-1]
	\arrow["{\bar d_0}"', from=1-2, to=1-1]
	\arrow["{\bar d'_0}", from=1-2, to=1-3]
	\arrow["{\apexX{\liftedX{d}}}"', from=1-2, to=2-2]
	\arrow["{\bar d}", from=2-2, to=2-1]
	\arrow["{\tmap_{i'}}", from=1-3, to=2-2]
\end{tikzcd}
}

\newcommand{\fibrMorphDiagToBe}{
\begin{tikzcd}[ampersand replacement=\&,sep=small]
	\bullet \& \circ \\
	{G_c} \& {G_{c'}} \\
	c \& {c'}
	\arrow["{\tmap_i}"', from=1-1, to=2-1]
	\arrow["{\liftedX{d^\arfun}}"', dashed, from=1-2, to=1-1]
	\arrow["{\tmap_{i'}}", dashed, from=1-2, to=2-2]
	\arrow["{d^\arfun}", from=2-2, to=2-1]
	\arrow["{\text{[pb]}}"{description}, draw=none, from=1-2, to=2-1]
	\arrow["d"', from=3-1, to=3-2]
\end{tikzcd}
}

\newcommand{\provingCosoundPrelimDiags}{
\begin{tikzcd}[ampersand replacement=\&,sep=small]
	{} \& {} \& {} \& {} \& {} \& {} \& {} \\
	{i.d^*} \& {i'} \& {c'} \& {G_{c'}} \& G \& {t'} \& t \\
	\& i \&\& c \& {G_c} \& t \\
	\&\&\& \liftedTriangleDom \\
	\&\&\& {x'} \& x
	\arrow["{d^\arfun}"', from=2-4, to=3-5]
	\arrow["\liftedTriangle", from=4-4, to=5-5]
	\arrow["y"', from=5-4, to=5-5]
	\arrow["{\bar d}"', from=2-1, to=3-2]
	\arrow["{y_E}", from=2-2, to=3-2]
	\arrow["{!_{y_E}}"', dashed, from=2-2, to=2-1]
	\arrow["d"', from=2-3, to=3-4]
	\arrow[""{name=0, anchor=center, inner sep=0}, "{\carrop{\csig}}", draw=none, from=1-3, to=1-4]
	\arrow[""{name=1, anchor=center, inner sep=0}, "\Instcatsig", draw=none, from=1-1, to=1-2]
	\arrow[""{name=2, anchor=center, inner sep=0}, "\slicecatdeltaGGG", draw=none, from=1-6, to=1-7]
	\arrow["{!_y}", dashed, from=5-4, to=4-4]
	\arrow[""{name=3, anchor=center, inner sep=0}, "\bmap"', from=3-5, to=2-5]
	\arrow[""{name=4, anchor=center, inner sep=0}, "{\bmap'}", from=2-4, to=2-5]
	\arrow["{y_M}"', from=2-6, to=3-6]
	\arrow["{1_t}", from=2-7, to=3-6]
	\arrow["{y_M}", dashed, from=2-6, to=2-7]
	\arrow[""{name=5, anchor=center, inner sep=0}, "{\LDx{G}}", draw=none, from=1-4, to=1-5]
	\arrow["{\triangleMorX{d^\arfun}}"{description}, curve={height=6pt}, draw=none, from=4, to=3]
	\arrow["\ellsig"', curve={height=-18pt}, shorten <=17pt, shorten >=17pt, from=1, to=0]
	\arrow["{\ellX{G}}", curve={height=18pt}, shorten <=14pt, shorten >=14pt, from=5, to=0]
	\arrow[""{name=6, anchor=center, inner sep=0}, "{\modelsd_{G}}"{description}, curve={height=24pt}, shorten <=19pt, shorten >=19pt, tail reversed, from=2, to=5]
	\arrow[shift right=2, curve={height=30pt}, shorten >=19pt, from=6, to=1]
\end{tikzcd}
}

\newcommand{\provingCosoundFinalDiag}{
\begin{tikzcd}[ampersand replacement=\&,sep=scriptsize]
	{} \&\&\&\& {} \\
	{(c',\bmap')} \&\&\&\& {(c,\bmap)} \\
	{{}} \&\& {\liftedTriangleDom{:}\modelsd^{i.d^*}} \&\& {{}} \\
	{t'} \&\&\&\& t
	\arrow["{\modelsd_G}", draw=none, from=1-1, to=1-5]
	\arrow["\liftedTriangle", from=2-1, to=2-5]
	\arrow[""{name=0, anchor=center, inner sep=0}, "{x'{:}\modelsd^{i'}}"{description}, dotted, tail reversed, from=2-1, to=4-1]
	\arrow[""{name=1, anchor=center, inner sep=0}, "{x{:}\modelsd^{i}}"{description}, dotted, tail reversed, from=2-5, to=4-5]
	\arrow["{y_M}"', from=4-1, to=4-5]
	\arrow[dotted, from=3-3, to=2-1]
	\arrow[dotted, from=3-3, to=4-5]
	\arrow["\liftedTriangle"', shorten <=6pt, shorten >=11pt, from=3-3, to=3-5]
	\arrow["{!_y}"'{pos=0.6}, color={rgb,255:red,243;green,63;blue,162}, shorten <=17pt, dashed, from=3-1, to=3-3]
	\arrow["y"{description}, curve={height=-24pt}, shorten <=25pt, shorten >=25pt, from=0, to=1]
\end{tikzcd}
}

\subsection{Constraint signatures and constraints}\label{sec:skeFormal-constraints}
\begin{figure}[h]
\centering
\begin{tabular}{c@{}c@{}c@{}c} 
\parbox{0.225\linewidth}{\centering    
			\includegraphics[width=0.975\linewidth]{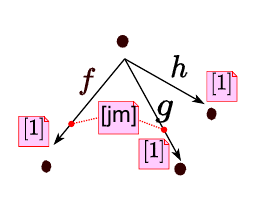}
		}
&\parbox{0.2\linewidth}{\centering  
			$\xymatrix @C=0.65em {
				[1] &  &	\jmpred \ar@<+0.5ex>[ll]^{d_1\,} \ar@<-0.5ex>[ll]_{d_2\,}	 
			}$ 
		} 
& \parbox{0.1\linewidth}{\centering  
			\begin{tikzcd}[row sep=3ex, column sep=1ex, ampersand replacement = \&]
				0 \ar[d, "01"] \\ 1
			\end{tikzcd}	
		} 
& \parbox{0.2\linewidth}{\centering   
		\begin{tikzcd}[row sep=1.25ex, column sep=4ex, ampersand replacement = \&]
				\& 1 \\
				0 \ar[ur, "01"] \ar[dr, "02" '] \& \\
				\& 2
			\end{tikzcd}
	} 
\\ 
a) sketch $S$
& b) Category $\predsig$
& c1) $[1]^\arfun$
& c2) $\jmpred^\arfun$
\end{tabular}
\caption{A sample signature  $(\carr{\predsig}, \arfun)$: $d_1^\arfun(01)=01$, $d_2^\arfun(01)=02$, 
} 
\label{fig:signature4jm}
\end{figure}

\ActTR{}{
\subsubsection{Hierarchical categories}
\zdzd{I'll write a short motivating piece later}
\renewcommand{\HH}{\enma{H}}
\begin{mydef}[Hierarchical categories]\label{def:hcat}
	We consider a natural number $N$ as a category 
	$$\NN= \{ 0\rightarrow 1\rightarrow \ldots\rightarrow (N-1)\}$$
	consisting of $N$ objects and $N-1$ non-identity arrows.%
	\footnote{The reason we start counting at 0 rather than 1 is to make the generalization for the first infinite ordinal $\omega$ straightforward: category $\ovr{\omega}$ has all natural numbers $n<\omega$ as objects.} 
	
	A {\em \hical\ category (\hcaty\ {\em in brief}) of height $N$} is given by a small category \HH\ along with a surjective on objects functor \frar{\heifun}{\HH}{\NNop} so that all {\em layers} $\Ob_n H:=\heifun^{-1}(n)$ for $n<N$ are not empty.  We refer to $N$ by $\Hei(\HH)$. Note that if $N=0$, then category \NN\ is empty and we have an empty \hical\ \caty\ of height 0. Thus, $\Ob_0\HH=\varnothing$ iff $\Hei(\HH)=0$. 
	
	Moreover, the following {\bf\em acyclicity} condition holds for any two objects $x,y$ in \HH: if $\HH(x,y)\noteq\varnothing$, then either $\heifun(x) >\heifun(y)$ or $x=y$, but in the latter case we require $\HH(x,x)=\{\id_x \}$. Thus, all \HH-arrows go down from higher to lower layers, and the only arrows within a layer are identities. 
	\zdzd{there was also the Parenting condition but it is now commented in the text}
	%
	For an h-\caty\ \HH, we call elements of layer $\Ob_n\HH$ {\em n-cell types} or {\em n-(hyper)edge  types} and typically denote them by $H$. We will also call 0-cell types  {\em node types}. 
\end{mydef}
\begin{mycor}\label{cor:hcat}
	If \HH\ is an h-\caty, then for any $n<N$, the full subcategory $\HH_n\subset\HH$ generated by objects of height less than $n$ is a h-\caty\ as well. We have a cumulative chain of h-\cats
	$$
	\HH_0 \subset \HH_1 \subset\ldots\subset\HH_{N-1}=\HH
	$$
	and for any object $x\in\HH$, $\heifun(x)=\nfont{min}\compr{n}{H\in\HH_n}$.
\end{mycor}
\begin{mydef}[Discrete h-\cats]\label{def:hcat-disc}
	A \hical\ \caty\ is called {\em discrete} if its only arrows are identities. However, a discrete \hcaty\ is more than a set: it is a set equipped with a height function, \ie, a layered set  $\HH=\bigcup_{n<N}H_n$. We will also write $L_{\le n}(\HH)$ for the set $\bigcup_{0\le i\le n}L_i(\HH)$
\end{mydef}
\begin{mydef}[Finitary hierarchical categories]\label{def:hcatfin}
	A \hcaty\ is called {\em finitary} if for any object $x$ the set $\HH(x,\_)$ is finite. \sizeXno{In this paper, we will assume all \hcats\ to be finitary.} 
\end{mydef}
\begin{mycor}[Finite h-\cats]
	\label{cor:finiteHCats}
	A finitary  \hcaty\ \HH\ is finite as soon as its object set $\Ob{\HH}$ is finite
\end{mycor}
\begin{myrem}[Finitarity conditions]\label{rem:finty}
	\defref{hcatfin} is the first in the family of similar definitions, which require finiteness of some construct assigned to the notion at hand. If an h-\caty\ \HH\ is finite, then it is finitary, but the converse is obviously not true, and we will often deal with infinite but finitary objects (signatures, sketches, graphs). 
\end{myrem}

}
%
Thinking discretely, the syntactical part of a constraint signature  \predsig\ is to be  a set \carr{\predsig} of  constraint (predicate) names/symbols with their assigned arity graphs, which makes arity a function  \frar{\arfun}{\carr{\predsig}}{\Ob\GG} (below we will denote graph $\arfun(c)$ by $c^\arfun$). However, we often deal with dependencies between constraints, for example, consider the property of a span of functions to be jointly monic (jm). As, in general, semantics of an arrow $f$ in a schema is a span \semm{f}, declaring a span of arrows $(f,g)$ to be jm should assume 
that arrows $f$ and $g$ are to be functions, \ie, have multiplicity [1], as is shown in \cref{fig:signature4jm}(a). An intelligent tool would itself infer constraints [1] for arrows $f$, $g$ as soon as a jm-constraint is declared for span $(f,g)$. 
Hence, it makes sense to record the dependency by adding to the signature two dependency arrows (see diagram b) below) together with graph \mor s \frar{d_i^\arfun}{[1]^\arfun}{\jmpred^\arfun}, $i=1,2$. 
This makes a signature a category and arity a functor \frar{\arfun}{\opX{\carr{\predsig}}}{\GG}. 
%

%
This syntactical arrangement only makes sense if it is aligned with the corresponding semantic mapping: pulling back a valid \jmpred-instance \frar{t}{X}{G_\jmpred} along $d_i^\arfun$ ($i=1,2$)  results in valid instances of [1] so that mapping \semm{.} can be seen as a functor $\predsig\rightarrow\catcat$ (compositionality is by the pullback lemma). It is equivalent to an opfibration \frarxy{\int\semm{.}}{\predsig} 
or, equivalently, a fibration \frar{\ellX{\csig}}{\Instcat}{\predsigop}, where $\Instcat=\opX{(\int\semm{.})}$ 
\begin{defN}[Dag-\caties, Lawvere-Makkai \cite{makkai-ske}(p.62) ]
A {\em dag} category is a category whose underlying graph is a dag, \ie, a directed acyclic (multi-)graph. 
Then $Hom(x,x){=}\{\id_x\}$ for all objects $x$. %
\end{defN}
\begin{defN}[Constraint Signatures]\label{def:predsig}
Let \GG\ be an \FCC\  whose objects are called {\em graphs}. A {\em constraint} 
{\em signature} over \GG\ is a commutative square diagram \predsig\ in \catcat\ shown as the outer rectangle in diagram  \cref{eqdiag:predsigDef}a), in which \carr{\predsig} is a dag-\caty\ called the {\em carrier}, the left vertical arrow is a fibration (whereas the right one is always such due to \cref{lemma:codfibr4deltas}
) 
and the pair of functors $(\tmapfunX{\predsig}, \arfun)$ is a fibration \mor\ (below we will often omit script \predsig). 
 
 Objects of \caty\ \carrpredsig\ are called {\em constraint} names/symbols, and arrows are {\em dependencies}. Functor \arfun\ provides constraint names with their {\em arity} graphs (we will write $G_c$ for graph  $\arfun(c)$)  and maps dependencies \frar{d}{c}{c'} to \GG-\mor s \frar{d^\arfun}{G_{c'}}{G_c} where we write $d^\arfun$ for $\arfun(d)$. 
\begin{equation}\label{eqdiag:predsigDef}
\begin{array}{c@{\qquad}c} 
	\predsigDefDiag
	&  \fibrMorphDiagToBe
	\\ a) & b) 
\end{array}
\end{equation}
Objects of \caty\ \Instcatsig\ are {\em indexes} of  {\em valid} {\em instances}: 
an index  $i\in\Instcatsig$ with $\ellX{\csig}(i)=c$ 
gives us a valid instance  \frar{\tmap^\predsig_i}{X_i}{G_c}.

 The \req\ to pair $(\tmapfunsig, \arfun)$ to be a fibration \mor\ means that when a dependency \flar{d}{c}{c'} in \carrop{\predsig}\ is lifted to 
\flar{\bar d}{i}{\,i'{=}d^*(i)} in \InstcatX{\predsig}, its image $\tmap^\csig_{\bar d}$ in \arrowcatdeltaGG\ is 
the lift of $d^\arfun$ along $\coddelta$ (see \cref{lemma:codfibr4deltas}), \ie, the diagram b) above. 
\end{defN}
\begin{constrN}[Preinstance classifier]
Taking PB of span $(\cod,\arfun)$ gives us cospan \okTargetX{\csig} (note the label [pb]).  Objects of \caty\ \okTargetsig\ are called {\em \csig-preinstances}: they are all possible pairs $(c,t)$ with $t\in\GG/G_c$ without any concerns of $t$'s validity (hence, the question mark script). PB-universality ensures a unique functor $\ok_\csig$, which selects valid instances amongst all possible candidates. Instances we considered in previous sections were uniquely identified by such pairs, which makes span $(\ellsig, \tmapfunsig)$ jointly monic, and functor \oksig\  an embedding. The general definition above admits different indexes with the same (label, instances)-values of \oksig\ as it is discussed in \cref{sec:ske4systems}. 
\end{constrN}
\zd{To be completed!!!
\begin{example}[Vehicles-Drivers database]\label{ex:veh-drivers-predsig}
	Play with the vehicle ontology example in sect.2 
\end{example}
}

The following notion is fundamental for DCL.
%
%
\begin{mygroup}
\renewcommand{\projfunUD}[2]{\enma{p_#1}}
\begin{constrN}[Constraint satisfaction] \label{constr:catOfLabelledDiags}
	Let $G$ be an arbitrary object/graph $G\in\Obb\GG$. Then we have the following commutative diagram of \caties\ and functors \cref{eqdiag:cartInst-homogenDiags}. The left square (including its internal PB shown triangularly) has been discussed above. The middle square is PB of cospan $(\arfun, \dom_{G'})$. Objects of \caty\ \LDx{G}\ are {\em constraint declarations} over $G$,  \ie, pairs $(c, \bmap_c)$ of a constraint name $c\in \Obb\carrpredsig$ and a {\em binding mapping} (diagram) \frar{\bmap_c}{G_c}{G} (and so constraints over $G$ are nothing but {\em labelled diagrams} over $G$). Arrows \frarxy{(c,\bmap)}{(c',\bmpa')} are pairs $(d, d^\arfun)$ 	with  \flar{d}{c}{c'} a dependency  in {\carrpredsig} (note the reversal)
	and \frar{d^\arfun}{\bmap}{\bmap'} an arrow in $\GG/G$, \ie, $d^\arfun.\bmap'=\bmap$ (recall that $d^\arfun$ denotes  \frar{\arfun(d)}{G_{c}}{G_{c'}}).  Functor \ellX{G} is a discrete fibration as $\dom_G$ is such and PBs preserve them. 
	Nodes $\_\timm\_$ and \PreinstcatX{G} are products, and $!_G$ is a unique arrow by the universality. 
\begin{equation}\label{eqdiag:cartInst-homogenDiags}
	\cartInstHomogenDiag
\end{equation}
	The commuting square $\pbfundeltavX{G}.\coddelta = h^0_G.\dom_G$  was built above in \cref{lemma:twistedSubst2-Deltas}. We thus have a span with apex \PreinstcatX{G} and feet \arrowcatdeltaGG\ and \carrop{\csig}, which commutes with cospan $(\coddelta,\arfun)$ (\ie, together they make a commuting square). Hence we have a unique arrow $\#_G$ by the universality of pullback [pb]$_\csig$. Finally, we take PB of cospan $(\oksig, \#_G)$ and obtain a monic span $(\modelsd_G, \pEx{G}, \liftedX{\oksig})$, for which we also designate its diagonal \pSx{G} and projection $\pMx{G}=\liftedX{\oksig}.v_G$. We consider the ternary span $(\modelsd_G, \pSx{G}, \pMx{G}, \pEx{G})$ as a eSat span over base $(\catcat, \Instcatsig)$.  It is also monic as \PreinstcatX{G} is the product of the two feet. 
	
	For a triple $(t, (c,\bmap), i)\in \Ob\slicecatdeltaGGG\timm\Ob\slicecatcsigGGG\timm\Ob\InstcatSig$, we will write $t \modelsd_G^i (c,\bmap)$ if $\ellsig(i)=c$ and  $\pbfunvdeltaX{G}(t, \bmap)=\tmap_i^\csig$ in $\GG/G_c$.
\end{constrN}
\begin{propo}[coSoundness]\label{prop:cartInst-homo-Soundness} For any $G\in\Ob\GG$, functor \pSx{G} in diagram above is a left fibration wrt. $\pMx{G}$, \ie, it is a fibration whose lifts are \pMx{G}-vertical. 
\end{propo}
{\em Proof.} We will prove that $p_S$ is a fibration by, first, building weakly Cartesian lifting of an arrow in \slicecatcsigGGG, and then showing that the lifting construction is compositional. Let $x\in\Ob\modelsd_G$, whose projections on the three feet are $t=p_M(x)$, $(c,\bmap)=p_S(x)$ and $i=p_E(x)$ so that $t\modelsd^i (c,\bmap)$ (\ie, $\pbfundeltavX{G}(t, \bmap) = \tmap_i^\csig\in \GG/G_c$). Arrows in \slicecatcsigGGG\ are commutative triangles generated by dependencies in \carrop{\predsig}, and let \flar{\triangleMorX{d^\arfun}}{(c, \bmap)}{(c',\bmap')} be such, \ie, we have \frar{d}{c}{c'} in \carrpredsig\ and  $d^\arfun.\bmap=\bmap'$ as shown in the middle part of diagram \cref{eqdiag:provingCosound}a) (\ie, the shown arrow \frar{d}{c'}{c} is an arrow in \carrop{\predsig}). 

We want to lift  arrow \triangleMorX{d^\arfun} to a weakly Cartesian arrow \flar{\liftedTriangle}{x}{\cdot} in $\modelsd_G$, whose domain will be denoted by {\liftedTriangleDom} (as is customary for fibrations).  Clearly, $\ellX{G}(\triangleMorX{d^\arfun})=d$, and as arrow \ellX{\csig} is a fibration (by the definition of a constraint signature), $d$ is lifted to arrow \flar{\bar d}{i}{i.d^*} in \Instcatsig, and the fibration \mor\ condition implies that $t\modelsd_G^{i.d^*} (c',\bmap')$. 
No00w recall that span $(p_M, p_S, p_E)$ is monic, and hence the triple of arrows $(1_t, \triangleMorX{d^\arfun}, \bar d)$ targeting at, resp, $p_M(x), p_S(x)$ and $p_E(x)$, gives us a unique arrows $\liftedTriangle$ 
into $x$. These data are specified in the right-upper triangle of diagram \cref{eqdiag:provingCosound}b) above (ignore arrow $y$ for a moment). 

Now we prove the universal property of the lift: for any arrow \frar{y}{x'}{x} with $p_S(y)=\triangleMorX{d^\arfun }$, there is a unique arrow  $!_y$ over $G_{c'}$ (\ie, $p_S(!_y)=1_{G_{c'}}$) such that the bottom triangle in diagram \cref{eqdiag:provingCosound}a) commutes. 
\begin{figure}
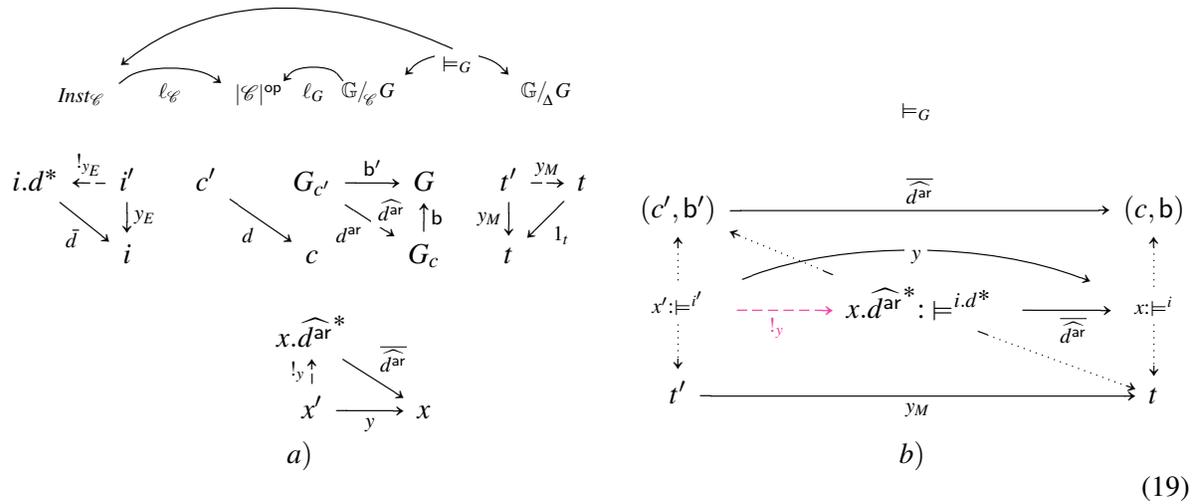

\begin{equation}\label{eqdiag:provingCosound}
	\begin{array}{cc}
		\provingCosoundPrelimDiags
		& \provingCosoundFinalDiag
		\\ a) & b)
	\end{array}
\end{equation}
\caption{Lifting dependency \frar{d}{c}{c'} and proving that it is (weakly) Cartesian\label{figdiag:provingSoundness}}
\end{figure}
Given data are shown in diagram \cref{eqdiag:provingCosound}b) by solid and dotted arrows, whereas dashed arrow $!_y$ is to be built. Note in the feet of span $\modelsd_G$ in diagram a) projection arrows $y_E= p_E(y)$ on the left and $y_M= p_M(y)$ on the right. As \ellsig\ is a fibration, there is a unique arrow $!_{y_E}$ making the leftmost triangle in diagram a) to commute. (The rightmost triangle can be seen similarly  with $y_M$ playing the role of a bang arrow.) Now the triple of arrows in the feet, $(1_{(c', \bmap')}, !_{}y_E, y_M)$, constitutes a unique arrow $!_y$ that factorizes $y$ into composition  $1_y.\liftedTriangle$ because span $\models_G$ is monic. 

The last step is to show compositionality of so built lifting:
$ \liftedX{\triangleMorX{d^\arfun}.\triangleMorX{d'^\arfun}} =
\liftedX{\triangleMorX{d^\arfun}}.\, \liftedX{\triangleMorX{d'^\arfun}}
$. As \ellX{G} is a functor, and \ellsig\ is a fibration and hence compositional, we can rewrite the left part as 
 $$
(1_t, \triangleMorX{(d.d')'^\arfun}, \liftedX{d.d'})
= (1_t.1_t, \triangleMorX{d^\arfun}.\triangleMorX{d'^\arfun}, \bar d.\bar d' )
= (1_t, \triangleMorX{d^\arfun}, \liftedX{d}).\,
	(1_t, \triangleMorX{d'^\arfun}, \liftedX{d'})
= \liftedX{\triangleMorX{d^\arfun}}.\, 
	\liftedX{\triangleMorX{d'^\arfun}} \qqquad \qed
$$
\end{mygroup}

\subsection{
Cartesian interoperability 3: Slice categories with labelled objects
}
We need to generalize \cref{lemma:twistedSubst2-Deltas} for the case when h-arrows (diagrams) are labelled with constraint symbols from \csig. 
\begin{lemma}[Twisted substitution 3: Horizontal arrows are labelled diagrams]\label{lemma:twistedSubst3-labels}
Given a constraint signature $(\GG, \csig)$, any \mor\ \frar{f}{G}{G'} gives rise to the commutative diagram of \caties\ and functors in the left part of diagram \cref{eqdiag:cattyDiag3--withOk} (ignore for a while  the structure associated with arrows \ok\ with and without bars). Green arrow $\dom_{G'}$ (the counterpart of dashed orange $\dom_G$) and orange arrow \ellX{G} (the counterpart of dashed green \ellX{G'}) are not shown to avoid clutter. 
\newcommand{\horzX}[1]{\slicecatcsigXY{\GG}{#1}}
\newcommand{\vertX}[1]{\slicecatdeltaXY{\GG}{#1}}

\renewcommand{\LDx}[1]{\horzX{#1}}
\newcommand{\PsubobjX}[1]{\enma{\vDash_{#1}^{}}}
\newcommand{\PsubobjtwiX}[1]{\enma{\vDash_{\twixSymbol#1}}}

\renewcommand{\rightCsigU}{\enma{f^\csig_*\timm 1}}

\renewcommand{\PobjX}[1]{\PreinstcatX{#1}}
\renewcommand{\PobjtwiX}[1]{\PreinstcatX{\twixSymbol#1}}

\newcommand{\QobjDLR}[3]{\enma{  
		\begin{array}{c}
			\PobjX{#1}=
			\\
			\horzX{#2}\timm\vertX{#3}.
		\end{array}
}}

\newcommand{\cattyOkayedFull}{
\begin{tikzcd}[ampersand replacement=\&,row sep=small]
	\textcolor{rgb,255:red,214;green,92;blue,92}{\horzX{G}} \& \textcolor{rgb,255:red,214;green,92;blue,92}{\vertX{G}} \&\& \textcolor{rgb,255:red,41;green,163;blue,41}{\horzX{G'}} \& \textcolor{rgb,255:red,41;green,163;blue,41}{\vertX{G'}} \\
	\textcolor{rgb,255:red,214;green,92;blue,92}{\PobjX{G}} \&\& {\PobjX{f}} \&\& \textcolor{rgb,255:red,41;green,163;blue,41}{\PobjX{G'}} \&\& \textcolor{rgb,255:red,214;green,92;blue,92}{\PsubobjX{G}} \& {\PsubobjX{f}} \& \textcolor{rgb,255:red,41;green,163;blue,41}{\PsubobjX{G'}} \\
	\\
	\&\& {\okTargetX{\csig}} \&\&\&\&\& \OKobj \\
	\& \GG \&\& {\carrop{\csig}}
	\arrow["{f^*_\Delta}"', curve={height=18pt}, from=1-5, to=1-2]
	\arrow["{f_*^\csig}", curve={height=-18pt}, from=1-1, to=1-4]
	\arrow["h"', color={rgb,255:red,214;green,92;blue,92}, from=2-1, to=1-1]
	\arrow["{\twix{v}}"{pos=0.2}, color={rgb,255:red,41;green,163;blue,41}, dotted, from=2-3, to=1-5]
	\arrow["{v'}"', color={rgb,255:red,41;green,163;blue,41}, from=2-5, to=1-5]
	\arrow["{h'}"'{pos=0.2}, color={rgb,255:red,41;green,163;blue,41}, from=2-5, to=1-4]
	\arrow["\leftDeltaU"{description}, color={rgb,255:red,214;green,92;blue,92}, from=2-3, to=2-1]
	\arrow["\rightCsigU"{description}, color={rgb,255:red,41;green,163;blue,41}, from=2-3, to=2-5]
	\arrow["{\#_G}", color={rgb,255:red,214;green,92;blue,92}, from=2-1, to=4-3]
	\arrow["{\#_{G'}}"', color={rgb,255:red,41;green,163;blue,41}, from=2-5, to=4-3]
	\arrow["v"{pos=0.3}, color={rgb,255:red,214;green,92;blue,92}, from=2-1, to=1-2]
	\arrow["{\twix{h}}"'{pos=0.2}, color={rgb,255:red,214;green,92;blue,92}, dotted, from=2-3, to=1-1]
	\arrow["{\#_f}"', dashed, from=2-3, to=4-3]
	\arrow["\arfun"{description}, curve={height=-12pt}, from=5-4, to=5-2]
	\arrow["{\dom_G}"{description}, color={rgb,255:red,214;green,92;blue,92}, curve={height=6pt}, dashed, from=2-1, to=5-2]
	\arrow["{\ellX{G'}}"{description, pos=0.3}, color={rgb,255:red,41;green,163;blue,41}, curve={height=-6pt}, dashed, from=2-5, to=5-4]
	\arrow["{\cod^?}"', from=4-3, to=5-2]
	\arrow["{\ell^?_\csig}"{pos=0.7}, from=4-3, to=5-4]
	\arrow["\ok_\csig"', curve={height=12pt}, from=4-8, to=4-3]
	\arrow["{\liftedX{\#_G}}"'{pos=0.3}, color={rgb,255:red,214;green,92;blue,92}, from=2-7, to=4-8]
	\arrow["{q_f}"', from=2-8, to=4-8]
	\arrow["{\liftedX{\#_{G'}}}"{pos=0.3}, color={rgb,255:red,41;green,163;blue,41}, from=2-9, to=4-8]
	\arrow["{\rho_f}", color={rgb,255:red,214;green,92;blue,92}, from=2-8, to=2-7]
	\arrow["{\rho'_f}"', color={rgb,255:red,41;green,163;blue,41}, from=2-8, to=2-9]
	\arrow["{\ellX{\csig}}"', from=4-8, to=5-4]
	\arrow["{\liftedX{\okArrow}_{G'}}"'{pos=0.2}, color={rgb,255:red,41;green,163;blue,41}, curve={height=12pt}, dashed, from=2-9, to=2-5]
	\arrow["{\liftedX{\okArrow}_f}"'{pos=0.4}, curve={height=12pt}, dotted, from=2-8, to=2-3]
\end{tikzcd}
}

\begin{equation}\label{eqdiag:cattyDiag3--withOk}
	\cattyOkayedFull
\end{equation}
\end{lemma}
{\em Proof.} Reasoning is similar to that in the proof of \cref{prop:cartInst-1}, but there is an essential difference in the input data: now h-arrows (constraints/sentences) are labelled, and the relevant constructs should respect labelling. Diagram \cref{eqdiag:mainThProof-chasing} combines two versions of diagram \cref{eqdiag:cartInst-homogenDiags}: for carrier objects $G$ (the upper half, orange) and $G'$ (lower. green). (the identity arrow ID between two occurrences of \GG\ is not shown to avoid clutter).  
%
\newcommand{\horzX}[1]{\slicecatcsigXY{\GG}{#1}}
\newcommand{\vertX}[1]{\slicecatdeltaXY{\GG}{#1}}

\renewcommand{\rightCsigU}{\enma{f^\csig_*\timm 1}}

\newcommand{\PsubobjX}[1]{\enma{\vDash_{#1}^{}}}
\newcommand{\PsubobjtwiX}[1]{\enma{\vDash_{\twixSymbol#1}}}

\renewcommand{\PobjX}[1]{\PreinstcatX{#1}}
\renewcommand{\PobjtwiX}[1]{\PreinstcatX{\twixSymbol#1}}

\newcommand{\QobjDLR}[3]{\enma{  
		\begin{array}{c}
			\PobjX{#1}=
			\\
			\horzX{#2}\timm\vertX{#3}.
		\end{array}
}}

\renewcommand{\QobjDLR}[3]{\PreinstcatX{#1}}

\newcommand{\mainThProofChasingNewest}{
\begin{tikzcd}[ampersand replacement=\&,sep=small]
	\arrowcatdeltaGG \&\& \GG \&\& \textcolor{rgb,255:red,205;green,100;blue,76}{\GG/G} \& \textcolor{rgb,255:red,205;green,100;blue,76}{\_{\times}\_} \& \textcolor{rgb,255:red,205;green,100;blue,76}{\vertX{G}} \\
	\&\&\&\&\&\&\& {} \\
	\&\&\&\&\& \textcolor{rgb,255:red,205;green,100;blue,76}{\QobjDLR{G}{G}{G}} \\
	\&\&\&\& \textcolor{rgb,255:red,205;green,100;blue,76}{\LDx{G}} \\
	\OKobj \& \textcolor{rgb,255:red,33;green,87;blue,212}{\okTargetsig} \&\& {\carrop{\csig}} \&\& {\PreinstcatXupYdn{\csig}{f}} \& {\_{\times}\_} \\
	\&\&\&\& \textcolor{rgb,255:red,36;green,143;blue,36}{\LDx{G'}} \\
	\&\&\&\&\& \textcolor{rgb,255:red,36;green,143;blue,36}{\QobjDLR{G'}{G'}{G'}} \\
	\\
	\arrowcatdeltaGG \&\& \GG \&\& {\GG/G'} \& \textcolor{rgb,255:red,36;green,143;blue,36}{\_{\times}\_} \& \textcolor{rgb,255:red,36;green,143;blue,36}{\vertX{G'}}
	\arrow["\tmapfunsig"{description}, from=5-1, to=1-1]
	\arrow["\cod", from=1-1, to=1-3]
	\arrow["{\ellX{\csig}}", curve={height=30pt}, from=5-1, to=5-4]
	\arrow["{\ok_\csig}", from=5-1, to=5-2]
	\arrow[dashed, from=5-2, to=1-1]
	\arrow[dashed, from=5-2, to=5-4]
	\arrow["{\text{[pb]}_\csig}"{description, pos=0.3}, draw=none, from=5-2, to=1-3]
	\arrow["{\dom_{G}}"', color={rgb,255:red,205;green,100;blue,76}, from=1-5, to=1-3]
	\arrow["{\bmapfunX{G}}"{pos=0.7}, color={rgb,255:red,205;green,100;blue,76}, dashed, from=4-5, to=1-5]
	\arrow["{h_0}"', color={rgb,255:red,205;green,100;blue,76}, dotted, from=1-6, to=1-5]
	\arrow["v"{description, pos=0.4}, color={rgb,255:red,205;green,100;blue,76}, from=3-6, to=1-7]
	\arrow["{(h.\bmapfun^G,v)}"{pos=0.7}, color={rgb,255:red,205;green,100;blue,76}, dashed, from=3-6, to=1-6]
	\arrow["{\pbfunvdeltaX{G}}"', color={rgb,255:red,205;green,100;blue,76}, curve={height=24pt}, from=1-6, to=1-1]
	\arrow["{\#_{G}}"'{pos=0.3}, color={rgb,255:red,205;green,100;blue,76}, curve={height=12pt}, Rightarrow, from=3-6, to=5-2]
	\arrow["{\text{[pb]}_{G}}"{description, pos=0.4}, color={rgb,255:red,205;green,100;blue,76}, draw=none, from=4-5, to=1-3]
	\arrow["{v_0}", color={rgb,255:red,205;green,100;blue,76}, dotted, from=1-6, to=1-7]
	\arrow["h", color={rgb,255:red,205;green,100;blue,76}, from=3-6, to=4-5]
	\arrow["{\twixD{v}}"{description, pos=0.3}, color={rgb,255:red,36;green,143;blue,36}, from=5-6, to=9-7]
	\arrow["{f^*_\Delta}"', curve={height=30pt}, from=9-7, to=1-7]
	\arrow["\leftDeltaU"', Rightarrow, from=5-6, to=3-6]
	\arrow["{h'}"', color={rgb,255:red,36;green,143;blue,36}, from=7-6, to=6-5]
	\arrow["{v'}"{description, pos=0.7}, color={rgb,255:red,36;green,143;blue,36}, from=7-6, to=9-7]
	\arrow["\rightCsigU"', color={rgb,255:red,36;green,143;blue,36}, Rightarrow, from=5-6, to=7-6]
	\arrow["{f_*^\csig}"', from=4-5, to=6-5]
	\arrow[dotted, from=9-6, to=9-7]
	\arrow["{(h'.\bmapfun^{G'},v')}"', color={rgb,255:red,36;green,143;blue,36}, from=7-6, to=9-6]
	\arrow[draw={rgb,255:red,36;green,143;blue,36}, dotted, from=9-6, to=9-5]
	\arrow["{\bmapfun^{G'}}"'{pos=0.7}, dashed, from=6-5, to=9-5]
	\arrow["{\twixD{h}}"'{pos=0.3}, from=5-6, to=4-5]
	\arrow["{v_{0f}}", from=5-7, to=9-7]
	\arrow["{!_{f/\csig}}", from=5-6, to=5-7]
	\arrow["{h_{0f}}"{description, pos=0.2}, curve={height=30pt}, dotted, from=5-7, to=1-5]
	\arrow["\leftDeltaU"{description}, color={rgb,255:red,205;green,100;blue,76}, curve={height=30pt}, dashed, from=5-7, to=1-6]
	\arrow["\tmapfunsig"{description}, from=5-1, to=9-1]
	\arrow["{\#_{G'}}"{pos=0.3}, color={rgb,255:red,36;green,143;blue,36}, curve={height=-12pt}, Rightarrow, from=7-6, to=5-2]
	\arrow["{\ellX{G}}"', color={rgb,255:red,205;green,100;blue,76}, dashed, from=4-5, to=5-4]
	\arrow["{\ellX{G'}}", color={rgb,255:red,36;green,143;blue,36}, dashed, from=6-5, to=5-4]
	\arrow["\arfun"{description, pos=0.6}, from=5-4, to=1-3]
	\arrow["{\pbfunvX{G'}}", color={rgb,255:red,36;green,143;blue,36}, curve={height=-18pt}, from=9-6, to=9-1]
	\arrow["{\mathsf{ID}}"{description}, curve={height=30pt}, Rightarrow, dotted, no head, from=1-1, to=9-1]
	\arrow["\cod"{description}, from=9-1, to=9-3]
	\arrow["\arfun"{description}, from=5-4, to=9-3]
	\arrow["{\text{[pb]}_\csig}"{description, pos=0.4}, draw=none, from=5-2, to=9-3]
	\arrow[dashed, from=5-2, to=9-1]
	\arrow["{f_*}"{description}, curve={height=-30pt}, dotted, from=1-5, to=9-5]
	\arrow[draw={rgb,255:red,36;green,143;blue,36}, curve={height=-30pt}, dotted, from=5-7, to=9-5]
	\arrow["{\dom_{G'}}"', color={rgb,255:red,36;green,143;blue,36}, from=9-5, to=9-3]
	\arrow["\rightU"{description}, color={rgb,255:red,36;green,143;blue,36}, curve={height=-18pt}, dashed, from=5-7, to=9-6]
	\arrow["{\text{[pb]}_{G'}}"{description, pos=0.4}, color={rgb,255:red,36;green,143;blue,36}, draw=none, from=6-5, to=9-3]
\end{tikzcd}
}

\begin{figure}[h]  
\begin{equation}\label{eqdiag:mainThProof-chasing}
		\mainThProofChasingNewest 
\end{equation}
\end{figure}


%
%
The two halves are connected by arrows $f^*_\Delta$ (see \cref{lemma:twistedSubst2-Deltas}) and its constraint/sentence counterpart $f_*^\csig$ provided by the universality of [pb]$_{G'}$. 

Now we take two twisted products: $\PreinstcatXupYdn{\csig}{f}= \slicecatcsigGGG\timm\slicecatdeltaGGG$ with projections $(\twixD{h}, \twixD{v})$, and $\GG/G\timm\slicecatdeltaGGGprim$ with projections $(h_{0f}, v_{0f})$ as specified in the diagram (we need the subscript $/\csig$ near the former product only to distinguish it from the latter; after the lemma is proved, we will omit it).  
Universality of products provides arrow \bangXdn{f/\csig}, and thus there are two spans with apex  \PreinstcatXupYdn{\csig}{f} and feet \arrowcatdeltaGG\ and \carrop{\csig}: one is via the upper (orange) structure and the other is via the lower (green) structure. Both these spans make a commutative square with cospan $(\cod,\arfun)$ and thus universality of [pb]$_\csig$ provides a unique ``orange'' and a unique ``green'' arrows from \PreinstcatXupYdn{\csig}{f} to \Preinstcatsig. Hence, the  uniqueness implies that these `orange unique' and 'green unique' arrows coincide and make the diagonal of the commuting square formed by double-body arrows in the diagram. 
In a compact way, the construction is specified by the commutative ``catty'' diagram \cref{eqdiag:cattyDiag3--withOk} (the left half without the OK-structure), in which arrow $\#_f$ is the diagonal of the double-body arrow square just specified  (and the additional subscript $/\csig$ near \PreinstcatX{f} is omitted).     
\qed  
\begin{propo}[Cartesian institutions 2]\label{prop:cartInst-hetero-eSat}
	Let \GG\ be a PBC and \csig\ a constraint signature over \GG. These data produce a functor 
	\frar{\modelsd_\csig}{\GG}{\twistedSatcat(\catcat, \Instcatsig)}, \ie, a categorical e-institution, whose components are described below. 
\end{propo}
{\em Proof.}  The institution's ingredients are as follows.  The \caty\ of signatures is \GG. 
Functors \frar{\modfun}{\GG}{\opX{\catcat}} and \frar{\senfun}{\GG}{\catcat} are given by  $\modfun(G)=\deltaslicecatGGG$ with $\modfun(f)=\fstardelta$ and $\senfun(G)=\slicecatcsigGGG$ with $\senfun(f)=\fstarsig$. 
Given $G$, the eSat-span $\modelsd_G$ is given by (see diagram \cref{eqdiag:cattyDiag3--withOk}) pulling back arrow $\#_G$ along \oksig. This gives us a family of eSat-spans $\modelsd_G$, $G\in\Ob\GG$. 

Now let \frar{f}{\Sigmaa}{\Sigmaa'} be a signature \mor. It produces a commutative square with diagonal $\#_f$ (in diagram \cref{eqdiag:cattyDiag3--withOk},  note a wide triangle in the middle of the diagram)  and the \pblemma\ ensures the corresponding commutative square with diagonal $q_f$ on the right, which is obtained by pulling back along \oksig\  (the lifted arrow  \liftedX{\ok_G} is not shown to avoid clutter). Comparison of this diagram with diagram \cref{eqdiag:twistedMorphDef--MStoE}a) on p.~\pageref{eqdiag:twistedMorphDef--MStoE} shows that we mapped arrow \frar{f}{G}{G'} to  a twisted \mor\ from eSat-span $(\liftedX{\ok}_G, \#_G)$ to eSat-span $(\liftedX{\ok}_{G'}, \#_{G'})$: the eSat-axiom, \ie, conditions i) and ii) of \cref{def:twistedMorph}, hold due to the very construction of our eSat-spans.
%

Compositionality of so built mapping
$f\mapsto (\fstarsig, \fstardelta)\in \twistedSatcat(\modelsd_G, \modelsd_{G'})$ 
is ensured by \cref{prop:twistedMorph--Compose}: composition of twisted pairs that happen to be twisted \mor s of eSat-spans, is a twisted \mor.  \qed

Now Propositions \ref{prop:cartInst-homo-Soundness} and \ref{prop:cartInst-hetero-eSat} imply the following major result. 
\begin{theorem}[Cartesian DCL-logic is e-institutional and sound]\label{thm:mainTh}
		Let \GG\ be a \caty\ with pullbacks and \csig\ a constraint signature over \GG. These data produce an institution with evidence \frar{\modelsd_\csig}{\GG}{\twistedSatcat(\catcat, \Instcatsig)} with the model functor  \frar{\modfun= \slicecatdeltaGGdot}{\GG}{\opX{\catcat}} and the sentence functor  \frar{\senfun = \slicecatsigGGdot }{\GG}{{\catcat}}.    Moreover, this institution is cosound. 
\end{theorem}




\renewcommand{\CC}{\enma{\mathbf{C}}}
\newcommand{\coprodNode}{\enma{\modelsdCS\,{:}=\coprod_{c\in C_S} \modelsd_c}}
\newcommand{\coprodToInstsig}{\enma{\e^S}}
\newcommand{\coprodToSig}{\enma{\ell^S}}
\newcommand{\coprodToLD}{\enma{\bmapfun^S}}

\newcommand{\ind}{\enma{\mathrm{ind}}}

\newcommand{\skeInstcatDef}{
\begin{tikzcd}[ampersand replacement=\&,column sep=scriptsize]
	\slicecatdeltaGGGS \&\&\& {\Instcat_S\,{:}= \lim_{c\in C_S}\tmapfun^c} \\
	\Instcatsig \& {\modelsd_{G_S}} \& \circ \& {\modelsd_c} \\
	\carrsigop \& \slicecatsigGGGS \& {C_S} \& \tcat
	\arrow["p", from=2-2, to=3-2]
	\arrow["q"{description}, from=2-2, to=1-1]
	\arrow["\e"{description}, from=2-2, to=2-1]
	\arrow["{\#_S}", from=3-3, to=3-2]
	\arrow[dashed, from=2-3, to=3-3]
	\arrow[dashed, from=2-3, to=2-2]
	\arrow[""{name=0, anchor=center, inner sep=0}, "c", from=3-4, to=3-3]
	\arrow[dashed, from=2-4, to=3-4]
	\arrow[dashed, from=2-4, to=2-3]
	\arrow["{\tmapfun^c}"{description, pos=0.2}, dashed, from=2-4, to=1-1]
	\arrow["\lrcorner"{anchor=center, pos=0.125, rotate=-90}, draw=none, from=2-3, to=3-2]
	\arrow["{\tmapfun^S}"', dashed, from=1-4, to=1-1]
	\arrow["{p_c}", dashed, from=1-4, to=2-4]
	\arrow["\ellGS", from=3-2, to=3-1]
	\arrow["\ellsig", from=2-1, to=3-1]
	\arrow["{\e_c}"{description, pos=0.8}, dotted, tail reversed, no head, from=2-1, to=1-4]
	\arrow["\lrcorner"{anchor=center, pos=0.125, rotate=-90}, draw=none, from=2-4, to=0]
\end{tikzcd}
}
%
\newcommand{\skeInstcatInter}{
	\begin{tikzcd}[ampersand replacement=\&,sep=small]
		\& \slicecatdeltaGGGS \&\&\&\& {\Instcat_S} \\
		\&\& \modelsdGS \& \circ \& {\Instcat_c} \\
		\& \textcolor{rgb,255:red,214;green,92;blue,92}{\modelsd_f} \& \slicecatsigGGGS \& {C_S} \& \tcat \\
		\slicecatdeltaGGGSprim \&\&\&\&\& {\Instcat_{S'}} \\
		\modelsdGSprim \&\& \circ \& {\Instcat'_{c.f_C}} \\
		\& \slicecatsigGGGSprim \& {C_{S'}} \& \tcat
		\arrow["p"', from=2-3, to=3-3]
		\arrow["{\#_S}"', from=3-4, to=3-3]
		\arrow["c"', from=3-5, to=3-4]
		\arrow["{f_C}"{description}, color={rgb,255:red,214;green,92;blue,92}, from=3-4, to=6-3]
		\arrow["{f^\csig_*}"{description}, color={rgb,255:red,214;green,92;blue,92}, from=3-3, to=6-2]
		\arrow[dashed, from=2-5, to=2-4]
		\arrow[dashed, from=2-4, to=2-3]
		\arrow[dashed, from=2-4, to=3-4]
		\arrow[dashed, from=2-5, to=3-5]
		\arrow["{\tmapfun^S}"', dashed, from=1-6, to=1-2]
		\arrow["{q_c}"{description}, dashed, from=2-5, to=1-2]
		\arrow["q", from=2-3, to=1-2]
		\arrow["{p_c}", dashed, from=1-6, to=2-5]
		\arrow["{p'}"', from=5-1, to=6-2]
		\arrow[""{name=0, anchor=center, inner sep=0}, "{\#_{S'}}"', from=6-3, to=6-2]
		\arrow[dashed, from=5-3, to=5-1]
		\arrow[dashed, from=5-3, to=6-3]
		\arrow["{c.f_C}"', color={rgb,255:red,8;green,2;blue,2}, from=6-4, to=6-3]
		\arrow["ID"{description, pos=0.2}, Rightarrow, dotted, no head, from=3-5, to=6-4]
		\arrow["{f^*_\Delta}", color={rgb,255:red,214;green,92;blue,92}, from=4-1, to=1-2]
		\arrow["{q'}", from=5-1, to=4-1]
		\arrow["{q'_f}"', color={rgb,255:red,214;green,92;blue,92}, from=3-2, to=4-1]
		\arrow["{p_f}", color={rgb,255:red,214;green,92;blue,92}, from=3-2, to=3-3]
		\arrow["\lrcorner"{anchor=center, pos=0.125, rotate=-90}, draw=none, from=2-4, to=3-3]
		\arrow["\lrcorner"{anchor=center, pos=0.125, rotate=-90}, draw=none, from=2-5, to=3-4]
		\arrow[dashed, from=5-4, to=5-3]
		\arrow[dashed, from=5-4, to=6-4]
		\arrow["\lrcorner"{anchor=center, pos=0.125, rotate=-90}, draw=none, from=5-4, to=6-3]
		\arrow["{p_{c.f_C}}", dashed, from=4-6, to=5-4]
		\arrow["{\rho_f}", color={rgb,255:red,214;green,92;blue,92}, dashed, from=3-2, to=2-3]
		\arrow["{\text{[pb]}}"{description}, color={rgb,255:red,214;green,92;blue,92}, draw=none, from=3-2, to=1-2]
		\arrow["{\rho'_f}"{pos=0.7}, color={rgb,255:red,214;green,92;blue,92}, dashed, from=3-2, to=5-1]
		\arrow["{\text{[pb]}}"{description}, color={rgb,255:red,214;green,92;blue,92}, draw=none, from=3-2, to=6-2]
		\arrow["{!!^c_f}"{description}, color={rgb,255:red,214;green,92;blue,92}, dashed, from=5-4, to=2-5]
		\arrow["{f^*}"', color={rgb,255:red,214;green,92;blue,92}, dashed, from=4-6, to=1-6]
		\arrow["{!^c_f}"{description}, color={rgb,255:red,214;green,92;blue,92}, curve={height=18pt}, dashed, from=5-4, to=3-2]
		\arrow["{\tmapfun^{S'}}"{pos=0.6}, dashed, from=4-6, to=4-1]
		\arrow["{p'^c_f}"', color={rgb,255:red,214;green,92;blue,92}, dashed, from=4-6, to=2-5]
		\arrow["\lrcorner"{anchor=center, pos=0.125, rotate=-90}, draw=none, from=5-3, to=0]
	\end{tikzcd}
}

\newcommand{\skeSettingForDatamod}{
\begin{tikzcd}[ampersand replacement=\&,column sep=small,row sep=scriptsize]
	\arrowcatdeltaGG \&\& \Instcat \&\& {\catcat/\Instcatsig} \\
	\&\&\& \catcat \& {\catcat/\carrsigop} \\
	\GG \&\& \skecatsigGG \&\& {\catcat/\arrowcatsigGG}
	\arrow["\globalGS", from=3-3, to=3-1]
	\arrow["\coddelta"', from=1-1, to=3-1]
	\arrow["\globaltmap"', from=1-3, to=1-1]
	\arrow["\schemaind"', from=1-3, to=3-3]
	\arrow["{\text{[fib]}}"'{pos=0.3}, shift right=1, color={rgb,255:red,246;green,60;blue,122}, draw=none, from=1-3, to=3-3]
	\arrow["{\text{[fib]}}"'{pos=0.3}, shift right=1, color={rgb,255:red,92;green,92;blue,214}, draw=none, from=1-1, to=3-1]
	\arrow["\globalecurry", from=1-3, to=1-5]
	\arrow["\globalCS", dashed, from=3-3, to=2-4]
	\arrow["\dom"', color={rgb,255:red,92;green,92;blue,214}, dotted, from=1-5, to=2-4]
	\arrow["{\_.\ellsig}", from=1-5, to=2-5]
	\arrow["{\_.\globalellGS}"', from=3-5, to=2-5]
	\arrow["{\#_{\_}}", from=3-3, to=3-5]
	\arrow[color={rgb,255:red,92;green,92;blue,214}, dotted, from=2-5, to=2-4]
	\arrow[color={rgb,255:red,92;green,92;blue,214}, dotted, from=3-5, to=2-4]
\end{tikzcd}
}

\section{
	Generalized sketches} \label{sec:skeFormal-sketches}
Simplest logical theories are built from conjunctions of atomic formulas, \eg, $P(x,y) \wedge P(y,z)$ or $\bigwedge\{P(x,y), P(y,z)\}$. A diagrammatic counterpart of conjunctive theories is the notion of a {\em sketch}: a set of atomic formulas over a ``graph of variables'', which is closed w.r.t. all dependency arrows in the signature.

\subsection{Basic notions and constructs}

\begin{defN}[Sketches] 
\label{def:skedef}
Let $\GG, \predsig$ 
be a constraint signature (Def. \ref{def:predsig}).  

a) A {\em (generalized) sketch} over \predsig\ 
is a pair $S=(G_S, C_S)$ of a graph $G_S\in \GG$ called the {\em carrier}, and $C_S=(\apexX{C_S}, \labfun_S, \diagrfunSke)$  a span in \catcat,
which makes a commutative square in diagram \cref{eqdiag:skedef}a) below  
We will omit sub/subscripts $S$ near functors in the diagram if it is clear.  
%
\begin{figure}[ht]
\begin{equation}\label{eqdiag:skedef}
\begin{array}{c@{\qquad}c}
\begin{tikzcd}[row sep=2ex, column sep = 3ex]
\ConstrX{S} 
		\ar[rr, "\gapnamegap{0}{\bmapfunX{S}}{0}"] 
		\ar[dd, "\labfun_S" ',  thick ] 
		\ar[dd, phantom, "\gapnamegap{2.5}{\dfiblabelRed}{-2.} ",  thick ]
		\ar[rd, derived, "\#_S"]
&& \GG/G_S 
			\ar[dd, "\gapnamegap{0.5}{\dom_{G_S}}{-0.5}" ', thick ] 
			\ar[dd, phantom, "\gapnamegap{2.5}{\dfiblabelBlack}{-2.} ",  thick ]
\\ 
	& \dernode{\LDx{G_S}}
			\ar[ru, derived]  
			\ar[ld, derived]
			\ar[rd, phantom, gray,  "\text{[pb]}", pos=0.25]
	&
\\   
	\carrop{\predsig} 
			\ar[rr, "\arfun"]  && \GG
\end{tikzcd}
&  
\begin{tikzcd}[row sep=2.75ex, column sep = 5ex]
	&\ConstrX{S'}
					\ar[r, "\#_{S'}"]
					\ar[ld, <-, "f_C" ']
	& \slicecatsigGGGSprim
				\ar[ld, <-,  "\gapnamegap{1}{f_*^\csig}{-0.5}" ', pos=0.75, xshift=0.7ex]
\\  
	\ConstrX{S} 
			\ar[r, "\#_S", pos=0.75] 
		& \slicecatsigGGGS
\end{tikzcd}
\\
\text{a) Sketch def. } & \text{b) Sketch \mor}
\end{array}
\end{equation}
\vspace{-5mm}
\end{figure}

If $c$ is an object of the apex \caty\ \ConstrX{S} with $\labfun(c)=\clabelX{c}$ and \frar{\bmapfun(c)=\bmap_c}{G_{\clabelX{c}}}{G_S}, we say that $c$ is a {\em constraint declared in} $S$, constraint symbol \clabelX{c} is its {\em label}, and  $\bmap_c$ is its {\em binding} 
mapping.  We then write $(c, \clabelX{c},\bmap_c)\in C_S$ or sometimes $c\in C_S$ and use $C_S$ to denote both the span and its apex \caty.   
A sketch is called {\em mono-sketch} if span $(\labfunSke,\bmapfun)$ is jm and hence functor $\#$ is an embedding but in general, we allow for  $c'\noteq c$ with $\clabelX{c}=\clabelX{c'}$ and $\bmap_c=\bmap_{c'}$ (and may refer to general sketches as to {\em multi-sketches}).

We also require: a1)  functor $\labfun_S$ to be a discrete fibration so that all arrows in \apexX{C_S} are lifts of \carr{\predsig}-arrows: this is the closure condition mentioned above and discussed in detail in \cite{DiskinWolter-entcs08}; %
and a2) pair $(\bmapfunX{S}, \arfun)$ to be a fibration \mor\ (indeed, $\dom_{G_S}$ is always a discrete fibration by  \cref{lemma:arrcatfibr}(ii) on p.\pageref{lemma:arrcatfibr}).%
\footnote{We call this definition {\em fibrational} to distinguish it from a typical {\em indexed} sketch definition, in which discrete fibration \plabel\ is replaced by copresheaf \frar{\semm{.}}{\carr{\PP}}{\setcat}---
this is how sketches were defined in \cite{DiskinWolter-entcs08} \shortlong{}{assuming also that \diagr\ is injective}.
\zd{it'd go to the remark multi-v vs. mono-sketches  below}
}
%

b) 	A sketch {\em \mor}  \frar{f}{S}{S'} is a pair $(f_\GG, f_C)$ of a \GG-\mor\ \frar{f_\GG}{G_{S}}{G_{S'}} and  functor \frar{f_C}{C_{S}}{C_{S'}} 
such that $\#_S.f^\csig_* = f_C.\#_{S'}$ (see diagram \cref{eqdiag:skedef}b), where $f^\csig_*$ denotes  \frar{(f_\GG)^\csig_*}{\slicecatsigGGGS}{\slicecatsigGGGSprim}
%
Obviously, sketch \mor s are associatively composable, and $(\id_{G_S},\id_{C_S})$ is the unit, which gives us the category of \csig-sketches \skecatsigGG\ endowed with functor \frar{\carrcdot}{\skecatsigGG}{\GG} giving the {\em carrier (object)}.
\end{defN}
%
\shortlong{}{
\textcolor{blue}{
	\begin{defN}[Finitarity]\zd{make it a remark}
		A sketch is {\em finitary} if the signature is finitary. A sketch is {\em finite} if its carrier is a finite graph and category \ConstrX{S} is finite. A sketch $S=(\carrr{S}, C_S)$ is {\em carrier-wise finite} if the graph \carrr{S} is finite, and {\em constraint-wise finite} if category \ConstrX{S} 
		is finite. A sketch is {\em finite} if both finiteness conditions hold. 
	\end{defN}
		\begin{corol}[Sketches are fibration morphisms]
			\label{cor:skearefibr}
			As $\frar{\dom_{\carrr{S}}}{\GG/\carrr{S}}{\GG}$ is a discrete fibration (\cref{fact:arrcatfibr} in \cref{sec:arrowCats}), then \frar{\domop_{\carrr{S}}}{(\GG/\carrr{S})^\opp}{\GGop} is a discrete opfibration, but  $(\GG/\carrr{S})^\opp = \GGop/\carrr{S}$  and then commutativity of the outer square in diagram \cref{eq:skedef} makes the pair $(\arfun, \bind)$ an opfibration morphism. This is an exact formulation of the major requirements for sketches with dependencies: the set of diagrams is to be closed under ``inference rules'' arising from dependencies. 
		\end{corol}
\begin{remark}[(Multi)sketches vs. mono-sketches]
			\label{def:multiVsMonoSketches}\zd{\Large to be shortened}
			Executing pullback in \Catcat\ gives us the lower triangle in the diagram, whose apex   \LDiagrcat{} 
			consists of all possible labelled diagrams into $G_S$, \ie, all pairs $(c,\bmap)$ with \frar{\bmap}{G_c}{G_S}
\\			
			\medskip
			Let the internal square 
			in diagram \cref{eqdiag:skedef} be a pullback in \catcat, then its apex \LDiagrcat{S} is the category whose objects are all \predsig-labelled diagrams over graph \carrr{S} (\ie, pairs $(P,d)$ with $P$ a predicate symbol and \frar{d}{P^\arfun}{\carr{S}} a diagram),  and arrows are lifts of dependency arrows in \PP\ (recall that pulling back a split fibration results in a split fibration).  	The diagonal mapping $\#$ is provided by the universality of pullbacks. 
			By an abuse of notation and terminology, we will often identify a constraint $c\in\iConstr{S}$ with the corresponding labelled diagram $c^\#=(\plabel(c),\bind(c))$ also called a constraint.
\\			
			A sketch is called {\em mono-sketch} if the pair of functors $(\labell,\bind)$ is jointly monic and hence mapping $\#$ is monic.  Note that if the target sketch is a mono sketch, then the second component of a sketch morphism becomes redundant as it can be derived from \frar{\carr{f}}{\carr{S}}{\carr{S'}}: given $c\in\iConstr{S}$ with $P=\plabel(c)$ and \frar{d=\bind(c)}{P^\arfun}{\carrr{S}}, we define $\constrfun{f}(c)$ to be the constraint over \carrr{S'} determined by the pair $(P, d\comp\carr{f})$.
		\end{remark}
}}
%
\begin{constrN}[Category of instances of a sketch] Intuitively, 
	an {\em instance} of sketch $S=(G_S, C_S)$ is a preinstance \frar{t}{X}{G_S} satisfying all constraints in $S$, which gives rise to a \caty\ \InstcatS\ of sketch instances and their deltas. 
More formally, the definition is given by diagram \cref{eqdiag:skeInstcatDef}:
\begin{equation}\label{eqdiag:skeInstcatDef}
	\skeInstcatDef
\end{equation}
In the bottom lane of this diagram, the middle and the rightmost squares are PBs, and the rightmost square is produced for each index $c\in \Ob C_S$. The wide span $(\InstcatS, p_c)_{c\in \Ob C_S}$ is produced by the multiPB of the wide cospan 
$(\tmapfun^c)_{c\in \Ob C_S}$. Functor $\tmapfun^S$ is the diagonal of the multiPB, which maps an index  in \InstcatS\ to a real instance in \slicecatdeltaGGGS. 

Dotted arrow $\e_c$ is produced by composition started with projection $p_c$ for each $c$, and we thus have functor \frar{\e_{\_}}{C_S}{\catcat(\Instcat_S,\,\Instcatsig)}. As \catcat\ is Cartesian closed, we can apply uncurrying followed by curring on the other arguemnt  to obtain functor \frar{\ecurry_S}{\Instcat_S}{\catcat(C_S,\,\Instcatsig)}. It maps a(n index of a) sketch instance $I$ to a ``tuple'' of pieces of evidence (one piece per a constraint index in $C_S$) that supports $I$'s conformance to all constraints declared in $S$.  
%
\end{constrN}
\begin{constrN}\label{constr:arrowCatSig} 
	Category \arrowcatsigGG\ has labelled diagrams $G_c \xrightarrow{\bmap_c}G$  as objects and pairs  
	$(
	G_c\xrightarrow{d^\arfun} G_{c'}, \;
	G \xrightarrow{f} G', 
	)$
	 (where \flar{d}{c}{c'} an arrow in \csig) that form a commutative square with $\bmap_c$ and $\bmap_{c'}$ as \mor s \frarxy{\bmap_c}{\bmap_{c'}}. (Note that we can consider this as a \mor\ \frar{d^\arfun}{f_*(\bmap_c)}{\bmap_{c'}} in $\GG/G'$. )
\end{constrN}
\begin{theorem}[Sketch setting for data and system modelling] \label{thm:mainTh2}
	The sketch construct gives rise to a commutative diagram \cref{eqdiag:skeSetting4datamod} of \caties\ and functors, in which the middle vertical arrow is a fibration, the pair $(\tmapfun, \carr{\cdot})$ is a fibration \mor, 
	arrows $\_.\ellsig$ and $\_.\globalellGS$ denote functors produced by postcomposition, and three dotted arrows denote \dom-functors \frar{\dom_X}{\CC/X}{\CC} defined for any \caty\ \CC.   
\begin{equation}\label{eqdiag:skeSetting4datamod}
	\skeSettingForDatamod
\end{equation}
The name of the theorem will be explained in \cref{sec:ske-at-work}. 
\end{theorem}
{\em Proof.} To build fibration \schemaind, we show that the \InstcatS-construct defined above  gives rise to a pseudo-functor \frar{\InstcatX{\_}}{\GG}{\opX{\catcat}}. The following elementwise reasoning provides the idea. Let \frar{f}{S}{S'} be a sketch \mor, $t\in\slicecatdeltaGGGSprim$ an instance of $S'$, and $c.\#_S$ a constraint in $S$ indexed by $c\in\Ob C_S$.  Then $f^*_\Delta(t)\modelsd_{G_S}^i c.\#_S$ iff $t\modesld_{G_{S'}}^i f_*^\csig(c.\#_S)$ by \cref{thm:mainTh}, while the right-hand side of the `iff' above holds as i) $f$ is a sketch \mor\ (and hence $f_*^\csig(c.\#_S)$ is declared in $S'$) and ii) $t$ is an $S'$-instance.  A formal proof based on this idea is specified by diagram \cref{eqdiag:skeInstcatInter}, in which arrows derived from having a sketch \mor\ $f$ are shown orange. 
\begin{equation}\label{eqdiag:skeInstcatInter}
 \skeInstcatInter
\end{equation}
As $(f^\csig_*, f^*_\Delta)$ is a twisted \mor\ by \cref{thm:mainTh}, we add to the diagram the twisted Sat category \modelsdF\ with its projection functors forming two pullback squares. As $f$ is a sketch \mor, the span with apex $\Instcat'_{c.f_C}$ and feet \modelsdGSprim\ and \slicecatdeltaGGGS\ commutes with cospan $(p',f^\csig_*)$, and hence universality of the lower pullback ensures a unique arrow $!^c_f$ as shown. Now universality of $\Instcat_c$ ensures a unique arrow $!!^c_f$, whose composition with $p_{c.f_C}$ gives us an arrow \frar{p'^c_f}{\Instcat_{S'}}{\Instcat_c} for each $c$ in $\Ob C_S$. Moreover, wide span $(p'^c_f)_{c\in\Ob C_S}$ commutes with wide cospan $(q_c)_{c\in\Ob C_S}$, and hence universality of \InstcatS\ ensures a unique functor  $f^*$ as shown that keeps the entire diagram commutative. 
Compositionality (up to coherent iso\mor s) of the construct, \ie, $(f.f')^* \cong f'^*.f^*$, is straightforward.  Now the standard argument of the Grothendieck construction provides us with a fibration, whose projection functor  is denoted \schemaind. 

Functor \globalGS\ is obvious. 
Mapping \globaltmap\ is defined on objects  by setting $\globaltmap(S,I) :=\tmap_I^S$ and on arrows by \frar{\globaltmap(f,u):= \tmap_u^S.\liftedX{f}}{I}{I'}, where \frar{f}{S}{S'}, \frar{u}{I}{f^*(I')} and \liftedX{f} is the lift of $f$ (we silently use a standard notation for the Grothendieck construction).  Checking functoriality of \globaltmap\ is straightforward. 
 It is also seen from the Grothendieck construction for building \schemaind\ that pair $(\globaltmap, \carrcdot)$ is a fibration \mor. 

Now consider the right half of diagram \cref{eqdiag:skeSetting4datamod}. 
Functor \globalecurry\ is defined based on functor $\ecurry_S$ similarly to how functor \globaltmap\ was defined based on functor  $\tmapfun^S$. 
Arrows \globalbmap\ and \globalellGS\ are derived from diagram \cref{eqdiag:skeInstcatDef}: for sketch $S$, define $\globalbmap(S):=\#_S$, and for arrow $\bmap_c\in\Ob\arrowcatsigGG$, define $\globalellGS(\bmap_c):=\ellX{\cod(\bmap_c)}(\bmap_c) = c$. Commutativity is obvious from diagram \cref{eqdiag:skeInstcatDef} and construction of \globalecurry. 
\qed
\begin{remark}
	Diagram \cref{eqdiag:skeSetting4datamod} does not show all important relationship between the \caties\ involved, \eg, the codomains of diagram embedded into \caties\ in the right half are exactly the carrier objects from the left half. The diagram presents a special view of the entire structure, and this view is conceptually and practically important as discussed in the next section. 
\end{remark}

\subsection{Sketches at work: Discussion}\label{sec:ske-at-work}

\subsubsection{A general stage}
\begin{figure}[h]
\centering
\begin{tabular}{cc}
	\includegraphics[width=0.475\linewidth]{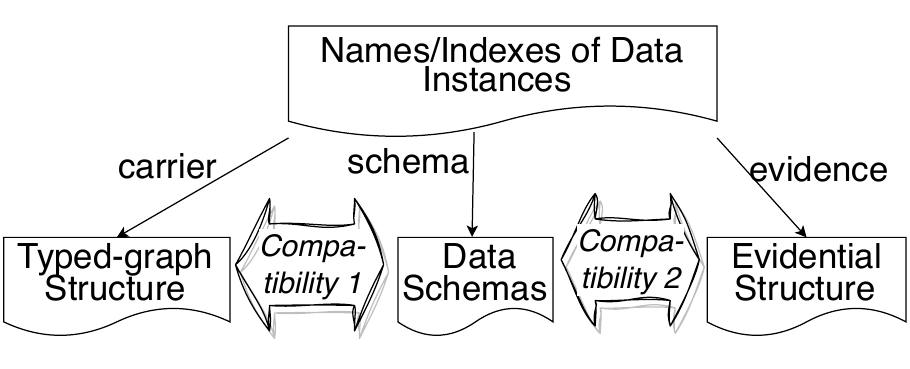}  
& 
	\includegraphics[width=0.475\linewidth]{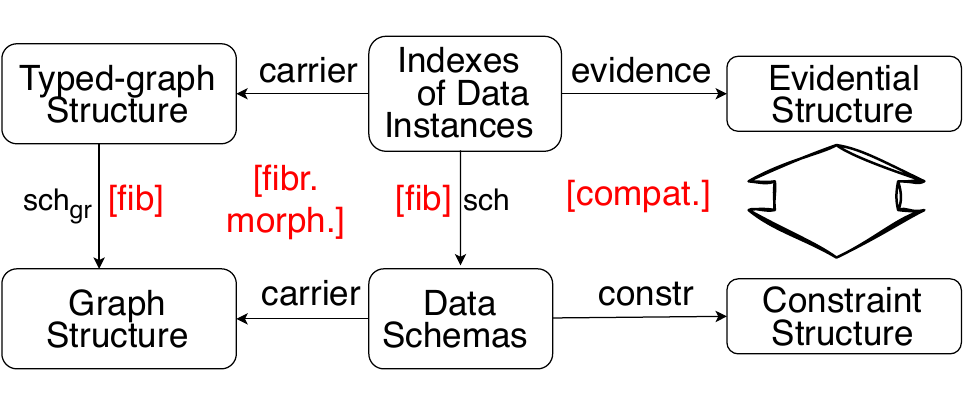}
\\ 
	a) & b) 
\end{tabular}
\caption{Conceptual schemas for data \& system modelling
}
\label{fig:dataMod-abstract}
\vspace{-2ex}
\end{figure}

Diagram a) in \cref{fig:dataMod-abstract} shows a very general arrangement of the data modelling scene. 
Its nodes denote collections of objects named in the node
, and arrows are mappings. Objects of the top node are not data instances themselves: they are their names or indexes, and three mappings provides an instance index with its three basic ingredients: its data schema providing  types for data elements, its carrier structure (that can typically be modelled by a typed graph structure), and (crucially in the context of SE and assurance) a structure of pieces of evidence supporting the claim that the instance does conform to the schema. Sketchy block arrows Compatibility 1,2 refer to unspecified interconnections between the three components. 

Diagram b) is a refinement of a) suggested by diagram \cref{eqdiag:skeSetting4datamod}. 
%
Indeed, the \req\ to mapping \schemaind\ to be a fibration ensures there is a model reduct functor of the institution theory for instances, and hence is a must \req\ as soon as we want patterns of the abstract model theory to be applicable. 
A typical data schema consists of two parts. The left one is the carrier graphical structure of {\em types} so that any instance is a typed graph, which gives us a projection functor \schemaindgra. View computation only works with the carrier structure, which corresponds to the \req\ to the left square to be a fibration \mor\ (note the \corring\ red labels). The second component of a data schema is a collection of constraints it declares. Then each valid instance of a schema must demonstrate sufficient (or at least some) evidence that all constraints in the schema are satisfied. This amounts to having some correspondence between the two structures, say, a span or a cospan, and the \req\ to the right square to be commutative. 

Overall, schema b)  can be seen as a version of the notion of an institution with evidence, in which schemas/signatures are just indexes while the logical ``mechanics'' is given by the carrier structure (on the left) and the constraint structure (on the right). Then we can interpret \cref{thm:mainTh2} as a statement showing that sketches correctly implement the general schema. 

The idea of data fibred over schemas is fundamental for many data management applications in software engineering.  In such applications, schema \mor s can be seen as {\em view definitions} and then their lifting (ensuring the reduct functor) corresponds to executing the view definition (see, \eg, \cite{me-sle13} or \cite{me-fase18} for details). 
Thus, requiring \schemaind\ to be a fibration amounts to having a view execution mechanism. 

\newcommand{\SkecatGG}{\enma{\Skecat(\GG)}}
\newcommand{\schfun}{\enma{\mathsf{sch}}}
\newcommand{\InstcatSprim}{\enma{\Instcat_{S'}}}

\subsubsection{Delta lenses and sketches}

We will briefly sketch several ideas, whose accurate formalization is still to be done.  A central issue for views is the view update problem addressed in the delta lens \fwk. A delta lens from \caty\ \spA\ to \caty\ \spB\ is a pair $L=(\get, \putl)$ of a functor \frar{\get}{\spA}{\spB} and cofunctor \flar{\putl}{\spA}{\spB}, which share the same object mapping 
\frar{\get_0}{\Ob\spA}{\Ob\spB}. In the sketch \fwk, the source and target \caties\ are categories of sketch instances (instance indexes to be precise), $\Instcat_S$ and $\Instcat_{S'}$, whose carriers are slice \caties\ \slicecatdeltaGGGS\ and \slicecatdeltaGGGSprim\ (recall that subindex $\Delta$ refers to \mor s being spans, which model instance updates with deletions and insertions). A typical functor \get\ is our instance reduct functor \frar{f_\Delta^*}{\slicecatdeltaGGGS}{\slicecatdeltaGGGSprim} generated by a view definition (sketch \mor) \flar{f}{S}{S'}. A typical update policy is also defined over a view definition so that it is reasonable to write  a lens as a pair $L_f=(\get_f, \putl_f)$ indexed by a sketch \mor\ $f$. 

Lenses can be composed, which gives us \caty\ \lenscat, whose objects are \caties\ and arrows are lenses. In the sketch \fwk, $f$-indexing considered above gives rise to a functor \frar{\schfun}{\lenscat}{\SkecatGG} (over a given constraint signature \csig). Moreover, this functor is a fibration or, equivalently, we have a pseudo-functor \frar{\lenscat_{\_}}{\SkecatGG}{\catcat}, for which $L_S=\InstcatS$ and \frar{L_f}{\InstcatS}{\InstcatSprim} 
for \flar{f}{S}{S'} is defined as follows. Its 
 \get-functor is the model reduct functor we considered above, $\get_f= f_\Delta^*$. Its \putl-cofunctor, $\putl_f$, is the cofunctor freely generated by $\get_f$.  \cite{bryce-act23}[Bryce Clark's paper for ACT'23], Compositionality $L_{f,f'}=L_{f}.L_{f'} $ is obvious. 

Importantly, in the \fwk\ outlined above, update policies modelled by cofunctors $\putl_f$ appear to be compositional: $\putl_{f,f'}=\putl_{f'}.\putl_f$. In the usual setting for delta lenses, in which the syntactical schema part of the story is ignored, compositionality of update policies cannot even be formulated. We thus obtain an essential extension of the delta lens \fwk.

\section{Related Work 
        }\label{sec:related}
Sketches as a  graphical counterpart of logical theories were introduced by Ehresmann in 60s and became a classical subject of categorical logic (see \cite{Wells94} for the history and  \cite{BarrWells-TTT85,adamek-lpaCatsBook-94} about expressiveness). Their applications to data modelling can be traced back to 90s 
\cite{LellahiSpyratos1991,
Piessens94},
and more recently to  Johnson \& Rosebrugh's ERA-sketches \cite{JohnsonRD-audb01,JohnsonRosebrugh2002}. Several shortcomings of Ehresmann's sketches for data modelling applications are discussed in \cite{me-entcs08}. 

A suitable generalization of sketches for such applications was introduced by the author \cite{myTR-sketches96%
,Diskin1997} but even earlier, in a different context of an abstract approach to logic and completeness theorems, generalized sketches were invented by Makkai whose preprints had been circulating in early 90s but a journal publication appeared later \cite{makkai-ske}.%
\footnote{It is interesting that Makkai's completeness theorem is nothing but completeness of the injectivity calculus \cref{sec:injectivityPrimer}, which was independently reinvented by Adamek \etal).} 

Using generalized sketches as a foundation for visual object-oriented modelling \cite{
 Diskin-diagrams00} and software engineering \cite{me-entcs08} was proposed by Diskin and Wolter and further developed in a series of papers by Rutle \etal\  \cite{rutleRLW-jlamp12}; see also applications of sketches to an important Model-Driven Engineering problem of model consistency \cite{me-ecmfa17,patrick-faoc21}
 
Mathematical theory of generalized sketches did not attract much (if at all) attention of the categorical community. The author is not aware of any  work besides the early Makkai's work \cite{makkai-ske} and then Diskin \& Wolter's \cite{me-entcs08} and Wolter \cite{uwe-corr21}. The present paper subsumes all results of \cite{me-entcs08} in a much more general setting, while paper \cite{uwe-corr21} considers a different aspect of the sketch formalism. 


\section{Conclusions}

A vast diversity of constraint \specifin\ languages 
essentially contributes to the grand \interopy\ problem. Clearly, having an abstract unifying \fwk\ would help to manage the issue. A well known \fwk\ of this type is Goguen and Burstall's institutions, but they entirely ignore the locality feature of practical constraint languages. In this sense, sketches appear as a reasonable middle ground: they place constraints' locality at the centre but otherwise are fully abstract: a majority of constraint languages in use can be seen as implementations of sketches. 

The paper also showed that the sketch \fwk\ is readily extensible: two new important features of satisfaction's verification and instance (and earlier constraint) indexing were added and provided new application possibilities.  
Sketches are also an interesting mathematical subject. Especially intriguing is a series of results showing that if the category of graphs \GG\ (underlying sketches) is a presheaf topos, then the \caty\ of sketches over \GG\ is a presheaf topos as well. The first such result was discovered and proved by Makkai \cite{makkai-ske} for discrete signatures, and generalized for signatures with dependencies in \cite{myTR33-sketches}. Adapting these results for enriched sketches developed in the paper is left for future work.  
%

\bibliographystyle{eptcs}\label{sect:bib}
\bibliography{../library,../refsGrand17-21}

\appendix 
\renewcommand{\arrcatC}{\enma{\CC^\twocat}}
\renewcommand{\fancydom}[1]{\enma{\dom( #1)}}
\renewcommand{\fancycod}[1]{\enma{\cod( #1)}}

\bigskip
\noindent {\huge\bf Appendix}

\section{Injectivity primer}\label{sec:injectivityPrimer}

Injectivity is a far reaching generalization of regular logic, whose sentences are built according to the pattern $\forall \bar x (\exists \bar y \bigwedge_{i\le m}P_i(\bar x, \bar y) \rightarrow \exists \bar z \bigwedge_{j\le n}Q_j(\bar x, \bar z))$, where $P_i$, $Q_j$ are atomic predicates. Such an implication is modelled  by a \mor, and tuples of variables $\bar x$, $\bar y$, $\bar z$  are \mor s too (the idea goes back to Banashewsky and Herrlich's work on quasi-implications and quasi-variaties in universal algebra \cite{?}). 
\begin{defN}[Injectivity (well-known, \eg,  \cite{adamek-lpaCatsBook-94})]
	Let \frar{f}{P}{Q} be an arrow in an arbitrary category \CC. An object $A\in\Ob\CC$ is {\em injective} \wrt\ $f$ if for any \mor\ \frar{x}{P}{A} there is \mor\ \frar{y}{Q}{A} such that $x=f.y$. Then we write $A\modelsInjX{\CC}f$ and interpret $A$ as a 'model of formula' $f$. We will omit one or both scripts unless we deal with different carrier categories or different $\models$-relations in the same category. 
	
	A binary relation $\models\,\subset\Ob \CC\times\Arr \CC$ gives rise to two mappings: from a set of arrows (theory) $\Fset \subset\Arr\CC$ to its class of models 
	$\InjMod(\Fset) =\compr{A\in\Ob\CC}{A\models{} f \text{ for all } f\in \Fset}$, and from a set of objects (models) \nia\ to its theory $\InjTh(\nia)=\compr{f\in\Arr\CC}{A\models f \text{ for all } A\in\nia}$. 
	We write $\Fset\models f$ if $A\in\InjMod(\Fset)$ implies $A\models f$ for any object $A$.
	\qed\end{defN}
The 'model theory and logic' of injectivity are well studied. Closure Birkgoff-style operators for classes of objects are built by Rosicky \etal \cite{adamek-injBirkhoff-2002}. Sound and complete deductive calculi for injectivity (both finitary and infinitary)  are found by Ad{\' a}mek \etal \cite{adamek-injLogic-2007}. The calculus for finitary injectivity (when all formula-\mor s are finitary, \ie, their domains and codomains are finitely presented) is especially simple and elegant: it is  a set \dedSys\  (read it as {\em Injectivity Logic}) of the following four inference rules:
\begin{equation}\label{eq:injLogic-rules}
	\begin{tabular}[f]{cccc}
		$\dfrac{f_1,\; f_2}{f_1.f_2}$
		& $\dfrac{}{\id_A}$
		& $\dfrac{f_1.f_2}{f_1}$
		& $\dfrac{~f~}{~f'~}$ for every pushout $(g,f,g',f')$
		\\ [20pt] 
		a) {\sc Composition}
		& b) {\sc Idenity}
		& c) {\sc Cancellation}
		& d) {\sc PushOut}
		\\ &&&(validity of $g$ is not required)
	\end{tabular}
\end{equation}
The rules give rise to an entailment relation $\Fset\vdash^\dedSys f$, whose soundness is straightforward to check: if $\Fset\vdash^\dedSys f$ then $\Fset\modelsInj{}f$.  Specifically, {\sc Coproduct} rule specified in \cref{eqdiag:coproduct-derived}(a) can be derived as shown in diagrams\cref{eqdiag:coproduct-derived}(b$_1$), (b$_2$) and (c):  
\begin{equation}\label{eqdiag:coproduct-derived}
	\begin{tabular}{c@{\qquad}c@{\qquad}c}
		$\dfrac{f_1, f_2}{f_1+f_2}$
		&    
		\begin{tikzcd}
			P_1\ar[r, "f_1"] \ar[d, "\iota_{_{P_1}}" '] 
			& Q_1\ar[d, derived, "(\iota_{_{P_1}})' "]
			\\
			P_1+P_2 \ar[r, derived, "(f_1+P_2) "] & \dernode{Q_1+P_2}
		\end{tikzcd}
		& 
		\begin{tikzcd}
			P_1+P_2 \ar[r, "f_1+P_2"] \ar[d, "P_1+f_2" '] 
			& Q_1+P_2\ar[d, derived, "(P_1+f_2)' "]
			\\
			P_1+Q_2 \ar[r, derived, "(f_1+P_2)' "  '] 
			& Q_1+Q_2\ar[lu, {<-}, derived, "\gapnamegap{-1}{f_1+f_2}{0}" ', pos=0.35
			]
		\end{tikzcd}
		\\ [7.5pt]
		a) {\sc Coproduct} 
		&  b$_1$) apply {\sc Pushout}
		& c) apply {\sc Pushout} (twice) 
		\\
		& b$_2$) the same  for $P_1+f_2$
		&  followed by {\sc Composition}
	\end{tabular}
\end{equation}
The system \dedSys\ is also complete: semantic entailment $\Fset\modelsInj f$ implies syntactic derivability (inference) $\Fset\vdash^\dedSys f$, which is of course much more difficult to prove \cite{adamek-injLogic-2007}. 
\shortlong{}{\zd{I guess it's not needed but let it be here for the record. If $t$ is mono, then commutativity of the lower triangle implies comm. of the upper one. If $h_\tauu$ is epi, then the converse holds.}
}

\section{Arrow \caties}

\begin{mygroup}
\renewcommand{\CC}{\enma{\mathbf{C}}}

%

\subsection{Several facts about arrow \caties}\label{sec:arrowCats}
Let \twocat\ be a fixed \caty\  having two objects 0 and 1, and the only non-identity arrow \frar{01}{0}{1} (it is often called the {\em interval} category) and denoted $I$). 
\begin{constrN}[Arrow Categories]
\label{def:arrcats}
 Any category \CC\ gives rise to an {\em arrow \caty} \arrcatC\ whose objects are \CC-arrows and \mor s are commutative squares:
 given two \CC-arrows $a,b$, their morphism is a pair of \CC-arrows, \frar{f}{\fancydom{a}}{\fancydom{b}} and \frar{f'}{\fancycod{a}}{\fancycod{b}}, such that the square commutes: $a.f' = f.b$. 
 We call such a pair an {\em sq-arrow} and write \drar{(f,f')}{a}{b}.
 
\end{constrN}
The following facts can be proven easily for any category \CC:
\begin{lemma}
	\label{lemma:arrcatfibr}
 There are two obvious functors \frar{\domC}{\arrcatC}{\CC} and \frar{\codC}{\arrcatC}{\CC}, for which:
  \begin{enumerate}[label=\roman*)]
 \item Functor \domC\ is a {\em split} fibration (lifts are provided by  precomposition).
 \item The  restriction \frar{\dom_X}{\CC/X}{\CC} (for a given object $X$) is a {\em discrete} fibration.
\item Functor \codC\ is a {\em split} opfibration (lifts are provided by postcomposition)
\item The restriction \frar{\cod_X}{\cosliceCat{\CC}{X}}{\CC} is a {\em discrete} opfibration, 
 \end{enumerate}
\end{lemma}
\begin{lemma}
	\label{lemma:arrcatfibrPBPO} If, in addition, \CC\ has
\begin{enumerate}[label=\roman*)]
\item all pushouts, then functor \domC\ is also an opfibration, where lifts are constructed via pushouts.
\item all pullbacks, then functor  \codC\ is also a fibration, where lifts are constructed via pullbacks. 
\end{enumerate}
\end{lemma}

\end{mygroup}

\end{document}